\documentclass[opre,nonblindrev]{informs3}

\DoubleSpacedXI

\usepackage{endnotes}
\let\footnote=\endnote

\usepackage{amsmath, amssymb} 
\usepackage[prologue,svgnames]{xcolor} 
\usepackage[all,warning]{onlyamsmath}
\usepackage{graphics, graphicx} 
\usepackage{booktabs} 
\usepackage{units} 
\usepackage{microtype} 
\usepackage{enumitem} 
\usepackage[margin=1in]{geometry}

\usepackage{fancyhdr} 
\usepackage{url} 
\usepackage[normalem]{ulem}
\usepackage{subcaption} 
\usepackage{wrapfig} 
\usepackage{tikz} 
\usetikzlibrary{shapes,decorations}
\usetikzlibrary{arrows}
\usepackage{pgfplots} 
\pgfplotsset{compat=newest} 

\newcommand{\zero}{\ensuremath{\mathbf{0}}}

\newcommand{\ind}{\ensuremath{\mathbf{I}}}
\newcommand{\reals}{\ensuremath{\mathbb{R}}}

\newcommand{\defeq}{\ensuremath{\triangleq}}
\DeclareMathOperator{\prob}{\ensuremath{\mathbf{P}}}
\DeclareMathOperator{\expec}{\ensuremath{\mathbf{E}}}
\DeclareMathOperator{\bE}{\ensuremath{\mathbb{E}}}

\DeclareMathOperator{\var}{\ensuremath{\mathrm{Var}}}
\newcommand{\conv}{\ensuremath{\mathsf{Conv}}}
\newcommand{\cara}{\ensuremath{\mathsf{Cara}}}
\newcommand{\calP}{\ensuremath{\mathcal{P}}}
\newcommand{\join}{\ensuremath{\mathsf{Join}}}
\newcommand{\leave}{\ensuremath{\mathsf{Leave}}}

\usepackage{natbib}
 \bibpunct[, ]{(}{)}{,}{a}{}{,}%
 \def\bibfont{\small}%
\TheoremsNumberedThrough     
\ECRepeatTheorems

\EquationsNumberedThrough    

\begin{document}


\RUNAUTHOR{Anunrojwong, Iyer, and Lingenbrink}

\RUNTITLE{Persuading Risk-Conscious Agents}

\TITLE{Persuading Risk-Conscious Agents: A Geometric Approach}

\ARTICLEAUTHORS{
\AUTHOR{Jerry Anunrojwong}
\AFF{Columbia Business School, \EMAIL{jerryanunroj@gmail.com}}
\AUTHOR{Krishnamurthy Iyer}
\AFF{Industrial and Systems Engineering, University of Minnesota, Minneapolis, MN 55455, \EMAIL{kriyer@umn.edu}} 
\AUTHOR{David Lingenbrink}
\AFF{School of Operations Research and Information Engineering, Cornell
 University, Ithaca, NY 14853, \EMAIL{dal299@cornell.edu}}
}

\ABSTRACT{

We consider a persuasion problem between a sender and a receiver whose
utility may be nonlinear in her belief; we call such receivers {\em
  risk-conscious}. Such utility models arise when the receiver
exhibits systematic biases away from expected-utility-maximization,
such as uncertainty aversion (e.g., from sensitivity to the variance
of the waiting time for a service). Due to this nonlinearity, the
standard approach to finding the optimal persuasion mechanism using
revelation principle fails. To overcome this difficulty, we use the
underlying geometry of the problem to develop a convex optimization
framework to find the optimal persuasion mechanism. We define the
notion of {\em full persuasion\/} and use our framework to
characterize conditions under which full persuasion can be achieved.
We use our approach to study {\em binary persuasion}, where the
receiver has two actions and the sender strictly prefers one of them
at every state. Under a convexity assumption, we show that the binary
persuasion problem reduces to a linear program, and establish a
canonical set of signals where each signal either reveals the state or
induces in the receiver uncertainty between two states. Finally, we
discuss the broader applicability of our methods to more general
contexts, and illustrate our methodology by studying information
sharing of waiting times in service systems.



}

\AREAOFREVIEW{Revenue Management and Market Analytics}
\SUBJECTCLASS{Games: Bayesian persuasion, information design;
  Utility/preference: non-expected utility; Programming: convex
  optimization}

\maketitle

\section{Introduction}\label{sec:het}

Given the inherent informational asymmetries in online marketplaces
between a platform and its users, information design has an important
role to play in their design and operation. Building on the
methodological contributions of \citet{rayoS10}, \citet{kamenicaG11},
and \citet{bergemannM16}, information design has been applied in a
number of different application contexts, such as
engagement/misinformation in social networks~\citep{candoganD17},
service systems~\citep{lingenbrinkI19}, and online
retail~\citep{lingenbrinkI18a, drakopoulosJR18}.

In much of the previous work, the standard assumption is that the
agent being persuaded (the {\em receiver\/}), is an expected utility
maximizer (EUM). Although this assumption is well-supported
theoretically via axiomatic characterizations~\citep{savage54}, it is
empirically well-documented that human behavior is inadequately
explained by the central tenets of the theory~\citep{ellsberg61,
  allais79, rabin98, dellavigna09}. In particular, there is a long
line of work in economics studying the systematic biases in human
behavior leading to deviations from expected utility
maximization~\citep{kahnemanT72,kahnemanT79, machina82, tverskyK92}.
Because of these shortcomings, existing models of Bayesian persuasion
may not satisfactorily apply to information design problems in online
markets and other practical settings.

Motivated by this concern, our main goal in this paper is to extend
the methodology of Bayesian persuasion to settings where the receiver
may not be an expected utility maximizer. In our general model of
utility under uncertainty, given a finite state space $\Omega$ and a
set of actions $A$, we take as model primitive the utility function
$\rho(\mu, a)$ that specifies the receiver's utility for action
$a \in A$ under belief $\mu$. For EUM receivers, this function is
given by
$\rho(\mu, a) \defeq \sum_{\omega \in \Omega} \mu(\omega) u(\omega,a)$
for some function $u$, and thus $\rho$ is {\em linear\/} in the
belief. Our framework relaxes this linearity assumption, and allows
for the utility function $\rho$ to depend non-linearly on the belief.
We refer to such general receivers as {\em risk-conscious}.


\begin{figure}[t]
  \centering
  \includegraphics[scale=0.32]{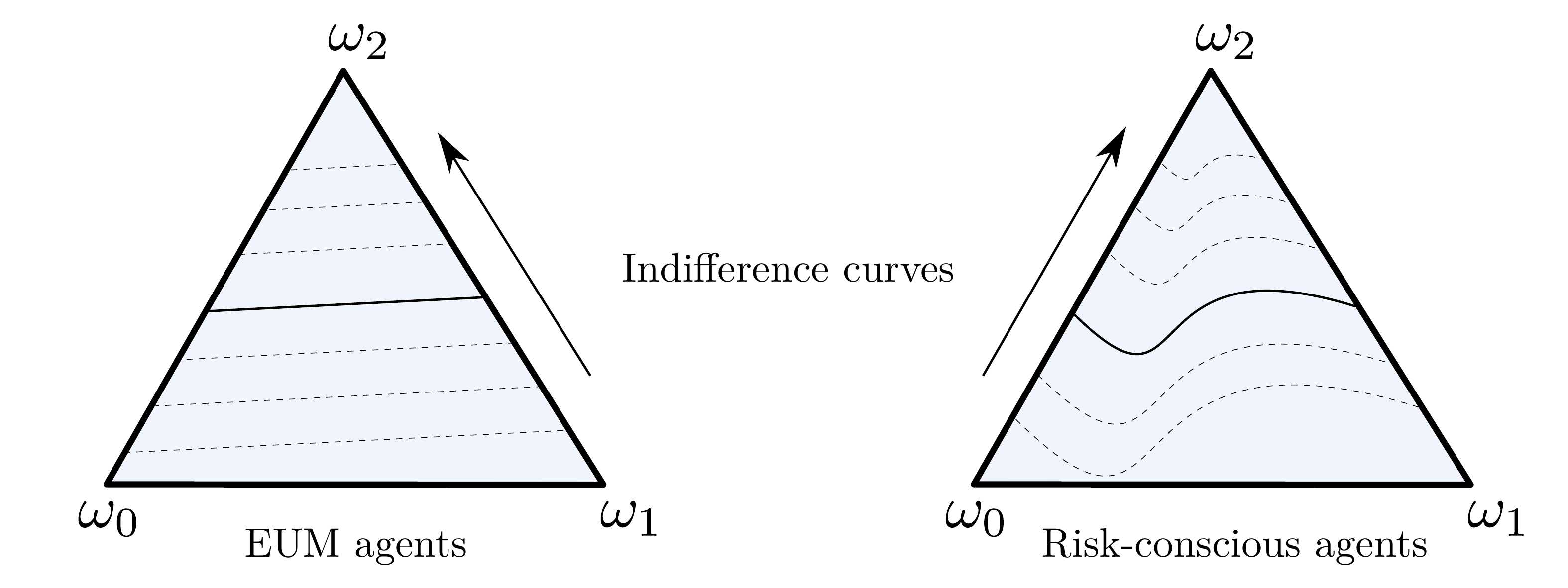}
  \caption{Indifference curves in the belief space. Here, the triangle
    denotes the simplex of beliefs over
    $\Omega = \{\omega_0, \omega_1, \omega_2\}$, and the lines/curves
    denote the beliefs with a fixed utility. The indifference curves
    for EUM receivers are necessarily hyperplanes, whereas they need
    not be so for risk-conscious
    receivers.}\label{fig:eum-vs-risk-conscious-indifference}
\end{figure}

\subsection{Motivation for
    Risk-consciousness}\label{subsec:motivate-risk-consciousness}

  While the standard model of expected utility maximization is
  subsumed in the risk-conscious framework, the latter allows for far
  more generality. Below we provide three motivation for studying
  risk-conscious behavior.

  \textbf{Deviations from EUM:} The literal interpretation of
  risk-consciousness is that the agent's behavior deviates
  systematically from expected utility maximization.  Such deviations
  are well-documented in the literature; we provide a few examples in
  Table~\ref{table:risk-conscious-examples} and discuss them in
  Appendix~\ref{ap:deviations-eum}. For instance, an agent with
  uncertainty-aversion may be modeled using mean-standard deviation
  utility. Thus, there is a need to develop a theory for the
  persuasion of such agents, both to better capture realistic behavior
  in operational settings and to gain qualitative insights into
  phenomena that do not arise with EUM receivers.

  \textbf{Dynamic decision making:} A second source of risk-conscious
  behavior comes from dynamic settings where the agent must make
  multiple decisions over time, with past decisions influencing the
  information available to her in the future. For any fixed action in
  the present, a change in the agent's belief may alter her actions in
  the future, thereby affecting her future (expected) payoffs. The net
  effect on the agent's utility of a change in belief may then be
  non-linear, thus leading to risk-consciousness. (We provide details
  in Section~\ref{sec:discussion}.) While such models arise in a
  number of operational settings, a concrete example that has received
  recent attention is the market for data and information
  products~\citep{bergemannBS18,bergemannB15,bergemannB19,zhengC21}. A
  buyer in such a market often uses the data to make subsequent
  decisions; thus, the buyer's utility for the data may depend
  non-linearly in her belief about the data quality. A seller who
  wishes to persuade such a buyer into purchasing the data then faces
  the problem of risk-conscious persuasion.

  \textbf{Modeling device:} Finally, risk-consciousness can be a
  useful modeling tool to analyze more complex settings.  In
  particular, using the risk-conscious framework, one can analyze
  instances of \textit{public persuasion}, where a sender seeks to
  persuade a group of receivers by sending a common public signal;
  such instances are common in markets, public services, platforms,
  and social networks, where practical concerns such as information
  leakage and fairness preclude private
  persuasion~\citep{candoganD17,candogan19,
    yangIF19,anunrojwongIM20}. Similarly, the framework can be applied
  to study {\em robust persuasion}, where a receiver with a private
  type is persuaded by a sender taking a worst-case view. Robust
  persuasion is important in market and platform services, where
  participants often have idiosyncratic components to their payoffs
  that are unknown to the
  platform~\citep{bimpikisP19bookchapter,candogan20tutorial}.

%

\begin{table}
  \caption{Examples of risk-conscious
    utility} \label{table:risk-conscious-examples} \centering
  \begin{tabular}{lccc}
    \toprule
    Risk-conscious utility & & Utility representation $\rho(\mu,a)$ &  \\ 
    \midrule
    Expected utility & & $\expec_{\mu}[u(\omega,a)]$ & \\ 
    Maximin utility & & $\min_{\theta} \expec_{\mu}[u(\omega,a;\theta)]$ & \\
    Mean-standard deviation utility & & $\expec_{\mu}[u(\omega,a)] - \beta \sqrt{\var_{\mu}(g(\omega,a))}$ & \\
    Value-at-Risk & & $-\min\{t \in \mathbb{R}: \prob_{\mu}(\ell(\omega,a) > t) \leq 1-\alpha\}$  & \\
    Conditional Value-at-Risk & &  $-\expec_{\mu}[\ell(\omega,a) | \ell(\omega,a) > \tau]$ &  \\
    Cumulative prospect theory & & $\sum_{\omega} f_{\omega}(\mu) u(\omega,a)$ &  \\
    \bottomrule
  \end{tabular}
\end{table}


\subsection{Challenges and Opportunities}

From a theoretical perspective, optimal persuasion of risk-conscious
agents presents new analytical challenges. When agents are expected
utility maximizers, a revelation-principle style argument is often
invoked to reduce the set of possible messages the sender might send
(i.e., {\em signals}) to the set of actions available to the receiver.
By pairing each signal directly with the action taken by the receiver,
this reduction simplifies the persuasion problem substantially, and
the resulting optimization problem can be written as a linear program
with one {\em obedience} constraint for each action. In contrast, with
risk-conscious agents, due to the nonlinearity of the receiver's
utility, a key step of the revelation principle argument fails (as we
describe in Section~\ref{sec:tractable}), rendering this approach to
finding an optimal {\em signaling scheme\/} ineffective. The lack of a
convenient set of signals, where each signal is paired with an optimal
action, presents analytical and computational challenges to
persuasion.

Despite these challenges, the setting with risk-conscious agents also
yields new possibilities for persuasion. We illustrate this aspect
with the following example.
\begin{example}\label{ex:full-persuade} Consider a sender who
    seeks to persuade a receiver into taking an action. Let there be
    four payoff-relevant states $\omega \in \Omega = \{0,1,2,3\}$. The
    sender and the receiver's prior belief is given by
    $\mu^* = \left( \frac{1}{4}, \frac{1}{4}, \frac{1}{4}, \frac{1}{4}
    \right)$.  For any belief $\mu = (\mu_0, \mu_1, \mu_2, \mu_3)$,
    the receiver's utility for taking the action equals
    $\hat{\rho}(\mu) \defeq \max_{\omega \in \{0,1,2\}} (\mu_{\omega}
    - \tfrac{2}{3})$; the receiver takes the action if and only if the
    utility is non-negative.
    
    Since $\hat{\rho}(\mu^*) < 0$, without persuasion the receiver
    will not take the action.  As
    $\hat{\rho}(e_\omega) = \tfrac{1}{3} > 0$ for
    $\omega \in \{0,1,2\}$ and $\hat{\rho}(e_3) = -\tfrac{2}{3} < 0$,
    under the full-information scheme that reveals the state, the
    receiver takes the action only in states $\omega \in \{0,1,2\}$.
    (Here, $e_\omega$ denotes the belief that puts all its weight on
    $\omega \in \Omega$.)

    Now, consider the following signaling scheme that uses three
    signals $\{s_0, s_1, s_2\}$: when the state is
    $\omega = i \in \{0,1,2\}$, the scheme sends the signal $s_i$, and
    when the state is $\omega = 3$, the scheme sends each signal with
    equal probability. One can verify that upon receiving the signal
    $s_i$, the receiver's posterior belief equals
    $\mu^i = \tfrac{3}{4} e_i + \tfrac{1}{4} e_3$, with
    $\hat{\rho}(\mu^i) = \tfrac{1}{12} > 0$. This implies that,
    regardless of the signal received, the receiver finds it optimal
    to take the action. Thus, the preceding signaling scheme is
    optimal, and \textit{fully} persuades the receiver into taking the
    action.

    Note that the optimal scheme requires three signals, which is
    strictly more than the number of choices available to
    receiver. (It can be verified that under any signaling scheme that
    uses $2$ signals, the probability of the receiver taking the
    action is at most $\tfrac{3}{8}$.)  This is in contrast with
    expected utility maximizing receivers, where the revelation
    principle states that action recommendations (and hence $2$
    signals) suffice for optimal persuasion.\Halmos
  \end{example}

In this example, with optimal persuasion, the receiver takes the
sender's most preferred action at each state. We refer to this outcome
as {\em full persuasion\/}, and study it in detail in
Section~\ref{sec:benefits-persuasion}. Observe that here, full
persuasion occurs despite the fact that under her prior the receiver
would choose otherwise. This is due to the nonlinearity inherent in
the receiver's utility, and in Theorem~\ref{thm:full-persuasion} we
show that this cannot occur when the receiver is an expected utility
maximizer.

\subsection{Main Contributions}

The main contributions of our paper are as follows: (1) we
  provide a convex optimization framework to overcome the analytical
  challenges in the persuasion of risk-conscious agents; (2) using
  this framework, we identify conditions under which persuasion is
  beneficial to the sender, and through the notion of full persuasion,
  we provide insights into the extent of this benefit; and (3) we
  demonstrate the structural properties of the resulting optimal
  persuasion mechanisms, by examining settings that impose additional
  regularity assumptions on the receiver's utility.

  The convex programming framework we develop in
  Section~\ref{sec:tractable} uses the underlying geometry of the
  persuasion problem to optimize directly over the joint distribution
  of the state and the receiver's actions. These variables are shown
  to lie in the convex hull of the set of beliefs for which a fixed
  action is optimal for the receiver. The optimal persuasion mechanism
  is then obtained through a convex decomposition of the optimal
  solution to the convex program. While related, our approach is
  different from the concavification approach~\citep{kamenicaG11},
  yielding a different convex program; we discuss the connection in
  detail in Appendix~\ref{ap:kg-comparison}.

  Using this framework, in Section~\ref{sec:benefits-persuasion}, we
  first characterize conditions under which the sender strictly
  benefits from the persuasion of a risk-conscious receiver. This
  result is analogous to, and extends, similar characterization for
  expected-utility-maximizers~\citep{kamenicaG11}. Specific to the
  risk-conscious framework, we formally define the notion of {\em full
    persuasion\/} as a measure of the extent of the benefits to
  persuasion, and provide necessary and sufficient conditions under
  which full persuasion is achievable.

  To obtain more insight into the structure of the optimal persuasion
  mechanism, we then study a specialized setting in
  Section~\ref{sec:binary-persuasion}, namely {\em binary persuasion},
  in which the receiver has two actions ($0$ and $1$), and the sender
  always prefers that the receiver choose action $1$ over action
  $0$. Under a convexity assumption on the receiver's utility
  function, we show that the convex program in fact reduces to a
  linear program whose solution can be efficiently computed. By
  analyzing this linear program, we establish a {\em canonical set\/}
  of signals for optimal persuasion. In other words, we show there
  exists an optimal signaling scheme that always sends signals in this
  canonical set for any prior belief of the receiver. This canonical
  set of signals consists of {\em pure} signals, which fully reveal
  the state to the receiver, and {\em binary mixed} signals, which
  induce uncertainty between two states. With additional monotonicity
  assumptions on the utility function, we show that the optimal
  signaling scheme induces a threshold structure in the receiver's
  action.

  In summary, our work provides a methodology to solve for the optimal
  persuasion mechanism with risk-conscious agents and demonstrates the
  fruitfulness of analyzing more realistic models of human behavior.
  In Section~\ref{sec:discussion}, we provide a brief discussion of
  the use of our approach in more general settings, beyond the
  persuasion of a risk-conscious receiver. Finally, as an illustration
  of our methodology, in Section~\ref{sec:signaling-queues} we analyze
  a model of a queueing system where the service provider seeks to
  persuade arriving risk-conscious customers to join an unobservable
  single server queue, and establish the intricate ``sandwich''
  structure of the optimal signaling scheme.

\subsection{Literature Review}\label{sec:lit-review}

Our work contributes to the literature on Bayesian persuasion
\citep{rayoS10, kamenicaG11, bergemannM16,bergemannM18, taneva19,
  kolotilinMZL17, dughmiX16}, where a sender \emph{commits} to a
mechanism of sharing payoff-relevant information with a receiver in
order to influence the latter's actions. For a recent review of the
literature, see \citet{kamenica18}. Our work particularly takes
influence from \citet{kamenicaG11}, who use a convex-analytic
\emph{concavification} approach to study the persuasion problem; we
discuss the close relation to our work in
Appendix~\ref{ap:kg-comparison}.

In the operations research literature, a number of authors have
applied the methodology of Bayesian persuasion to study varied
settings such as crowdsourced exploration~\citep{papanastasiouBS18},
spatial resource competitions~\citep{yangIF19},
engagement-misinformation trade-offs in online social
networks~\citep{candoganD17, candogan19}, warning policies for
disaster mitigation~\citep{alizamirVW18}, throughput maximization in
queues~\citep{lingenbrinkI19}, inventory/demand signaling in
retail~\citep{lingenbrinkI18a, drakopoulosJR18}, and quality of
matches in matching markets~\citep{romanyukA18}. Our work is inspired
by this stream of work, and seeks to broaden the domain of
applicability by incorporating more general utility models for the
receiver.

Our other source of inspiration is the economics literature on the
theory of individual preferences towards risk. Apart from the standard
expected utility hypothesis, there have been a number of theoretical
frameworks proposed to model preferences under uncertainty, including
the widely studied prospect theory~\citep{kahnemanT79} and the
cumulative prospect theory~\citep{tverskyK92}.  \citet{machina82,
  machina95} singles out the ``independence axiom'' in the expected
utility framework which leads to the utilities being ``linear in the
probabilities'', and study models that relax the independence
axiom. Our notion of risk-consciousness encompasses all of these
non-expected utility models, as it only requires the utility to be
continuous in beliefs. On the other hand, our notion does not capture
some non-expected utility models, such as the maxmin expected utility
with multiple priors~\citep{gilboaS89}.

The notion of risk-consciousness is related to the concept of {\em
  risk measures\/} in mathematical finance~\citep{artznerADEH01,
  follmerS11}. Commonly studied risk-measures, such as the variance of
the portfolio return~\citep{markowitz52}, the
value-at-risk~\citep{jorion06}, the expected
shortfall~\citep{acerbiT12}, and the entropic
value-at-risk~\citep{ahmadijavid12} are all nonlinear functions of the
distribution of the return. We note that to be a good measure of
financial risk, a risk measure needs to satisfy a number of properties
(e.g., \emph{coherence}~\citep{artznerADEH01}) that make sense in the
context of portfolio management but may not be relevant for capturing
aspects of human decision-making.

We end our discussion by mentioning two closely related recent
works. \citet{beaucheneBLL19} study a persuasion setting where a
sender uses an {\em ambiguous\/} communication device comprising
multiple ways of sending signals. The receiver is ambiguity-averse and
has the maxmin expected utility~\citep{gilboaS89} over the multiple
resulting posteriors. As mentioned earlier, such maxmin-utility
preferences are distinct from the class of risk-conscious receivers
considered in this paper. The authors analyze optimal persuasion, and
parallel to our result, note that full persuasion can be achieved
through the use of ambiguous communication device. In contrast, our
work demonstrates that full persuasion can be obtained from
unambiguous communication, purely due to risk-conscious preferences.

\citet{lipnowskiM18} consider persuasion of a receiver with similar
nonlinear preferences as in this paper, but focus on the setting where
the sender's preferences are perfectly aligned with those of the
receiver (for instance, the sender might be a trusted advisor). Noting
the failure of the revelation principle, the authors provide a
sufficient condition, namely that the receiver's payoffs are concave
in the belief for any fixed action, under which action recommendations
are optimal. In our intended applications, the sender is a platform or
a marketplace, and hence we model the sender as an expected utility
maximizer. Due to this, apart from trivial instances of our model, the
sender's preferences will not be aligned with those of the receiver.

\section{Model}\label{sec:model}

In the following, we present the model of Bayesian persuasion with
risk-conscious agents. Our development of the model follows closely to
that of the standard Bayesian persuasion
setting~\citep{kamenicaG11,kamenica18}.


\subsection{Setup}\label{subsec:setup}

We consider a persuasion problem with one \textit{sender} and one
\textit{receiver}. Let $X$ be a payoff-relevant random variable with
support on a known set $\mathcal{X}$. We assume that neither the
receiver nor the sender observes $X$. However, as we describe below,
the sender has more information about $X$ than the receiver, and seeks
to use this information to influence the receiver's actions.

Formally, we assume that the distribution of $X$ depends on the {\em
  state} of the world $\bar{\omega}$ which takes values in a finite
set $\Omega$ and which is observed by the sender but not the receiver.
We denote the distribution of $X$, conditional on
$\bar{\omega} =\omega$, by $F_\omega$. The distributions
$\{F_\omega : \omega \in \Omega\}$ are commonly known between the
sender and the receiver, and both share a common prior
$\mu^* \in \Delta(\Omega)$ about the state of the world
$\bar{\omega}$. (Throughout, for any set $S$, we let $\Delta(S)$
denote the set of probability measures over $S$. When $S$ is finite,
we consider $\Delta(S)$ as a subset of $\reals^{|S|}$, endowed with
the Euclidean topology.)
For each $\mu \in \Delta(\Omega)$, we let $F_\mu$ be the distribution
of $X$ when $\bar{\omega}$ is distributed as $\mu$: we have
$F_\mu = \sum_{\omega} \mu(\omega) F_{\omega}$.  Finally, we let
$X_\omega$ denote an independent random variable distributed as
$F_\omega$.

As in the standard Bayesian persuasion setting, we assume that the
receiver is Bayesian and that the sender can commit to a
\textit{signaling scheme} to influence receiver's choice of an action
(which is described below in detail). A signaling scheme $(S, \pi)$
consists of a signal space $S$ and a joint distribution
$\pi \in \Delta(\Omega \times S)$ such that the marginal of $\pi$ over
$\Omega$ equals $\mu^*$: for each $\omega \in \Omega$,
$\pi(\omega, S) = \mu^*(\omega)$. Specifically, under the signaling
scheme $(S, \pi)$, if the realized state is $\bar{\omega} = \omega$,
the sender draws a signal $\bar{s} \in S$ according to the conditional
distribution $\pi(\cdot| \bar{\omega} = \omega)$, and conveys it to
the receiver. For simplicity of notation, we denote a signaling scheme
$(S, \pi)$ by the joint distribution $\pi$.
Throughout, we assume that the sender commits to a signaling scheme
prior to observing the state $\bar{\omega}$, and that the sender's
choice of the signaling scheme $\pi$ is common knowledge between the
sender and the receiver.


As mentioned above, we assume that the receiver is Bayesian. Given the
sender's signaling scheme $\pi$, upon observing the signal
$\bar{s} = s$, the receiver uses Bayes' rule to update her belief from
her prior $\mu^*$ to the posterior $\mu_s \in \Delta(\Omega)$. In
particular, we have for all $\omega \in \Omega$,
\begin{align*}
  \mu_s(\omega ) = \frac{\pi(\omega, s)}{\sum_{\omega' \in \Omega} \pi(\omega',s)},
\end{align*}
whenever the denominator on the right-hand side is positive. (We let
$\mu_s \in \Delta(\Omega)$ be arbitrary if the denominator is zero.)
This implies that upon receiving the signal $\bar{s} = s$, the
receiver believes that the payoff-relevant variable $X$ is distributed
as $F_{\mu_s}$.

\subsection{Actions, Strategy and Utility}



Upon observing the signal $\bar{s}$, the receiver chooses an action
$a$ from a finite set $A$ of actions. Given a signaling scheme $\pi$,
the receiver's strategy $a(\cdot)$ specifies an action $a(s) \in A$
for each realization $s \in S$ of the signal $\bar{s}$. (Although our
definition implies a pure strategy, we can easily incorporate mixed
strategies where the receiver chooses an action at random. We suppress
this technicality for the sake of readability.)

We let $v(\omega, a)$ denote the sender's utility in state
$\bar{\omega} = \omega$ when the receiver chooses the action
$a \in A$. Furthermore, we assume that the sender is an expected
utility maximizer. (One can equivalently represent the utility
function $v(\omega, a)$ as an expectation of a utility function
$\hat{v}(X, a)$ over the payoff-relevant variable $X$ and the action
$a$, conditional on $\bar{\omega} =\omega$; we suppress the details
for brevity.)

Our point of departure from the standard persuasion framework is in
the definition of the receiver's utility. Specifically, we relax the
assumption that the receiver is an expected utility maximizer; as we
describe next, our setup allows for more general models of the
receiver's utility over the uncertain outcome $X$. We refer to such
receivers as being {\em risk-conscious}.

Formally, for any belief $\mu \in \Delta$ of the receiver, we assume
that the receiver's {\em utility} upon taking an action $a \in A$ is
given by $\breve{\rho}(F_\mu, a) \in \reals$. For notational simplicity,
we define the {\em utility function}
$\rho : \Delta(\Omega) \times A \to \reals$ as
$\rho(\mu,a) \defeq \breve{\rho}(F_\mu, a)$. Given a belief
$\mu \in \Delta(\Omega)$, we assume that the receiver chooses an
action $a \in A$ that achieves the highest utility $\rho(\mu, a)$.

Observe that a receiver is an expected utility maximizer if and only
if, for each $a \in A$, the utility function $\rho(\mu, a)$ is linear
in $\mu$. (Note that if $\rho(\mu, a)$ is a utility function of an
agent, then so is $g(\rho(\mu, a))$ for any increasing function
$g : \reals \to \reals$. Thus, this linearity holds only up to an
increasing transformation. We suppress such transformations for the
sake of clarity.) In particular, there exists a function
$u : \mathcal{X} \times \Omega \times A \to \reals$ such that
$\rho(\mu, a) = \expec\left[ u(X, \bar{\omega}, a) \right] =
\sum_{\omega \in \Omega} \mu(\omega) \expec[ u(X_\omega, \omega, a)]$
for all $\mu \in \Delta(\Omega)$ and $a \in A$, if and only if the
receiver is an expected utility maximizer. Our setup therefore
includes as a special case the standard Bayesian persuasion framework
with an expected utility maximizing
receiver. 
However, the generality of our setting allows us to capture a much
wider range of receiver behavior.  (We emphasize that the notion of
risk-consciousness is different from, and much more general than, the
conventional notion of risk-aversion, where utility is modeled as the
expectation of a function that is \emph{concave in the payoffs}. A
risk-averse agent is still an expected utility maximizer, and this
expected utility is necessarily \emph{linear in her belief}.)

From the sender's perspective, the only relevant aspect of the
  receiver's belief is the corresponding receiver's action induced by
  that belief. Thus, rather than modeling receiver's utilities
  directly, we could model the receiver as being characterized by the
  sets of beliefs for which she chooses each particular action. In
  other words, instead of the utility function $\rho$, our approach
  could equivalently take as model primitives the sets
  $\{ \calP_a : a \in A\}$, where $\calP_a$ is the set of posterior
  beliefs for which action $a \in A$ is optimal for the receiver:
  \begin{align}\label{eq:sets-pa}
    \calP_a \defeq \left\{ \mu \in \Delta(\Omega) : a \in \arg\max_{a' \in A}
    \rho(\mu, a') \right\}, \quad \text{for each $a \in A$.}
  \end{align}
  As we discuss in Section~\ref{sec:discussion}, this perspective
  allows us to apply our methods to more general settings beyond the
  context of risk-conscious receivers.

As an illustration, consider the setting of a customer deciding
whether or not to wait for service in an unobservable
queue. The receiver's utility depends on her unknown waiting time $X$,
and suppose the queue operator observes some correlated feature
$\omega$ (queue length, congestion, server availability, etc). A
natural risk-conscious customer model posits that the customer only
joins the queue and waits for service if, given her beliefs, the mean
of her waiting time plus a multiple of its standard deviation is below
a threshold~\citep{nikolovaS14, cominettiT16, lianeasNS18}. Such a
behavioral model may arise from the customer's requirement for service
reliability, or from an aversion to uncertainty due to a desire to
plan her day subsequent to service completion. This model can be
captured in our setting by letting
$A = \{\mathsf{join}, \mathsf{leave}\}$ and assuming, for example,
that
$\rho(\mu, \mathsf{join}) = \tau - \left(\expec_\mu[X] + \beta
  \sqrt{\var_\mu(X)}\right)$ for some $ \beta, \tau > 0$, and
$\rho(\mu, \mathsf{leave}) = 0$, implying
$\mathcal{P}_{\mathsf{join}} = \left\{ \mu \in \Delta(\Omega): \tau -
  \left(\expec_\mu[X] + \beta \sqrt{\var_\mu(X)}\right) \geq 0
\right\}$ and
$\mathcal{P}_{\mathsf{leave}} = \left\{ \mu \in \Delta(\Omega): \tau -
  \left(\expec_\mu[X] + \beta \sqrt{\var_\mu(X)}\right) \leq 0
\right\}$. (We use the notation
$\expec_\lambda$ to denote expectation with respect to a distribution
$\lambda$.) It is straightforward to check that
$\rho(\mu,\mathsf{join})$ is not linear in $\mu$.

Throughout this paper, we make the following assumption:
\begin{assumption}\label{as:closed} For each $a \in A$, the set $\calP_a$ is closed.
\end{assumption}
We remark that Assumption~\ref{as:closed} holds if, for each
$a \in A$, the utility function $\rho(\mu, a)$ is continuous in
$\mu$.

\subsection{Persuasion of Risk-conscious Agents}

We are now ready to describe the sender's persuasion problem. First,
we require that for any choice of the signaling scheme $\pi$, the
receiver's strategy maximizes her utility with respect to her
posterior beliefs: for each $s \in S$, we have
\begin{align}\label{eq:receiver-opt}
  a(s) \in \argmax_{a \in A} \rho(\mu_s, a).
\end{align}
We call any strategy that satisfies~\eqref{eq:receiver-opt} an optimal
strategy for the receiver. Given an optimal strategy $a(\cdot)$, the
sender's expected utility for choosing a signaling scheme $\pi$ is
given by $\expec_\pi \left[ v(\bar{\omega}, a(\bar{s}))\right]$, where
$\expec_\pi$ denotes the expectation over $(\bar{\omega}, \bar{s})$
with respect to $\pi$.  The sender seeks to choose a signaling scheme
$\pi$ that maximizes her expected utility, assuming that the receiver
responds with an optimal strategy. (When the receiver has multiple
optimal strategies, we assume that the sender chooses her most
preferred one; the literature refers to this as the sender-preferred
subgame-perfect equilibrium~\citep{kamenicaG11}.) Thus, the sender's
problem can be posed as
\begin{equation}\label{eq:sender-opt}
  \begin{aligned}
    \max_{\pi \in \Delta(\Omega \times S)}    & \expec_\pi [ v(\bar{\omega}, a(\bar{s})) ]\\
    \text{subject to, } \quad a(s) &\in \argmax_{a \in A} \rho(\mu_s,
    a), \quad \text{for all $s \in S$,}\\
    \pi(\omega, S) &= \mu^*(\omega), \quad \text{for all
      $\omega \in \Omega$.}
\end{aligned}
\end{equation}
Our main goal in this paper is to find and characterize the sender's
optimal signaling scheme to the persuasion
problem~\eqref{eq:sender-opt}. Note that the problem as posed is
computationally challenging, as it requires first choosing an optimal
set of signals $S$ and then a joint distribution $\pi$ over
$\Omega \times S$. Without an explicit handle on the set $S$ and the
resulting receiver actions, the persuasion problem seems
intractable. In the next section, we reframe the problem to obtain a
tractable formulation.

\section{Towards a Tractable Formulation}\label{sec:tractable}

When the receiver is treated as an expected utility maximizer, a
revelation-principle style argument is typically
invoked~\citep{bergemannM16} to restrict attention to signaling
schemes that use the set of actions $A$ as the signaling space $S$,
such that the receiver upon seeing a signal $a \in A$ finds it optimal
to take action $a$. Before we discuss our approach for general
risk-conscious agents, we provide a more detailed discussion of this
argument, and discuss why it fails in our setting.

\subsection{Failure of the Revelation Principle}

The revelation-principle style argument rests on the following
observation: when the receiver is an expected utility maximizer, if
two signals $s_1$ and $s_2$ both lead to the same optimal action
$a(s_1) = a(s_2) = a$, then $a$ is still an optimal action for the
receiver if the signaling scheme reveals only that
$\bar{s} \in \{s_1, s_2\}$ whenever it was supposed to reveal $s_1$ or
$s_2$. This property is straightforward to show using the linearity of
the utility functions $\rho(\cdot , a)$ for an expected utility
maximizer. One can then use this property to coalesce all signals that
lead to the same optimal action for the receiver into a single
signal. Such a coalesced signaling scheme has at most one signal per
action, which after identifying the signal with the corresponding
action, can be turned into an {\em action recommendation}. Moreover,
for such a signaling scheme, the agent's optimal strategy is {\em
  obedient}, i.e., it is optimal for the agent to follow the action
recommendation.

However, when the receiver is risk-conscious, the preceding argument
may no longer hold. This is because, when signals with the same
optimal action are coalesced, it may alter the posterior of the
receiver on the coalesced signal, and without linearity of
$\rho(\cdot,a)$, the receiver's optimal action may change. (To see
this, consider Example~\ref{ex:full-persuade} in reverse: under each
signal $s_i$, it is optimal for the receiver to take the
action. However, coalescing the three signals is equivalent to
providing no information, and not taking the action is uniquely
optimal for the receiver under her prior belief.)  Thus, it no longer
suffices to consider only those signaling schemes with action
recommendations.



\subsection{A Convex Programming Formulation}

Despite this difficulty, a version of the preceding argument, which we
term {\em coalescence}, continues to hold with a risk-conscious
receiver. To see this, observe that if two signals $\bar{s} = s_1$ and
$\bar{s} = s_2$ lead to the same posterior $\mu \in \Delta(\Omega)$
for the receiver, then the receiver's posterior is still $\mu$ if the
signaling scheme reveals only that $\bar{s} \in \{s_1, s_2\}$ whenever
it was supposed to reveal $s_1$ or $s_2$. This coalescence property
follows immediately from the fact that the receiver's posterior
belief, given $\bar{s} \in \{s_1, s_2\}$, is a convex combination of
beliefs under $\bar{s} = s_i$ for $i =1,2$. Thus, using the same
argument as before, the coalescence property allows us to coalesce all
signals that lead to the same posterior belief of the receiver into a
{\em belief recommendation}. In such a coalesced signaling scheme, we
can take the signal space $S$ to be $\Delta(\Omega)$, the set of
posteriors. Furthermore, in such a scheme, a property akin to
obedience holds: if the receiver is recommended a belief $\mu$, her
posterior belief is indeed $\mu$.

Summarizing the preceding discussion, we can write the sender's
persuasion problem \eqref{eq:sender-opt} as
\begin{equation}\label{eq:sender-opt-beliefs}
\begin{aligned}
  \max_{\pi \in \Delta(\Omega \times \Delta(\Omega))}     & \expec_\pi [ v(\bar{\omega}, a(\bar{s})) ]\\
  \text{subject to, } \quad a(s) &\in \argmax_{a \in A} \rho(\mu_s, a),
  \quad \text{for all $s \in \Delta(\Omega)$},\\
   \pi(\omega, \Delta(\Omega)) &= \mu^*(\omega), \quad \text{for each $\omega \in \Omega$,}\\
  \mu_s &= s, \quad \text{for all $s \in \Delta(\Omega)$}.
\end{aligned}
\end{equation}
Although we have characterized the set of signals, this is still a
challenging problem because of the complexity of the set
$\Delta(\Omega \times \Delta(\Omega))$. To make further progress, we
state the following lemma~\citep{aumannM95, kamenicaG11}, which gives
an equivalent formulation using the notion of {\em Bayes-plausible}
measures, which are probability measures
$\eta \in \Delta(\Delta(\Omega))$ over the set of beliefs
$\Delta(\Omega)$ with the property that their expectation equals the
prior belief $\mu^*$. (The proof of the results in this section are in
Appendix~\ref{ap:tractable}.)
\begin{lemma}[\citet{aumannM95, kamenicaG11}]\label{lem:splitting-lemma}
  A signaling scheme $\pi \in \Delta(\Omega \times \Delta(\Omega))$
  satisfies the condition $\mu_s = s$ for almost all
  $s \in \Delta(\Omega)$, only if the measure
  $\eta(\cdot) \defeq \pi(\Omega, \cdot) \in \Delta(\Delta(\Omega))$
  is {\em Bayes-plausible}. Conversely, for any Bayes-plausible
  measure $\eta$, the signaling scheme defined as
  $\pi(\omega, ds) = s(\omega)\eta(ds)$ satisfies $\mu_s = s$ for all
  $s \in \Delta(\Omega)$.
\end{lemma}
The preceding lemma allows us to
reformulate~\eqref{eq:sender-opt-beliefs} as an optimization over the
space of Bayes-plausible measures $\eta \in \Delta(\Delta(\Omega))$
with objective
$\sum_{\omega \in \Omega} \expec_\eta\left[ \bar{s}(\omega) v(\omega,
  a(\bar{s}))\right]$. (See~Lemma~\ref{lem:intermediate-reduction} in
Appendix~\ref{ap:tractable} for the details.) This reformulation is
still challenging, as it involves optimizing over a set of probability
measures on $\Delta(\Omega) \subseteq \reals^{|\Omega|}$. Our first
result allows us to overcome this difficulty, by establishing that one
can instead optimize over a much simpler space. To state our result,
we need some notation: for any set $H \subseteq \reals^m$, let
$\conv(H)$ denote the convex hull of $H$, defined as:
\begin{align*}
  \conv(H) = \left\{ y : y = \sum_{i=1}^j \lambda_i x_i, \text{for
  some $j\geq 1$, $\lambda_i \geq 0, x_i \in H$ for all $1 \leq i \leq
  j$ and $\sum_{i=1}^j \lambda_i = 1$}\right\}.
\end{align*}
In words, $\conv(H)$ is the set of all finite convex combinations of
elements in $H$. We have the following lemma which states that
corresponding to each Bayes-plausible measure $\eta$, there exists
$\{b_a \geq 0\}_{a \in A}$ and $\{m_a \in \Delta(\Omega)\}_{a \in A}$
such that the sender's expected utility under $\eta$, i.e., $\sum_{\omega \in \Omega} \expec_\eta\left[ \bar{s}(\omega) v(\omega,
  a(\bar{s}))\right]$,  can be written as a
bilinear function of $m_a$ and $b_a$. Thus, the lemma allows us to
directly optimize over $m_a$ and $b_a$, instead of over
Bayes-plausible measures $\eta$.
\begin{lemma}\label{lem:main-reduction}
  For any Bayes-plausible measure $\eta \in \Delta(\Delta(\Omega))$
  and optimal receiver strategy $a(\cdot)$, there exists
  ${\{(b_a, m_a)\}}_{a \in A}$, with $b_a\in [0,1]$ and
  $m_a \in \conv(\calP_a)$ for each $a \in A$, such that
  \begin{align}
    \sum_{a \in A} b_a m_a &= \mu^*, \label{eq:bayes-plausible}\\
    \expec_\eta \left[ \bar{s}(\omega) v(\omega,a(\bar{s})) \right]
                           &= \sum_{a \in A} b_am_a(\omega)v(\omega, a) \quad \text{for each
                             $\omega \in \Omega$.} \label{eq:objective}
  \end{align}
  Conversely, for any
  $\{(b_a, m_a) : b_a \in [0,1] \text{ and }m_a \in \conv(\calP_a)
  \text{ for each $a \in A$}\}$ satisfying \eqref{eq:bayes-plausible},
  there exists a Bayes-plausible measure $\eta$ and an optimal
  receiver strategy $a(\cdot)$ such that \eqref{eq:objective} holds.
\end{lemma}

\begin{figure}
	\begin{subfigure}[c]{0.33\textwidth}
		\centering
		\includegraphics[width = \textwidth]{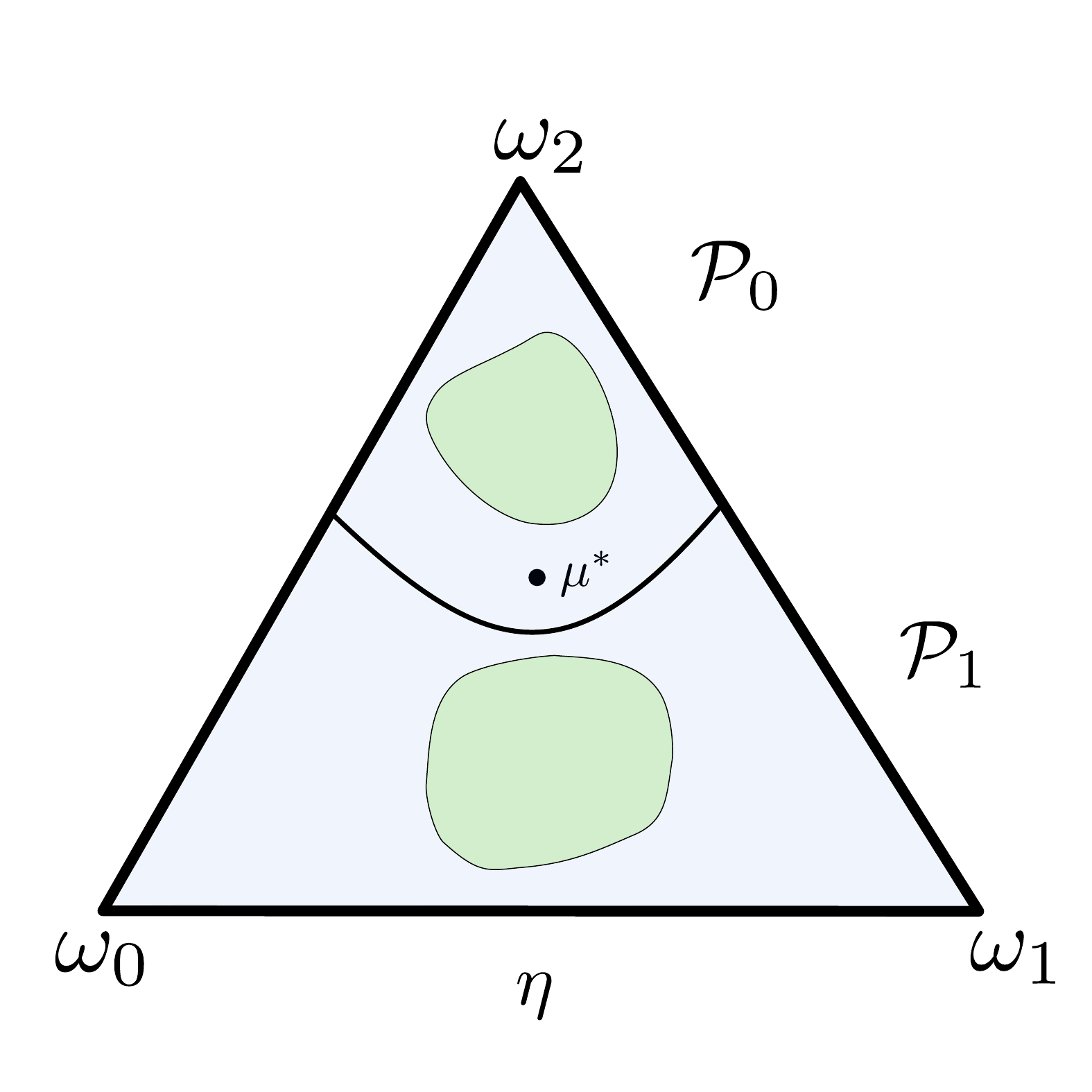}
		\label{fig:eta}
	\end{subfigure}\begin{subfigure}[c]{0.33\textwidth}
	\centering
	\includegraphics[width = \textwidth]{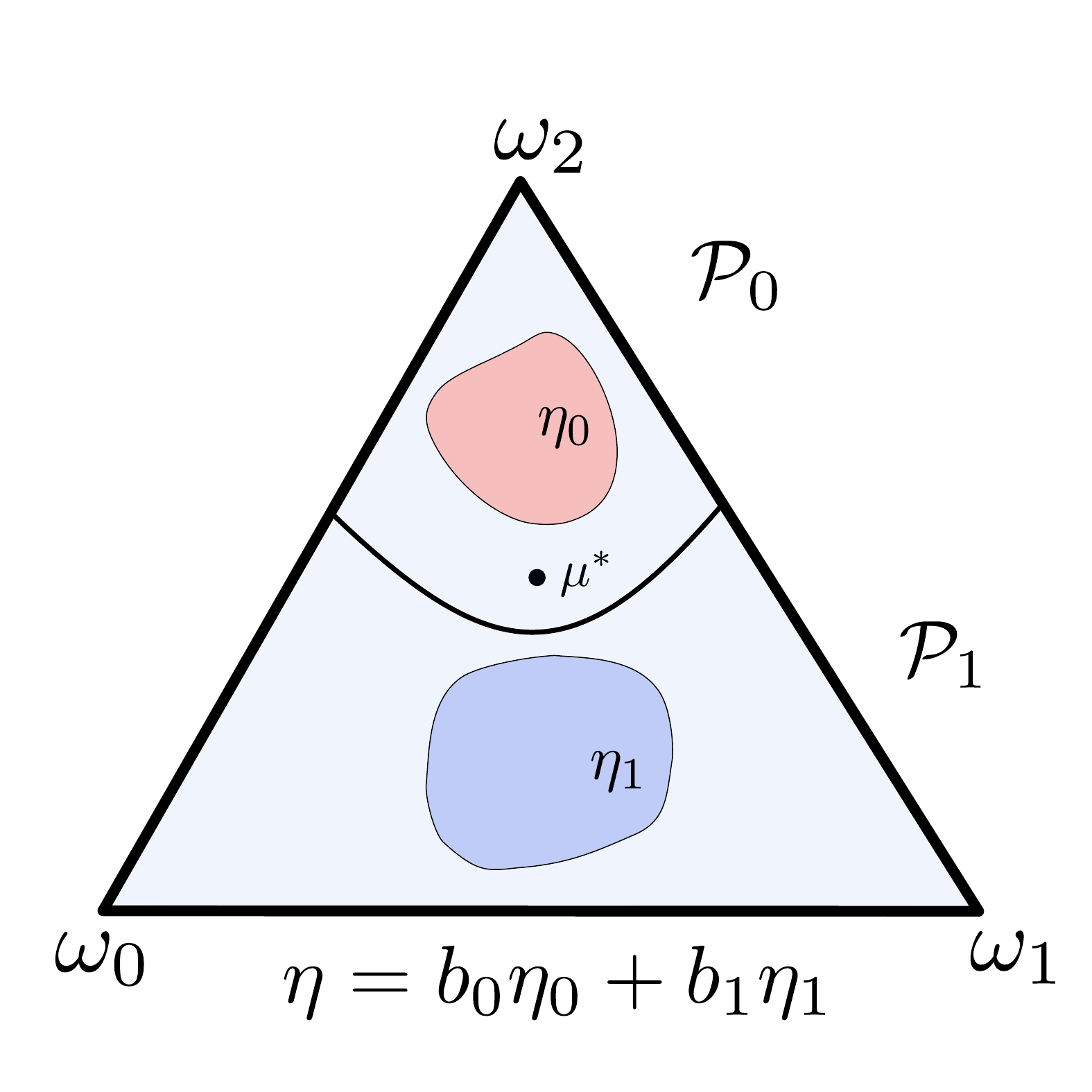}
	\label{fig:eta0}
\end{subfigure}\begin{subfigure}[c]{0.33\textwidth}
\centering
\includegraphics[width = \textwidth]{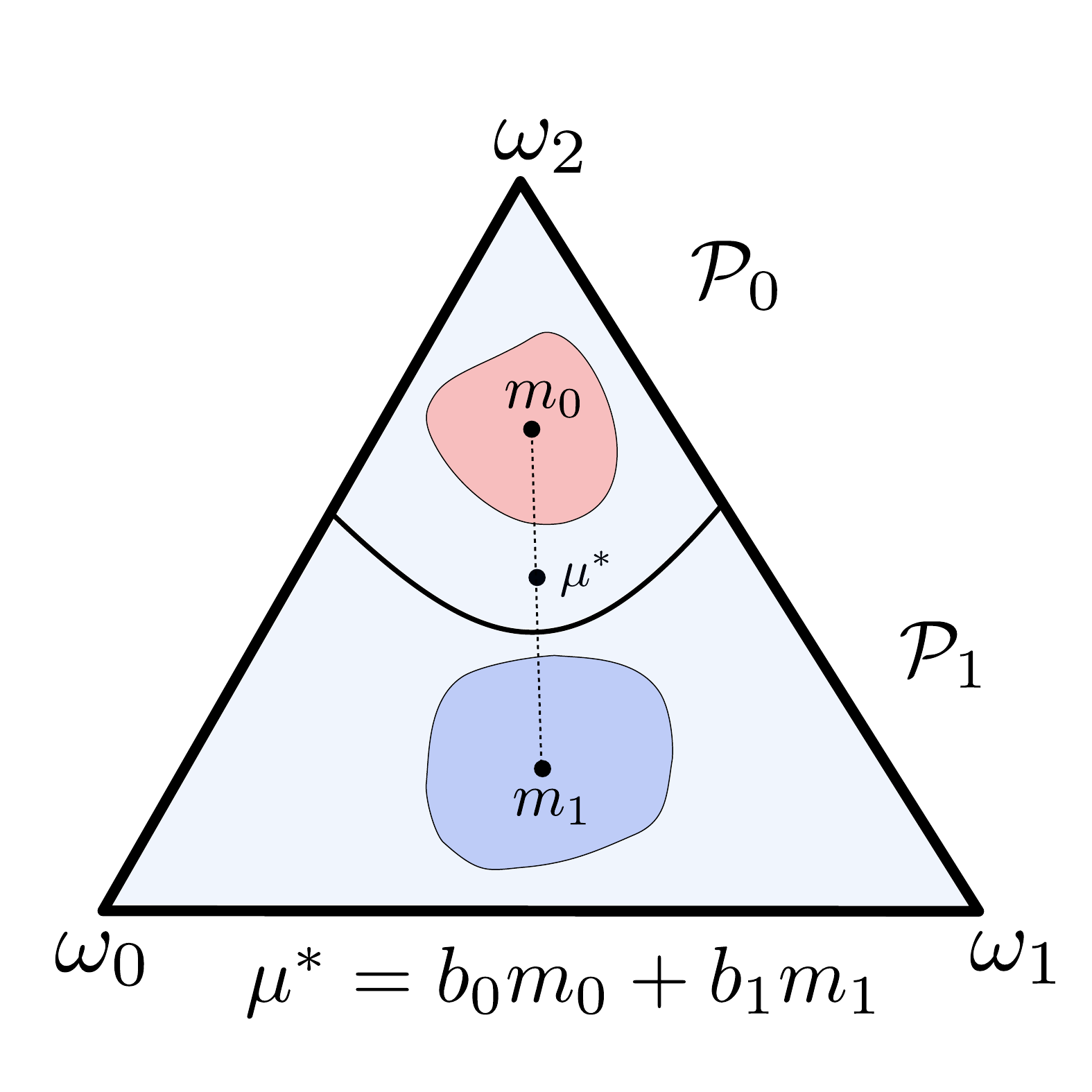}
\label{fig:m0}
\end{subfigure}
\caption{In the first figure, $\eta\in \Delta(\Delta(\Omega))$ has
  support in the green region.  Since $\eta$ is Bayes-plausible,
  $\expec_\eta [ \bar{s}(\omega) ] = \mu^*(\omega)$ for each
  $\omega \in \Omega$. In the second figure, we separate $\eta$ into
  $\eta_0$ and $\eta_1$, where $\eta_a$ has support in $\calP_a$.  In
  the third figure, we depict $m_0$ and $m_1$, where
  $\expec_{\eta_a}[ \bar{s}(\omega)] = m_a^*(\omega)$ and
  $m_a \in \conv(\calP_a)$. Finally, $b_a$ is defined as
  $\prob_\eta( a(\bar{s}) = a)$.  }\label{fig:lem2}
\end{figure}

The interpretation of the quantities $m_a$ and $b_a$ is as
follows. Given a Bayes-plausible measure $\eta$, the quantity $b_a$
denotes the probability that the receiver plays action $a \in A$ under
the optimal strategy $a(\cdot)$, when the sender uses the signaling
scheme corresponding to $\eta$; in other words,
$b_a = \prob_\eta( a(\bar{s}) = a)$. Similarly, $m_a$ denotes the
distribution of the state $\bar{\omega}$, conditioned on the receiver
choosing action $a$. By iterated expectation, we obtain
$m_a(\omega) = \expec_\eta [ \bar{s}(\omega) | a(\bar{s}) = a]$. Thus,
for any $a \in A$, the quantity $m_a$ denotes the mean of all
posterior beliefs the receiver holds, conditioned on her choosing
action $a$. Due to this reason, we refer to $m_a$ as the {\em
  mean-posterior} of the receiver corresponding to action $a \in A$.
Note that $m_a$ may not correspond to any actual posterior that the
receiver holds when she chooses action $a \in A$; in fact, the
mean-posterior $m_a \in \conv(\calP_a)$ may not even lie in the set
$\calP_a$.  Figure~\ref{fig:lem2} gives some geometric intuition for
these quantities, as well as for the distributions $\eta_a$ introduced
in the proof.

While the equations~\eqref{eq:bayes-plausible}
and~\eqref{eq:objective} are bilinear in $m_a$ and $b_a$, we can make
them linear by substituting $t_a(\omega) = b_a m_a(\omega)$ for each
$\omega \in \Omega$. Notice that $b_a \in [0,1]$ and
$m_a \in \conv(\calP_a)$ imply
$t_a \in \conv( \calP_a \cup \{\zero\})$, where
$\zero \in \reals^{|\Omega|}$ is the vector of all zeros. With this
substitution, we obtain our main theorem:
\begin{theorem}\label{thm:convex} The sender's persuasion
  problem~\eqref{eq:sender-opt}
  can be optimized by solving the following convex optimization
  problem:
  \begin{equation}
  \label{eq:sender-opt-convex}
  \begin{aligned}
    \max_{\{t_a : a  \in A\}} \ \sum_{\omega \in \Omega} \sum_{a \in A} & t_a(\omega) v(\omega, a)\\
    \text{subject to, }  \quad     \sum_{a \in A} t_a(\omega) &= \mu^*(\omega),\quad \text{ for each $\omega \in \Omega$.}\\
    t_a &\in \conv(\calP_a \cup \{\zero\}), \quad \text{ for each
      $a \in A$.}
\end{aligned}
\end{equation}
\end{theorem}
As $m_a$ denotes the receiver's mean posterior conditional on her
choosing action $a$, and $b_a$ denotes the probability she chooses
action $a$, we obtain that $t_a(\omega) = b_a m_a(\omega)$ denotes the
joint probability that the receiver takes action $a$ and the realized
state is $\bar{\omega} = \omega$. Thus, the
reformulation~\eqref{eq:sender-opt-convex} directly optimizes over the
joint probability distribution of the state and the receiver's
actions.

We next briefly remark on the complexity of solving the convex
program~\eqref{eq:sender-opt-convex}. First, note that relative to
optimizing over Bayes-plausible measures
$\eta \in \Delta(\Delta(\Omega))$, the convex program is extremely
simple: the optimization is over $|A|$ variables $t_a$ belonging to a
convex set $\conv(\calP_a \cup \{\zero\})$ in
$\reals^{|\Omega|}$. Thus, its computational complexity rests on
whether there exists an efficient characterization of the set
$\conv(\calP_a \cup \{\zero\})$ for each $a \in A$, which in turn
depends solely on the properties of the utility functions
$\rho(\cdot,a)$. While obtaining such an efficient characterization
can be hard in general, in certain settings with additional structure
on $\rho$, one can replace the convex sets $\conv(\calP_a \cup \{0\})$
by a convex polytope, yielding a linear program. We present and
analyze one such setting in Section~\ref{sec:binary-persuasion}.

\subsection{Optimal Signaling Schemes}

To conclude this section, we describe how to get an optimal signaling
scheme $\pi$ from the optimal solution $t = \{t_a\}_{a \in A}$ to the
problem~\eqref{eq:sender-opt-convex}. Since
$t_a \in \conv(\calP_a \cup \{\zero\})$ for each $a \in A$, let
$m_a \in \conv(\calP_a)$ and $b_a \in [0,1]$ be such that
$t_a = b_a m_a $. (Note that for $t_a \neq \zero$, the corresponding
$m_a$ is uniquely defined.) Since $m_a \in \conv(\mathcal{P}_a)$,
there exists a finite convex decomposition of $m_a$ in terms of the
elements of $\calP_a$. That is, there exists
$\{ (\mu_i^a, \lambda_i^a) : i = 1, \ldots, j_a\}$ for some
$j_a \geq 1$ with $\mu^a_i \in \calP_a$, $\lambda_i^a \geq 0$, and
$\sum_{i} \lambda_i^a =1$, such that
$m_a = \sum_{i=1}^{j_a} \lambda_i^a \mu_i^a$. An optimal signaling
scheme $\pi$ is then given by the (discrete) distribution over
$\Delta(\Omega \times \Delta(\Omega))$ that chooses
$(\bar{\omega} , \bar{s}) = (\omega, \mu_i^a)$ with probability
$b_a\lambda_i^a \mu_i^a(\omega)$. Observe that conditional on
$\bar{\omega} = \omega$, the optimal signaling scheme $\pi$ makes the
belief recommendation $\mu_i^a$ to the receiver with probability
\begin{align}\label{eq:scheme-from-t}
  \pi(\bar{s} = \mu_i^a | \bar{\omega} = \omega) &= \frac{b_a\lambda_i^a \mu_i^a(\omega)}{\sum_{a' \in A} b_{a'} \lambda_i^{a'} \mu_i^{a'}(\omega)}.
\end{align}
We note that in general, the representation of $m_a$ as a convex
combination of $\mu_i^a \in \calP_a$ need not be unique. Since the
preceding construction works for any convex decomposition of $m_a$, we
conclude that there may exist multiple optimal signaling schemes for
the receiver.

For any set $H \subseteq \reals^{d}$, let $\cara(H)$ denote the minimum
value of $j$ such that any point $x \in \conv(H)$ can be written as a
convex combination of at most $j$ points in $H$. The Caratheodory's
theorem~\citep{barany95} states that $\cara(H) \leq \dim(H) + 1$,
where $\dim(H)$ is the dimension of the smallest affine space
containing $H$. Thus, we obtain the following bound on the size of the
set of signals the sender needs to use to optimally persuade the
receiver:
\begin{proposition}\label{prop:signal-bound} There exists an optimal signaling scheme
  $\pi \in \Delta(\Omega \times S)$, where the set of signals $S$
  satisfies $|S| \leq \sum_{a \in A} \cara(\calP_a)$. Specifically,
  for any $a \in A$, the signaling scheme sends at most
  $\cara(\calP_a) \leq |\Omega|$ signals for which the receiver's
  optimal action is $a$.
\end{proposition}

In the case of expected utility maximizing agents, each set
  $\calP_a$ is convex, and hence the preceding proposition implies
  that at most $\cara(\calP_a) = 1$ signal per action suffices for
  optimal persuasion. This matches with the bound using the revelation
  principle, which implies the sufficiency of action
  recommendations. For risk-conscious receivers, the proposition
  states that in general, the sender must share more information than
  just action recommendations, and the additional information required
  to optimally induce any action may take up to $|\Omega|$ values.


  The bound we obtain here is related to the bound obtained by
  \citet{kamenicaG11}, who also use convex analytic arguments to show
  that at most $|\Omega|$ signals overall suffice for an optimal
  signaling scheme. In Appendix~\ref{ap:kg-comparison}, we discuss
  this connection in detail, and use their bound to provide an
  alternative approach to arrive at the convex
  problem~\eqref{eq:sender-opt-convex}.

\section{Benefits from Persuasion}\label{sec:benefits-persuasion}

Using Theorem~\ref{thm:convex}, we now turn to the question of
  determining the sender's benefits from persuasion. Let $V(\mu^*)$
  denote the optimal value of the program~\eqref{eq:sender-opt-convex}
  as a function of the prior $\mu^*$. Similarly, let
  $\hat{\nu}(\mu) \defeq \expec_{\mu}[v(\bar{\omega}, a_\mu)]$ denote
  the sender's expected payoff without persuasion as a function of the
  prior $\mu$, where $a_\mu \in \{ a \in A : \mu \in \calP_a\}$
  denotes the receiver's optimal action under the prior.

Note that $\hat{\nu}(\mu^*)$ is the objective value
of~\eqref{eq:sender-opt-convex} for the feasible solution
$t = \{t_a\}_{a \in A}$ with $t_a = \mu^*$ for $a = a_{\mu^*}$ and
$\zero$ otherwise. Since $V(\mu^*)$ is its optimal value, it is
immediate that the sender (strictly) benefits from persuasion if and
only if $V(\mu^*) > \hat{\nu}(\mu^*)$. The following result
characterizes when the latter condition holds; the result and its
proof in Appendix~\ref{ap:benefits-persuasion} are analogous to the
setting with expected-utility maximizing
receivers~\citep{kamenicaG11}. Using the terminology therein, we say
{\em there is information the sender would share\/} if there exists a
belief $\mu$ with
$\hat{\nu}(\mu) > \expec_\mu[ v(\bar{\omega}, a_{\mu^*})]$. For ease
of notation, let $\calP_\mu$ denote the set $\calP_a$ with
$a = a_\mu$.
\begin{proposition}\label{prop:benefits-persuasion}
  \begin{enumerate}
  \item If there is no information the sender would share, then sender
    does not benefit from persuasion.  If there is information sender
    would share, and $\mu^*$ lies in the interior of $\calP_{\mu^*}$,
    then sender benefits from persuasion.
  \item If for each $a \in A$ and $m_a \in \conv(\calP_a)$ we have
    $\sum_{\omega \in \Omega} m_a(\omega) \left( v(\omega, a) -
      v(\omega, a_{\mu^*})\right) \leq 0$, then the sender does not
    benefit from persuasion. If $\mu^*$ is in the interior of
    $\calP_{\mu^*}$ and there exists an $a \in A$ and an
    $m_a \in \conv(\calP_a)$ such that
    $\sum_{\omega \in \Omega} m_a(\omega) \left( v(\omega, a) -
      v(\omega, a_{\mu^*})\right) > 0$, then the sender benefits from
    persuasion.
  \end{enumerate}
\end{proposition}
The preceding result states that if the sender benefits from
persuasion, then there is an action $a$ and a belief
$\mu \in \conv(\calP_a)$ such that under prior $\mu$, the sender would
strictly prefer that the receiver take action $a$ over the action
$a_{\mu^*}$. The action $a$ need not be optimal for the receiver with
belief $\mu$; this is indeed the case if
$\mu \in \conv(\calP_a) \cap \calP_a^c$. This latter scenario is
impossible for an expected-utility maximizing receiver, for whom the
sets $\calP_a$ are convex, and hence
$\conv(\calP_a) \cap \calP_a^c = \calP_a \cap \calP_a^c = \emptyset$.

Having obtained the conditions under which the sender benefits
  from persuasion, we now turn to describing the extent of this
  benefit. To do this, we define the notion of {\em full persuasion}
  that we illustrated in the introduction.  Using
  Theorem~\ref{thm:convex}, we will then obtain a simple
  characterization of when full persuasion is possible with
  risk-conscious receivers.

  To formally define full persuasion, we start with some
  notations. For each action $a\in A$, let
  $\Upsilon_a \subseteq \Omega$ denote the set of pure states for
  which action $a$ is sender-optimal:
  \begin{align*}
    \Upsilon_a = \{ \omega \in \Omega : v(\omega, a) \geq v(\omega,
    a') \text{ for all $a' \in A$} \}.
  \end{align*}
  We say the sender {\em fully persuades\/} the receiver if there
  exists a signaling scheme such that at each state
  $\omega \in \Omega$, the receiver chooses a sender-optimal
  action. Formally, we have the following definition:
  \begin{definition}[Full persuasion]\label{def:full-persuasion} A
    signaling scheme $\pi \in \Delta(\Omega \times S)$ fully persuades
    the receiver if for each $\omega \in \Omega$ and $s \in S$ with
    $\pi(\omega, s) > 0$, if $a(s) = a$, then $\omega \in
    \Upsilon_a$. We say full persuasion is possible if there exists a
    signaling scheme that fully persuades the receiver.
  \end{definition}

  To state the main result of this section, we make the following
  simplifying assumption: at each state $\omega \in \Omega$ there
  exists a unique action that is sender-optimal. In other words, we
  assume that $\{ \Upsilon_a : a \in A\}$ forms a partition of the
  state space $\Omega$.  For each $\omega \in \Omega$ and $a \in A$,
  let
  $t^*_a(\omega) \defeq \mu^*(\omega) \ind\{ \omega \in \Upsilon_a\}$.
  We have the following theorem, whose proof is in
  Appendix~\ref{ap:benefits-persuasion}:
  \begin{theorem}\label{thm:full-persuasion}
    Suppose at each state there is a unique action that is
    sender-optimal. Then, the sender can fully persuade the receiver
    if and only if $t^* = (t_a^* : a \in A)$ is feasible for the
    convex program~\eqref{eq:sender-opt-convex}, i.e., for each
    $a \in A$, we have $t_a^* \in \conv(\calP_a \cup \{\zero\})$.
  \end{theorem}
  To build intuition about the theorem, consider the scenario where
  there is a fixed action $a$ that is uniquely sender-optimal at all
  states, i.e., $\Upsilon_a = \Omega$ and hence, $t_a^* = \mu^*$. Full
  persuasion in this context requires the sender to persuade the
  receiver to always take action $a$, irrespective of the realized
  state. The preceding theorem implies that this is possible if and
  only if $\mu^* \in \conv(\calP_a)$. Clearly, if $\mu^* \in \calP_a$,
  the action $a$ is optimal for the receiver under the prior belief,
  and the receiver needs no persuasion. In other words, if
  $\mu^* \in \calP_a$, the no-information scheme already fully
  persuades the receiver. The theorem implies that if
  $\mu^* \in \conv(\calP_a) \cap \calP_a^c$, there exists a signaling
  scheme that shares some state information and fully persuades the
  receiver. Once again, we note that this latter scenario cannot occur
  for an expected-utility maximizing receiver, for whom we have
  $\conv(\calP_a) \cap \calP_a^c = \emptyset$.

  More generally, Theorem~\ref{thm:full-persuasion} implies that an
  expected-utility maximizing receiver can be fully persuaded if and
  only if $t_a^*$, appropriately scaled, lies in the set $\calP_a$. In
  this case, the signaling scheme that ``reveals the partition'',
  i.e., the scheme that at each state $\omega$ reveals the set
  $\Upsilon_a$ containing $\omega$, is sufficient to fully persuade
  the receiver. In contrast, for a risk-conscious receiver, the
  signaling scheme that fully-persuades the receiver may need to
  reveal more information than just revealing the partition. We
  explore this point in more detail in the following section.


\section{Binary Persuasion}\label{sec:binary-persuasion}

We now focus on a setting of practical importance that we refer to as
{\em binary persuasion}. In this setting, the receiver's actions are
binary, i.e., $A = \{0,1\}$, and the sender's utility is always weakly
higher under action $1$, i.e., $v(\omega,1) \geq v(\omega, 0)$ for all
$\omega \in \Omega$.  This model matches settings where, independent
of the state, the sender seeks to persuade the receiver to take an
action, such as engage with social media
platforms~\citep{candoganD17}, wait in a queue~\citep{lingenbrinkI19},
or purchase a product~\citep{lingenbrinkI18a, drakopoulosJR18}.

To aid our discussion, we define the receiver's {\em differential
  utility} $\hat{\rho}(\cdot)$ as the difference in the utility
between choosing action $1$ and action $0$:
$\hat{\rho}(\mu) \defeq \rho(\mu, 1) - \rho(\mu, 0)$ for all
$\mu \in \Delta(\Omega)$. Note that action $a= 1$ is optimal for the
receiver at belief $\mu$ if and only if $\hat{\rho}(\mu) \geq 0$.

\subsection{Geometry of the Convex
  Program}\label{subsec:reduction-linear-program}

The convex program~\eqref{eq:sender-opt-convex} has variables defined
over the domain $\conv(\calP_a \cup \{\zero\})$ for each action
$a$. In this section, we show that the domain can be further
simplified under the following assumption:
\begin{assumption}\label{as:convex} The set
    $\calP_1^c = \{ \mu \in \Delta(\Omega) : \hat{\rho}(\mu) < 0\}$ is
    convex.
\end{assumption}


Intuitively, the assumption implies that the receiver is averse to
uncertainty when choosing action $1$: if action $1$ is not optimal
under beliefs $\mu$ and $\mu'$, then it cannot be optimal under a
belief $\gamma \mu + (1 - \gamma) \mu'$ that is obtained by inducing
uncertainty between $\mu$ and $\mu'$. Furthermore, the assumption
holds for a wide class of utility functions, as the following lemma
establishes.
\begin{lemma}\label{lem:quasiconvexity} Suppose the differential
  utility $\hat{\rho}(\cdot)$ is
  quasiconvex. Then Assumption~\ref{as:convex} holds.
\end{lemma}
Specifically, Assumption~\ref{as:convex} holds when $\rho(\mu,1)$ is
convex and $\rho(\mu, 0)$ is concave. At the same time, note that
Assumption~\ref{as:convex} is substantially weaker than requiring
quasiconvexity of $\hat{\rho}$, since the latter implies that every
level set $\{ \mu : \hat{\rho}(\mu) < c\}$ is convex.


Using Assumption~\ref{as:convex}, we now show that the convex
program~\eqref{eq:sender-opt-convex} can be simplified to a linear
program.  Recall that $e_\omega$ denotes the belief that assigns all
its weight to $\omega \in \Omega$. By a slight abuse of notation, we
identify $\omega \in \Omega$ with $e_\omega$ and consider $\Omega$ as
a subset of $\Delta(\Omega)$.

Let $K_1 \defeq \calP_1 \cap \Omega$ denote the set of states where
action $1$ is optimal for the receiver under
full-information. Similarly, let $K_0 \defeq \calP_0 \cap \Omega$ be
the set of states where action $0$ is optimal for the receiver under
full-information. Note, $K_0 \cap K_1$ may be non-empty if the
receiver finds both actions optimal at some state. We let
  $L_0 \defeq K_0 \cap K_1^c$ denote the set of states for which
  action $0$ is uniquely optimal for the receiver.

Next, for $\omega_0 \in L_0$ and $\omega_1 \in K_1$, consider the set
of beliefs obtained as the convex combination of $\omega_0$ and
$\omega_1$. In this set, we let $\chi(\omega_0, \omega_1)$ denote the
belief that puts the largest weight on $\omega_0$ while still
preserving the optimality of action $1$ for the receiver. (Since
$\calP_1$ is closed, such a maximal convex combination exists.)
Formally, for $\omega_0 \in L_0$ and $\omega_1 \in K_1$, we define
$\gamma(\omega_0, \omega_1) = \sup_{ \gamma \in [0,1]} \left\{ \gamma
  : \gamma \omega_0 + (1-\gamma)\omega_1 \in \calP_1 \right\}$, and
let
$\chi(\omega_0, \omega_1) = \gamma(\omega_0, \omega_1) \omega_0
+(1-\gamma(\omega_0, \omega_1)) \omega_1$. Since $\calP_0$ and
$\calP_1$ are closed, we obtain that
$\chi(\omega_0, \omega_1) \in \calP_0 \cap \calP_1$, and hence the
receiver is indifferent between action $0$ and $1$ at belief
$\chi(\omega_0, \omega_1)$. Finally, we define the set $K_{01}$ to be
a subset of such maximal convex combinations
$\chi(\omega_0, \omega_1)$:
\begin{align}\label{eq:k01}
K_{01} = \{\chi(\omega_0, \omega_1) \in \Delta(\Omega) \text{ for some
  $\omega_0\in L_0$ and $\omega_1\in K_1$}\}.   
\end{align}
We note that while $K_{01} \cap L_0$ must be empty, the sets
  $K_{01} \cap K_0$ and $K_{01} \cap K_1$ can be non-empty. We
  illustrate these sets pictorially in Fig.~\ref{fig:k0k1k01} (and in
  Fig.~\ref{fig:k0k1k01-1} in Appendix~\ref{ap:binary-persuasion}).
\begin{figure}
  \begin{subfigure}[c]{0.33\textwidth}
    \centering
    \includegraphics[width = \textwidth]{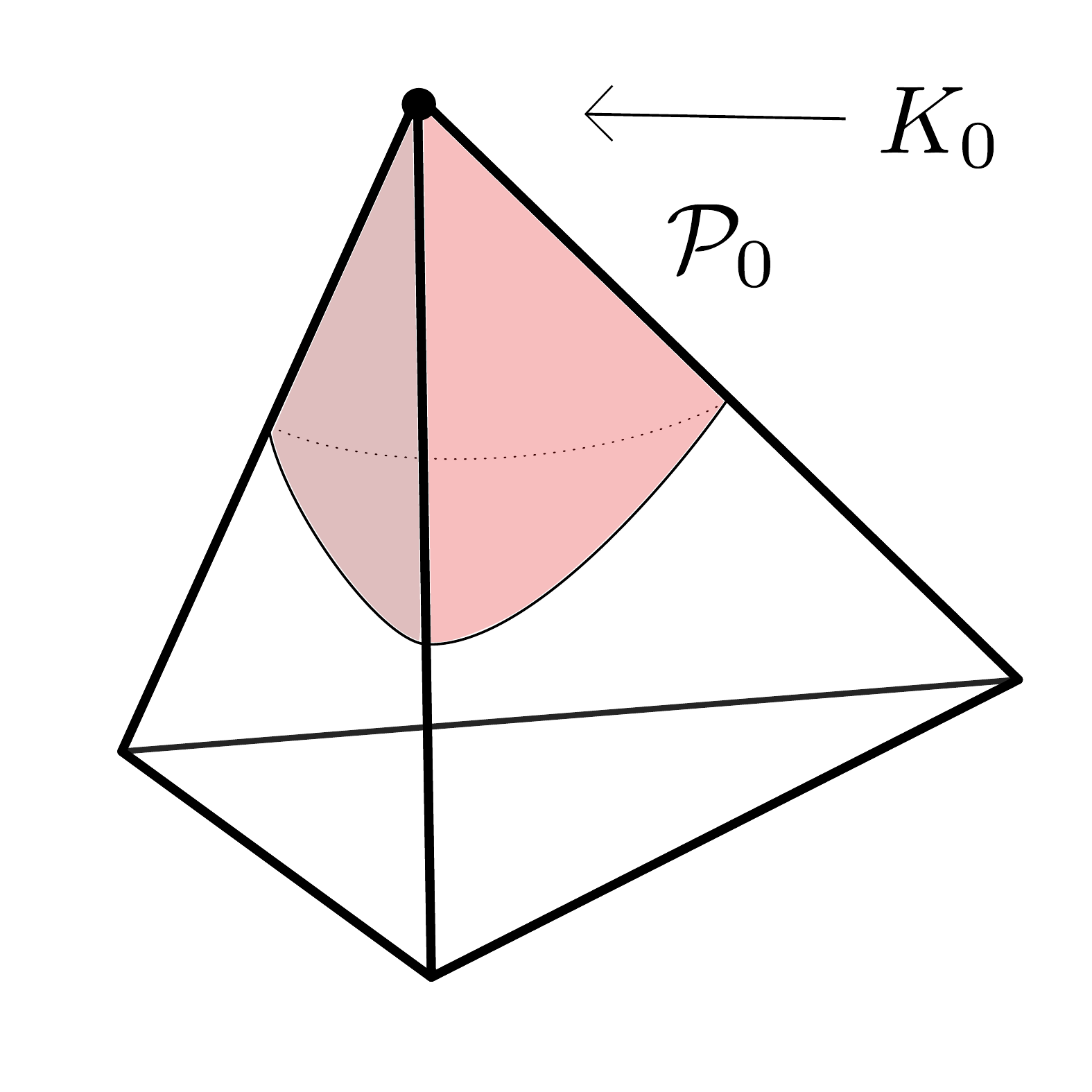}
    \caption{The set $K_0 = \Omega \cap \calP_0$.}%
    \label{fig:k0}
  \end{subfigure}%
  \begin{subfigure}[c]{0.33\textwidth}
    \centering
    \includegraphics[width = \textwidth]{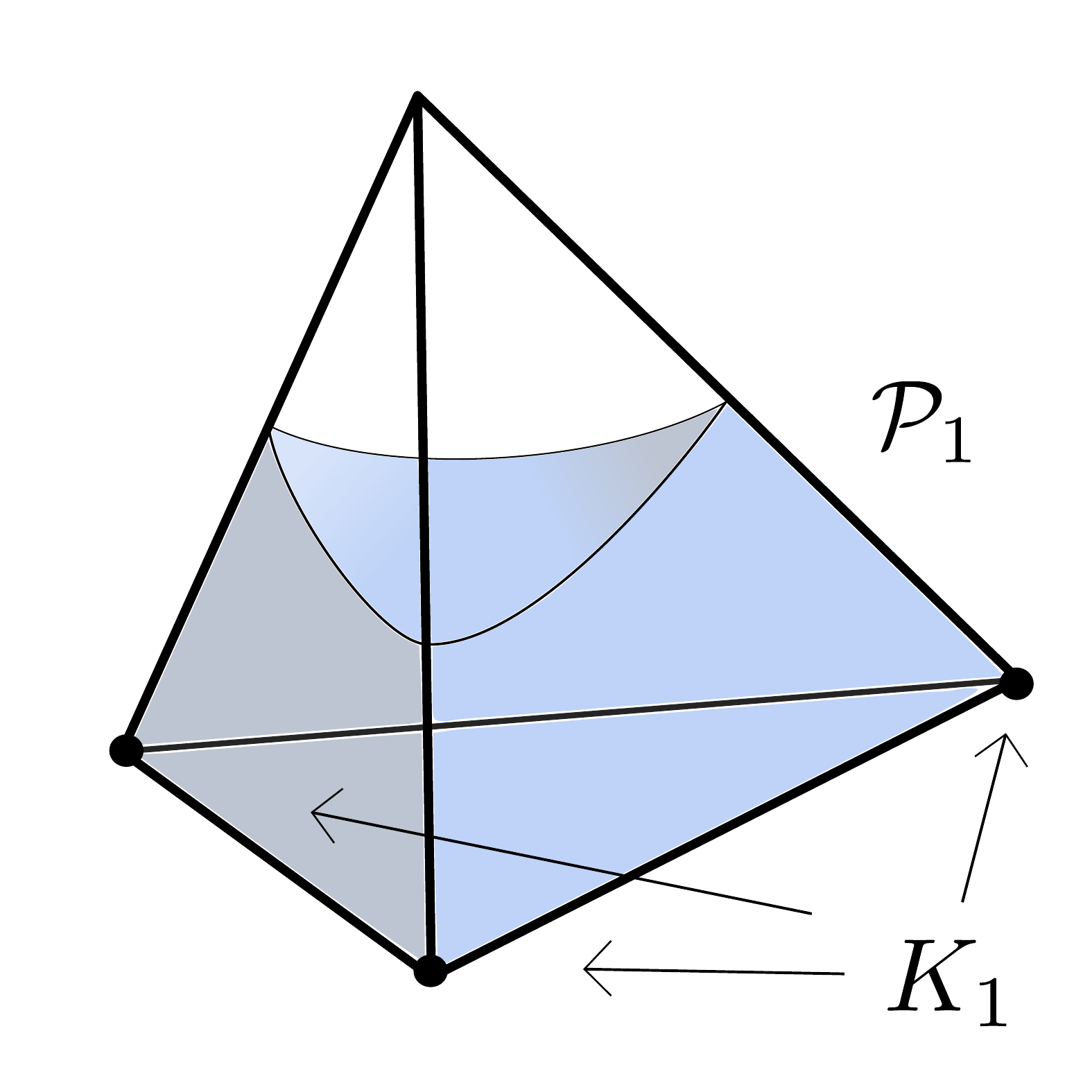}
    \caption{The set $K_1 = \Omega \cap \calP_1$.}%
    \label{fig:k1}
  \end{subfigure}%
  \begin{subfigure}[c]{0.33\textwidth}
    \centering
    \includegraphics[width = \textwidth]{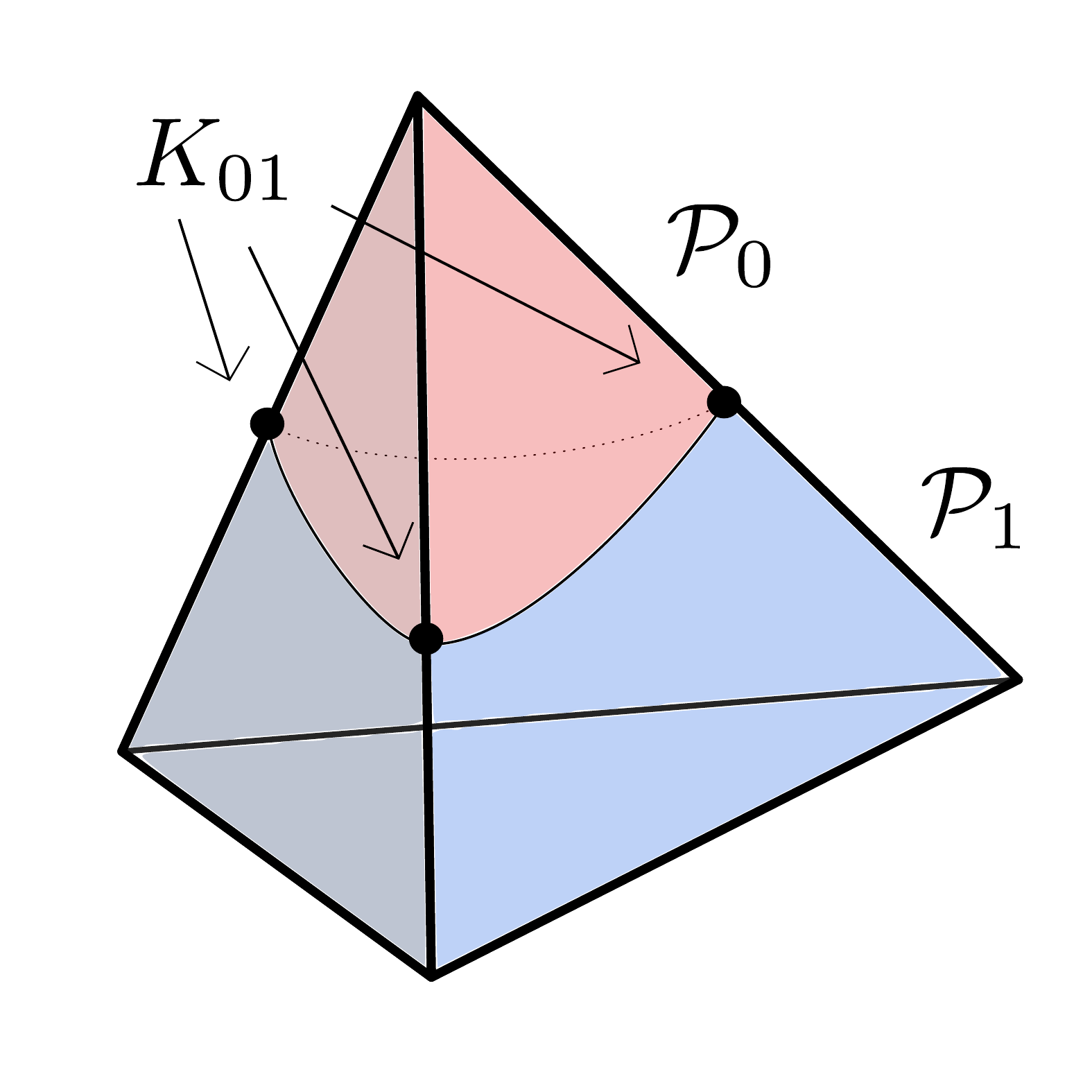}
    \caption{The set $K_{01}$ as defined in~\eqref{eq:k01}.}%
    \label{fig:k01}
  \end{subfigure}
  \caption{The sets $K_0, K_1$ and $K_{01}$. (Here, the red
      region is the set $\calP_0$ and the blue region is
      $\calP_1$. Furthermore, in this example,
      $L_0 = K_0$.)}\label{fig:k0k1k01}
\end{figure}

With these definitions, we can state the main theorem of this
section. The full proof is provided in
Appendix~\ref{ap:binary-persuasion}.
\begin{theorem}\label{thm:linear}
  Under Assumption~\ref{as:convex}, the sender's persuasion
  problem~\eqref{eq:sender-opt-convex} can be optimized by solving the
  following linear program:
  \begin{align}\label{eq:linear}
    \begin{aligned}
      \max_{t_0, t_1}& \sum_{\omega \in \Omega} v(\omega,1) t_1(\omega) + v(\omega,0)t_0(\omega) \\
      \text{subject to,}\quad t_1 &\in \conv(K_1 \cup K_{01} \cup \{\zero\}), \\
      t_0 &\in \conv(L_0 \cup \{\zero\}),\\
      t_0(\omega) + t_1(\omega) &= \mu^*(\omega) \quad \text{for each
      } \omega \in \Omega.
    \end{aligned}
  \end{align}
\end{theorem}
The proof intuition is as follows. First, under
Assumption~\ref{as:convex}, we prove in Lemma~\ref{lem:conv} that the
set $\conv(\calP_1)$ is a convex polytope with extreme points in
$K_1 \cup K_{01}$, and hence the constraint
$t_1 \in \conv(\calP_1 \cup \{\zero\})$
in~\eqref{eq:sender-opt-convex} is equivalent to the constraint
$t_1 \in \conv(K_1 \cup K_{01} \cup \{\zero\})$. Second, we establish
that any $t_0 \in \conv(\calP_0 \cup \{\zero\})$ that is not in
$\conv(L_0 \cup \{\zero\})$ cannot be part of an optimal solution, by
improving on any such feasible solution. See Fig.~\ref{fig:conv} (and
Fig.~\ref{fig:conv-alt} in Appendix~\ref{ap:binary-persuasion}) for
some geometric intuition.
\begin{figure}
  \begin{subfigure}[c]{0.5\textwidth}
    \centering
    \includegraphics[width = 0.6\textwidth]{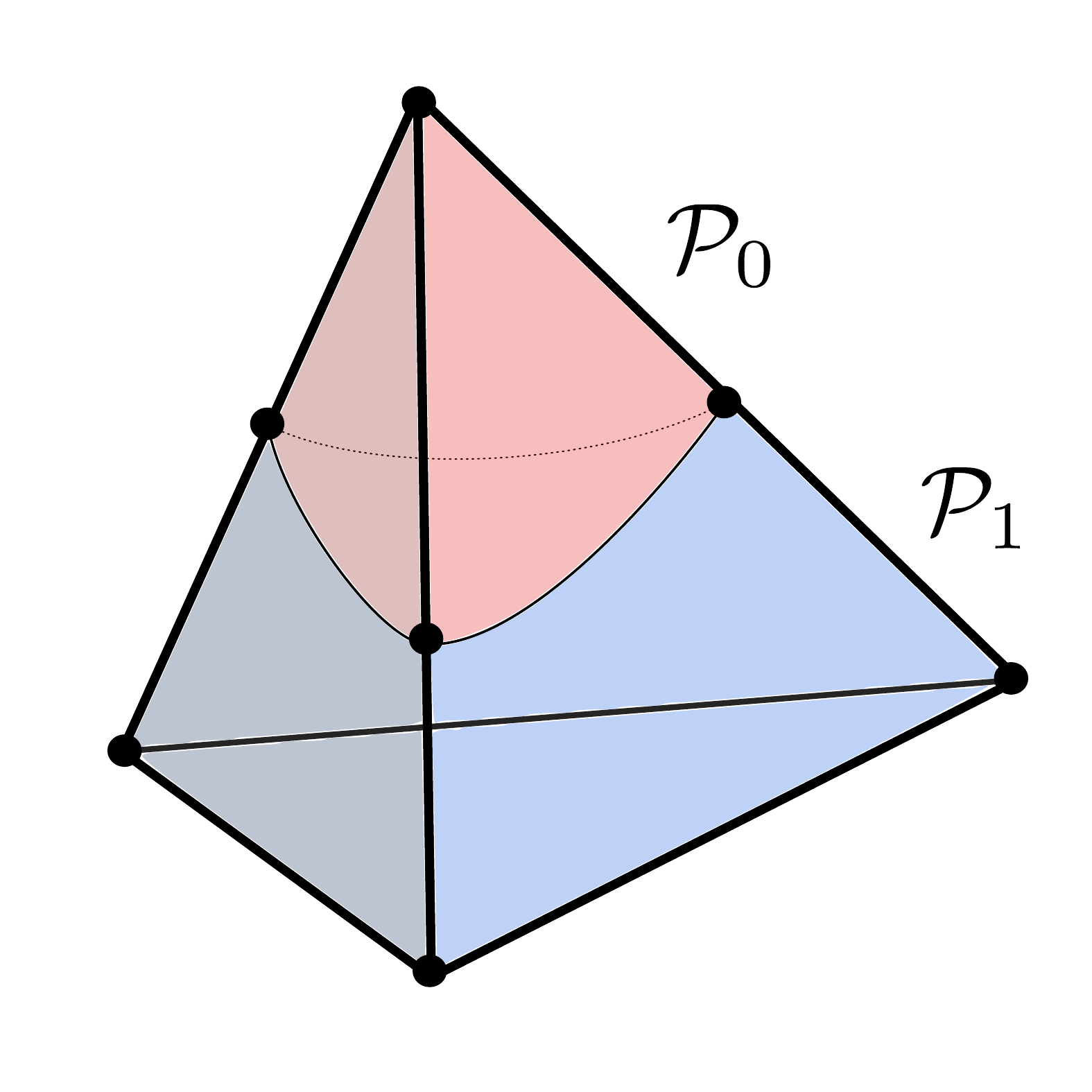}%
  \end{subfigure}
  \begin{subfigure}[c]{0.5\textwidth}
    \centering
    \includegraphics[width = 0.6\textwidth]{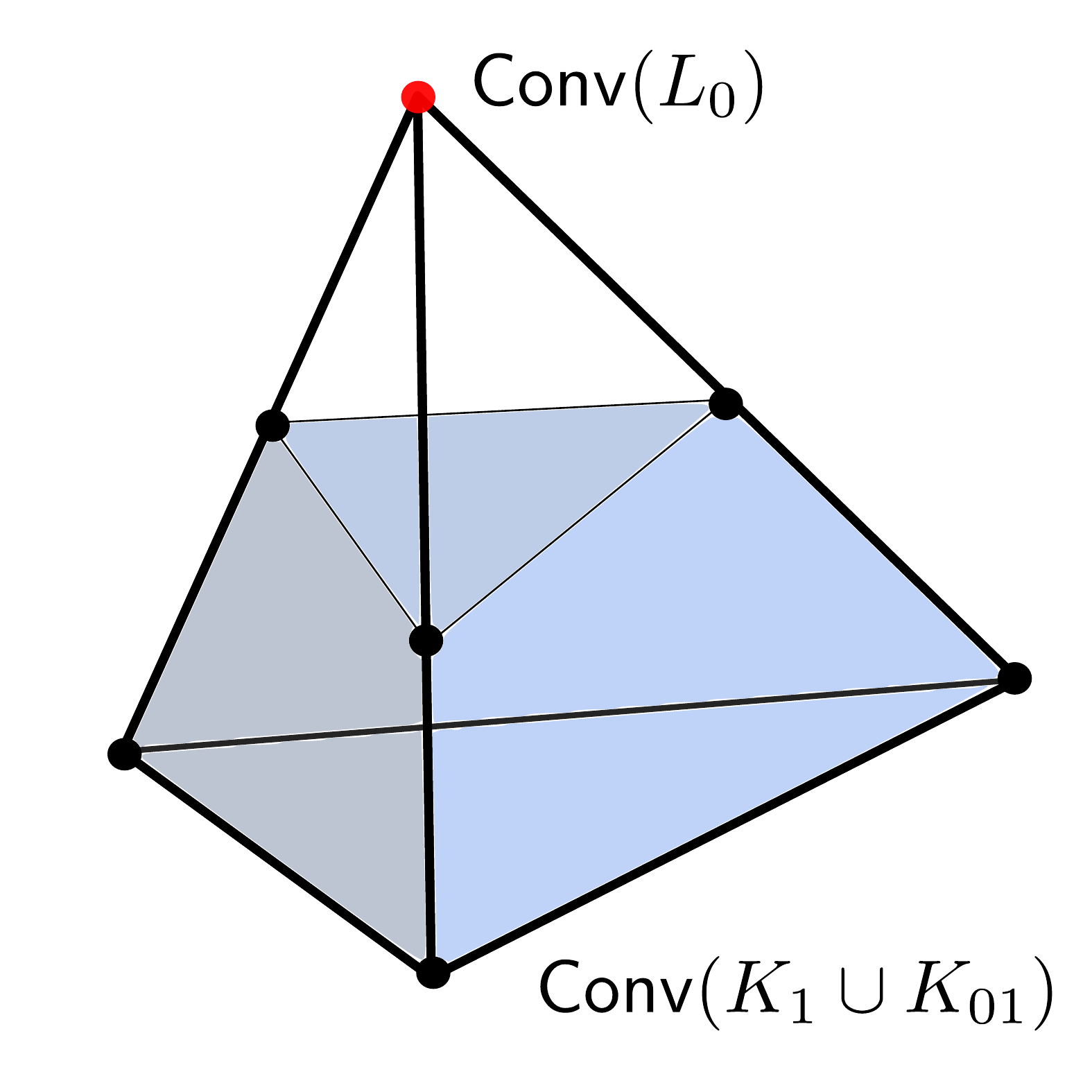}%
  \end{subfigure}
  \caption{Geometry of $\conv(L_0)$ and
      $\conv(K_1\cup K_{01})$. Here, $\conv(L_0)$ is the red vertex of
      the simplex.}\label{fig:conv}
\end{figure}

\subsection{Structural Characterizations}
The preceding theorem implies several results about the structure of
the optimal signaling scheme.

First, Theorem~\ref{thm:linear} establishes a {\em canonical set} of
signals for an optimal signaling scheme, namely the set
$\Omega \cup K_{01}$. In other words, for a given binary persuasion
setting, irrespective of the prior belief $\mu^*$, it suffices to only
send signals in the set $\Omega \cup K_{01}$.  This follows from the
fact that Theorem~\ref{thm:linear} implies that the mean-posteriors
$m_0$ and $m_1$ satisfy $m_0 \in \conv(L_0)$ and
$m_1 \in \conv(K_1 \cup K_{01})$. Hence, $m_0$ can be expressed as a
convex combination of beliefs in $L_0$, and $m_1$ can be expressed as
a convex combination of beliefs in $K_1 \cup K_{01}$. Thus, it follows
that inducing beliefs in the set
$L_0 \cup K_1 \cup K_{01} = \Omega \cup K_{01}$ suffices for optimal
persuasion.

Second, observe that the canonical set consists of {\em pure} signals
$\omega \in \Omega$ which fully reveal the state $\omega$, and {\em
  binary mixed} signals $\chi(\omega_0, \omega_1) \in K_{01}$ which
induce a belief over two states $\omega_0 \in L_0$ and
$\omega_1 \in K_1$. Thus, the optimal persuasion can always be
achieved by either fully revealing the state, or making the receiver
uncertain about two states. Of course, due to the convexity of
$\calP_1^c$, there also exists an optimal signaling scheme where all
posteriors that lead to the receiver taking action $0$ are replaced
with a single action recommendation ``take action $0$''.

Note that since $t_0 \in \conv(L_0 \cup \{\zero\})$, if the
  optimal signaling scheme induces the sender's least-preferred action
  $0$, then the receiver's belief only puts weight on states in the
  set $L_0$. At each of these states, the action $0$ is uniquely
  optimal for the receiver. This is analogous to a similar structural
  result for expected-utility maximizing
  receivers~\citep[Proposition~4]{kamenicaG11}. 

In the setting of binary persuasion, our earlier results
  regarding full persuasion simplify and yield a simple condition to
  determine if the receiver can be fully persuaded. Suppose at all
  states, the sender strictly prefers action $1$. Then, using the
  notation of Section~\ref{sec:benefits-persuasion}, we have
  $\Upsilon_1 = \Omega$, and thus Theorem~\ref{thm:full-persuasion}
  implies that the receiver can be fully persuaded if and only if
  $\mu^* \in \conv(\calP_1)$.  As we prove in Lemma~\ref{lem:conv},
  under Assumption~\ref{as:convex}, we have
  $\conv(\calP_1) = \conv(K_1 \cup K_{01})$. Thus we obtain a simple
  criterion, namely whether $\mu^* \in \conv(K_1 \cup K_{01})$, to
  determine if full persuasion is possible.


Finally, the linear programming formulation allows us to
  further characterize the structure of the optimal signaling scheme
  when the differential utility function $\hat{\rho}$ satisfies a
  monotonicity condition. For simplicity, we focus on the case where
  $v(\omega, 1) - v(\omega, 0) = v > 0$ for all $\omega \in
  \Omega$. We consider the following condition on the differential
  utility function:
  \begin{assumption}[Monotonicity]\label{as:monotone} There exists a
    strict total order $\prec$ on $\Omega$ such that if
    $\omega \prec \omega'$, then either (1) $\omega \in K_1$ or (2)
    $\omega, \omega' \in L_0$ and
    $\gamma(\omega, \hat{\omega}) > \gamma(\omega', \hat{\omega})$ for
    all $\hat{\omega} \in K_1$.
  \end{assumption}

  Recall that for $\omega_0 \in L_0$ and $\omega_1 \in K_1$,
  $\gamma(\omega_0, \omega_1)$ is the largest value of
  $\gamma \in [0,1]$ such that the belief
  $\gamma \omega_0 + (1- \gamma) \omega_1$ lies in $\calP_1$. Thus,
  the preceding monotonicity condition implies that for any
  $\omega, \omega' \in L_0$ with $\omega \prec \omega'$ and for any
  $\omega_1 \in K_1$, if $\gamma \omega' + (1- \gamma) \omega_1$ lies
  in $\calP_1$, then so does $\gamma \omega + (1- \gamma) \omega_1$.

A wide class of differential utility functions satisfies
  monotonicity, including the class of expected utility functions. In
  particular, given a strict total order on $\Omega$, the condition
  holds for any differential utility function $\hat{\rho}(\cdot)$ that
  is strictly decreasing under the first-order stochastic ordering on
  $\mu$.

We have the following result with proof in
  Appendix~\ref{ap:binary-persuasion}:
  \begin{proposition}\label{prop:threshold-binary} Suppose $v(\omega, 1) - v(\omega, 0) = v > 0$
    for all $\omega \in \Omega$, and the differential utility function
    $\hat{\rho}$ satisfies Assumptions~\ref{as:convex}
    and~\ref{as:monotone}. Then the optimal signaling scheme induces a
    {\em threshold\/} structure in the receiver's actions: there
    exists an $\omega_0 \in L_0$ such that
    \begin{align*}
      t_1(\omega) &= \begin{cases} \mu^*(\omega) & \text{ if $\omega \prec \omega_0$;}\\
        0 &  \text{if $\omega_0 \prec \omega$.}
      \end{cases}
    \end{align*}
    In particular, under the optimal signaling scheme, the receiver
    chooses action $1$ in states $\omega \prec \omega_0$, and chooses
    action $0$ in states $\omega$ with $\omega_0 \prec \omega$.
  \end{proposition}
  Finally, we emphasize that it is the induced action of the receiver
  that has a threshold structure; the optimal signaling scheme
  typically possesses a more intricate structure in order to induce
  beliefs in $\Omega \cup K_{01}$. (We illustrate this structure in
  more detail in an application in
  Section~\ref{sec:signaling-queues}.)

\section{Broader Applicability to other Contexts}\label{sec:discussion}


Next, we discuss the broader applicability of our approach to more
general contexts beyond the literal interpretation of persuading
risk-conscious receivers.

As mentioned in the introduction, risk-consciousness arises naturally
in dynamic decision-making settings even with
expected-utility-maximizers. To illustrate, consider a two-period
setting where the receiver, after choosing an action $a \in A$ in the
first period, must choose an action $d \in D$ in the second period.
After taking the first-period action $a$ but prior to the choice of
$d$, the receiver obtains additional state information whose quality
depends on the choice of the first-period action. Formally, the
receiver observes the realization of a random variable $Z$ whose
distribution depends on $(\omega, a)$. A natural model of utility that
captures this setting is given by
\begin{align*}
  \rho(\mu, a) \defeq \expec_{\mu}\left[ u(\omega, a) + \max_{d \in D} \expec_{\mu}\left[ r_a(\omega, d) \big| Z, a  \right] \right],
\end{align*}
where $u(\omega,a)$ is the receiver's first-period utility, and
$r_a(\omega, d)$ denotes her utility in the second-period. Letting
$d_\mu(Z,a)$ denote the decision rule that attains the inner
maximization, the receiver's utility can be written as
$\rho(\mu, a) = \expec_\mu \left[ u(\omega, a) + r_a(\omega,
  d_\mu(Z,a))\right]$. It follows that, in general, $\rho(\mu, a)$ is
non-linear in $\mu$. A special case of this general model arises in
data markets, where a buyer must choose whether to purchase relevant
data ($a=1$) or not ($a=0$) before making a
decision~\citep{bergemannBS18,bergemannB15,bergemannB19,zhengC21}.
Letting $Z$ denote the information content of the data, we have
$\rho(\mu, 0) = \max_{d \in D} \expec_\mu[r_0(\omega, d)]$ and
$\rho(\mu, 1) = \expec_{\mu}\left[ \max_{d \in D} \expec_\mu[
  r_1(\omega, d) | Z]\right]$. In such setting, a seller who wishes to
persuade the buyer to purchase the data faces the problem of
persuading a risk-conscious receiver.

Our methods can be used to study {\em public
  persuasion}~\citep{arieliB19,yangIF19,dasKM17,candoganD17,bimpikisEM19}
of a group of interacting agents, where the sender shares information
publicly with all the agents. For any public signal, the agents share
a {\em common posterior} and subsequently play an equilibrium of an
incomplete information game. The sender seeks to publicly share
payoff-relevant information to influence the agents' choice of the
equilibrium. To apply our methods, we view the group of agents as a
single risk-conscious receiver, and the equilibrium profile as the
action chosen by the receiver. Then, for any equilibrium $a$, the set
$\calP_a$ describes the set of common posteriors for which the agents
play the same equilibrium profile $a$. With this mapping, our results
can be used to find the optimal public signaling scheme, as long as
the set of (relevant) equilibria over all common posteriors is
finite~(see, e.g., \citep{yangIF19}).




Another application of our methods is to {\em robust
  persuasion}~\citep{huW18,inostrozaP21,ziegler20}, where a sender
persuades a single receiver with a private type $\theta \in \Theta$.
The sender takes a worst-case view, and seeks to persuade the receiver
irrespective of her type. Formally, suppose the utility of the
type-$\theta$ receiver with belief $\mu$ for action $a$ is given by
$\rho_\theta(\mu, a)$, and let
$a_\theta(\mu) \in \arg\max_a \rho_\theta(\mu,a)$ denote the optimal
action chosen by the receiver. Let $v(a)$ denote the sender's
(state-independent) utility when the receiver chooses action $a$.
Since the receiver's type $\theta$ is unknown to the sender, she
maximizes the (expectation) of the minimum of her utility across all
receiver types: $\expec [ \min_{\theta} v(a_\theta(\mu_s))]$. Here,
$\mu_s$ is the receiver's posterior belief subsequent to persuasion.
Such a setting of robust persuasion maps to our model, with the sets
$\calP_a$ given by
$\calP_a = \{ \mu \in \Delta(\Omega) : v(a) = \min_\theta
v(a_\theta(\mu))\}$ for each $a \in A$. Thus, our results yield a
robust signaling scheme through a convex program.

\section{Application: Signaling in Unobservable Queues}%
\label{sec:signaling-queues}

We conclude the paper with an illustrative application of our
methodology to study information sharing in a service system where
arriving customers must choose whether or not to join an unobservable
queue to obtain service. The model is based on that of
\citet{lingenbrinkI19}, with the difference being that here customers
are not expected utility maximizers. Instead, inspired by the
literature on the psychology of waiting in queues~\citep{maister84},
we consider customers who exhibit uncertainty aversion. Formally, the
customers have a mean-standard deviation utility \citep{nikolovaS14,
  cominettiT16, lianeasNS18}, where the disutility for joining the
queue is the sum of the mean waiting time and a multiple of its
standard deviation. Moreover, the setting has a key difference from
the model in Section~\ref{sec:model}: the customers' (receiver's)
prior belief is endogenously determined from the equilibrium queue
dynamics. We show that our theoretical results carry over to this
setting, and establish that the optimal signaling scheme has an
intricate ``sandwich'' structure.

We consider a service system modeled as an unobservable $M/M/1/C$ FIFO
queue, i.e. a single-server queue with Poisson arrivals with rate
$\lambda$, independent exponential service times with unit mean, and
queue capacity $C > 0$. Upon arrival, each customer chooses whether to
join the queue to receive service ($a=1$) or leave without obtaining
service ($a=0$); a customer cannot join the queue if there are $C$
customers already in queue. We assume that customers are averse to
waiting, but cannot observe the queue length before making joining
decision. Instead, the service provider can observe the queue length
and communicate this information to arriving customers. The service
provider aims to maximize the queue throughput; if service is offered
at a fixed price, this translates to maximizing the revenue rate.

As the customer cannot join the queue if the queue length is $C$, the
relevant state space is $\Omega = \{0, 1, \ldots, C-1\}$, where the
state $\bar{\omega} \in \Omega$ describes the queue length upon a customer
arrival. To focus on throughput-maximization, we set
$v(\omega, a) = a$ for $\omega \in \Omega$ and $a \in \{0, 1\}$.

For an arriving customer, the payoff-relevant variable $X$ is their
waiting time until service completion. When $\bar{\omega} = n \in \Omega$, the waiting time $X$ is distributed as the sum of
$n+1$ independent unit exponentials (the waiting times for $n$
customers in the queue plus the customer's own service time). To
capture uncertainty aversion on the part of the
customers~\citep{maister84}, we focus on the following differential
utility function for joining the queue:
\begin{align}\label{eq:mean-stdev-rho}
  \hat{\rho}(\mu) \defeq \tau - \left(\expec_\mu[X] + \beta \sqrt{\var_\mu[X]}\right),
\end{align}
where $\mu \in \Delta(\Omega)$ is the customer's belief about the
queue length, $\tau >0$ captures her value for service, and
$\beta \geq 0$ captures her degree of risk-consciousness. It is
straightforward to verify that $\hat{\rho}(\cdot)$ satisfies convexity
(Assumption~\ref{as:convex}) and monotonicity
(Assumption~\ref{as:monotone}); see Appendix~\ref{ap:signaling-queues}
for the details.

Using the results from Section~\ref{sec:binary-persuasion} and the
same approach as in \citet{lingenbrinkI19} to handle endogenous
priors, we obtain that the service provider's signaling problem can be
optimized by solving the following linear program:
\begin{subequations}\label{eq:queue-linear}
  \begin{align}
    & \max_{t_0, t_1} \sum_{\omega \in \Omega} t_1(\omega) \notag \\
    \text{subject to,}\quad &t_1 \in \conv(K_1 \cup K_{01} \cup \{\zero\}), \label{eq:queue-leave}\\
    &t_0 \in \conv(L_0 \cup \{\zero\}), \label{eq:queue-join}\\
    &t_0(\omega+1) + t_1(\omega+1) =\lambda t_1(\omega ),\quad \text{
      for each $\omega < C-1$,} \label{eq:queue-balance}\\
    &\sum_{\omega \in \Omega} t_0(\omega) + \sum_{\omega \in \Omega} t_1(\omega) + \lambda t_1(C-1)  = 1. \label{eq:queue-normalize}
  \end{align}
\end{subequations}
Here, the objective captures the probability that an arriving customer
joins the queue, whereas the constraint~\eqref{eq:queue-balance}
captures the detailed-balance conditions on the steady-state
distribution of the queue. The constraint~\eqref{eq:queue-normalize}
is the normalization condition for the steady-state distribution (with
$\lambda_1 t_1(C-1)$ being the probability that the queue is at
capacity).

Using the monotonicity of the differential utility function,
  our first result shows the optimal signaling scheme induces a
  threshold structure on the customers' actions. The proof uses a
  perturbation argument similar to \citet{lingenbrinkI19}, and is
  presented in Appendix~\ref{ap:proof-thresh}.
\begin{lemma}\label{lem:thresh}
  An optimal signaling scheme induces a \emph{threshold} structure in
  the customers' actions: there exists an $m \in \Omega$ such that an
  arriving customer joins the queue if the queue length is strictly
  less than $m$ and leaves if it is strictly greater than $m$.
\end{lemma}

While the preceding lemma provides insights into the structure
  of the customers' actions, it does not reveal the structure of an
  optimal signaling scheme. The following main result of this section
  characterizes the intricate structure of an optimal signaling scheme
  using the canonical set of signals (as described in
  Section~\ref{sec:binary-persuasion}). The proof is given in
  Appendix~\ref{ap:proof-sandwich}.
  \begin{proposition}\label{prop:sandwich}
    Suppose the optimal solution $t = (t_0, t_1)$
    to~\eqref{eq:queue-linear} satisfies $t_0 \neq \zero$. Then, there
    exists an optimal signaling scheme with signals
    $S = \{ \join_1, \join_2, \ldots, \join_J, \leave\}$ for some
    $J \leq |\Omega|$, such that joining the queue is optimal under
    each signal $\join_j$, and leaving is optimal under the signal
    $\leave$. Furthermore,
    \begin{enumerate}
    \item For each $j \leq J$, a customer's utility upon receiving the
      signal $\join_j$ is zero, i.e., $\hat{\rho}(\mu_j) = 0$, where
      $\mu_j$ is the induced belief upon receiving signal $\join_j$.
    \item For each $j \leq J$, the induced belief $\mu_j$ either puts
      all its weight on a state $\omega^j \in K_1 \cap K_0$, or there
      exists two states $\omega_0^j \in L_0$ and $\omega_1^j \in K_1$
      such that $\mu_j$ puts positive weight only on the states
      $\omega_0^j$ and $\omega_1^j$. (In the former case, we define
      $\omega_0^j = \omega_1^j = \omega^j$.) These states form a {\em
        sandwich} structure:
      $\omega_1^1 \leq \omega_1^2 \leq \cdots \leq \omega_1^J \leq
      \omega_0^J \leq \omega_0^{J-1} \leq \cdots \leq \omega_0^1$.
    \item For each $j \leq J$, we have
      $\expec[X|\join_{j-1}] \leq \expec[X| \join_j]$ and
      $\var(X|\join_{j-1}) \geq \var(X|\join_j)$.
    \end{enumerate}
  \end{proposition}

\begin{figure}
  \begin{subfigure}[c]{0.5\textwidth}
    \centering
    \includegraphics[width = 0.95\textwidth]{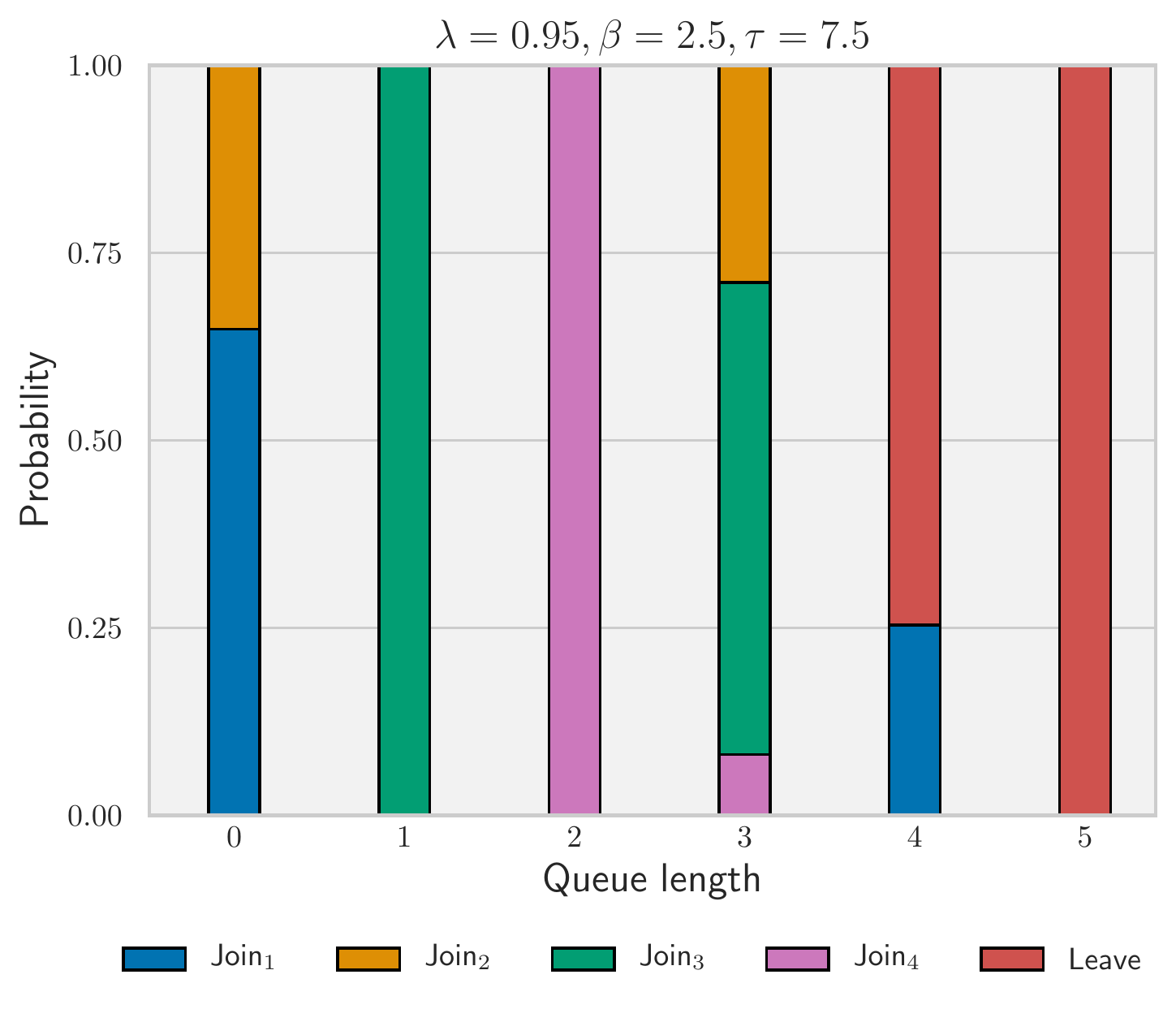} 
  \end{subfigure}%
  \begin{subfigure}[c]{0.5\textwidth}
\centering
\begin{tabular}{ccccc} 
  \toprule
  Signal $s$ & & Posterior belief & & $\expec[X|s]$  \\
  \midrule
  $\join_1$ & & $0.76 \cdot e_0 + 0.24 \cdot e_4$ & & 1.97   \\
  $\join_2$ & & $0.59 \cdot e_0 + 0.41 \cdot e_3$ & & 2.24   \\
  $\join_3$ & & $0.64 \cdot e_1 + 0.36 \cdot e_3$ & & 2.72   \\
  $\join_4$ & & $0.93 \cdot e_2 + 0.07 \cdot e_3$ & & 3.07  \\
  $\leave$  & &  $0.76 \cdot e_4 + 0.24 \cdot e_5$ & & 5.24 \\
  \bottomrule
\end{tabular}%
  \end{subfigure}%
\caption{Optimal signaling schemes}%
\label{fig:signal-example}
\end{figure}

%

The preceding proposition states implies that, assuming not all
customers join the queue, the customer's utility upon receiving any
$\join_j$ signal is zero, and hence
$\expec_\mu[X | \join_j] + \beta \sqrt{\var_\mu[X|\join_j]} = \tau$
for all $j \leq J$. Thus, the optimal signaling mechanism compensates
for the higher expected waiting times under some signals with lower
variance, leading to the sandwich structure in the induced posterior
beliefs.

We numerically illustrate the preceding result in
Figure~\ref{fig:signal-example}, for
$(\lambda, \beta, \tau) = (0.95, 2.5, 7.5)$ and $C = 100$. We depict
an optimal signaling scheme that uses $5$ signals ($\join_1$,
$\join_2$, $\join_3$, $\join_4$ and $\leave$), the first four of which
persuade the customer to join. The plot on the left shows the
probabilities of sending each signal conditional on the queue
length. For instance, if the queue length is $\omega = 3$ upon a
customer arrival, the signals $\join_2$, $\join_3$, $\join_4$ are sent
with probabilities $0.29$, $0.63$, and $0.08$, respectively.  As
established in Lemma~\ref{lem:thresh}, the plot reveals the threshold
structure induced on the customers' actions: a customer joins for
queue lengths smaller than $4$, and leaves for queue lengths larger
than $4$. The table on the right shows the induced posterior belief
upon receiving each signal. We see that each signal $\join_j$ puts
weight only on two states and these states form a sandwich structure.
The sandwich structure also orders the signals in the increasing order
of mean waiting times; since the customer's differential utility upon
receiving any $\join_j$ signal is zero, the variance of the waiting time
is decreasing. This ordering provides a convenient vocabulary for the
service provider to communicate with the customers: rather than
sending an abstract signal such as $\join_j$, the service provider can
directly convey the induced expected waiting times $\expec[X| \join_j]$
(or the induced uncertainty $\sqrt{\var(X|\join_j)}$).



\newpage

\begin{APPENDICES}
  \section{Relation to the Concavification Approach}%
\label{ap:kg-comparison}

In their seminal work, \citet{kamenicaG11} (hereafter [KG11]) present
a convex analytic concavification approach to find the optimal
signaling scheme in persuasion problems. Furthermore, in Proposition~4
of the online appendix of [KG11], the authors establish that the
number of signals in an optimal signaling scheme can be upper-bounded
by the cardinality $|\Omega|$ of the state space. While the focus in
[KG11] is on the case of an expected utility maximizing receiver, the
approach presented therein also applies to the case of risk-conscious
receiver.  In this appendix, we discuss how the concavification
approach relates to ours, and show how their results provide an
alternative path to arrive at the convex programming
formulation~\eqref{eq:sender-opt-convex}.

\subsection{Convex Analytic Argument of \citet{kamenicaG11}}

To describe their approach in detail, define
$\hat{v}(s) \defeq \sum_{\omega \in \Omega} s(\omega) v(\omega, a(s))$
for $s \in \Delta(\Omega)$, where
$a(s) \in \argmax_{a \in A} \rho(s, a)$ denotes the receiver's optimal
action under belief $s \in \Delta(\Omega)$. Using this definition and
Lemma~\ref{lem:splitting-lemma}, the persuasion
problem~\eqref{eq:sender-opt-beliefs} can be written as the following
problem:
\begin{align*}
  \begin{aligned}
   \max_{\eta \in \Delta(\Delta(\Omega))}     & \expec_\eta \left[  \hat{v}(\bar{s})\right] 
  \quad \text{subject to} \quad  \expec_{\eta}[ \bar{s}] = \mu^*.
\end{aligned}
\end{align*}
Thus, letting
$V(\mu) \defeq \sup_{ \eta \in \Delta(\Delta(\Omega)) } \{ \expec_\eta
\left[ \hat{v}(\bar{s}) \right] : \expec_\eta [ \bar{s} ] = \mu \}$,
the sender's largest payoff from persuasion is given by $V(\mu^*)$.

The main result of [KG11] is that $V(\cdot)$ is the smallest concave
function that dominates $\hat{v}$. In particular, the authors
establish that, for all $\mu \in \Delta(\Omega)$,
\begin{align}\label{eq:opt-charac}
  V(\mu) &= \sup \{ x : (x, \mu) \in \conv(\mathsf{hyp}(\hat{v}))\},
\end{align}
where
$\mathsf{hyp}(\hat{v}) \defeq \{ (x,s) : x \in \reals, s \in
\Delta(\Omega), x \leq \hat{v}(s)\}$ denotes the hypograph of
$\hat{v}$. Furthermore, the authors show that any representation of
$(\mu^*, V(\mu^*))$ as a convex combination of elements of
$\mathsf{hyp}(\hat{v})$ yields an optimal signaling scheme. Note
that~\eqref{eq:opt-charac} yields a convex program to compute
$V(\mu^*)$.

To obtain the bound on the number of signals in the optimal signaling
scheme, the authors apply the Fenchel-Bunt theorem (see the online
appendix of [KG11] for the detailed argument) to
$\mathsf{hyp}(\hat{v})$ to show that any
$(x,\mu) \in \conv(\mathsf{hyp}(\hat{v}))$ can be written as a convex
combination of at most $|\Omega|$ elements of $\mathsf{hyp}(\hat{v})$.
From this, the authors deduce the existence of an optimal signaling
scheme with at most $|\Omega|$ signals.

Our approach has many parallels with the preceding approach. First,
our approach requires us to compute the set $\calP_a$ for each
$a \in A$, and compute the convex hull
$\conv(\calP_a \cup \{\zero\})$. In contrast, the preceding approach
requires computing the set $\conv(\mathsf{hyp}(\hat{v}))$. Second, we
formulate the convex optimization problem~\eqref{eq:sender-opt-convex}
with variables $t_a \in \conv(\calP_a \cup \{\zero\})$ for each
$a \in A$. In contrast, the characterization~\eqref{eq:opt-charac} of
$V(\mu^*)$ suggests a convex optimization formulation with variables
$(x,\mu) \in \conv(\mathsf{hyp}(\hat{v}))$. Finally, we use the
Caratheodory theorem to split each optimal $t_a$ into at most
$\cara(\calP_a) \leq |\Omega|$ signals (see
Proposition~\ref{prop:signal-bound}). In contrast, the preceding
approach directly splits the point $(\mu^*, V(\mu^*))$ into at most
$|\Omega|$ signals using the Fenchel-Bunt theorem.

We note that since our splitting argument applies to each $t_a$
separately, our argument furnishes an upper-bound of at most
$\cara(\calP_a) \leq |\Omega|$ signals per action, and
$\sum_a \cara(\calP_a) \leq |\Omega| \cdot |A|$ signals in total, in
the optimal signaling scheme. In comparison, the argument of [KG11]
provides an upper-bound of at most $|\Omega|$ signals in total in the
optimal signaling scheme. Thus, when the state space is large and the
action space is small, our bound might be better, whereas the bound of
$|\Omega|$ might be better if the action space is large. We further
note that the advantage of our approach is that in some cases, it
provides a canonical set of signals, as we show in the case of binary
persuasion in Section~\ref{sec:binary-persuasion}.



\subsection{Alternative Argument Yielding Convex
  Formulation~\eqref{eq:sender-opt-convex}}

In this section, we present an alternative argument to arrive at the
convex programming formulation~\eqref{eq:sender-opt-convex} using the
upper-bound result in [KG11]. In fact, we show that this formulation
can be arrived at using any finite bound larger than $|\Omega|$ on the
number of signals in the optimal signaling scheme.

To begin, let $B \geq |\Omega|$ be a bound on the number of signals
required in the optimal signaling scheme.  Hence, for any receiver
action $a \in A$, there are at most $B$ signals that induce it under
the optimal signaling scheme. Thus, it suffices to consider a
signaling scheme, with signals
$S = \{ (a,i) : a \in A, 1 \leq i \leq B\}$, where each signal
$(a,i) \in S$ induces the receiver to take action $a \in A$. Thus, the
sender's optimization problem~\eqref{eq:sender-opt} can be
written as
\begin{align*}
  \max_{\pi} &\sum_{\omega \in \Omega} \sum_{ (a, i) \in S}  \pi( \omega, (a, i) ) v(\omega, a)\\
  \text{subject to, } & a \in \arg\max_{a' \in A} \rho(\mu_{(a,i)}, a'), \quad \text{for all $(a,i) \in S$,}\\
             &\pi(\omega, S) = \mu^*(\omega), \quad \text{for all $\omega \in \Omega$.}
\end{align*}
where
$\mu_{(a,i)}(\omega) = \frac{\pi(\omega, (a,i)) }{\sum_{\omega'}
  \pi(\omega', (a,i))}$ denotes the receiver's belief after receiving
the signal $(a,i) \in S$. Note that the constraint
$a \in \arg\max_{a' \in A} \rho(\mu_{(a,i)}, a')$ is essentially an
{\em obedience constraint}, which requires that the receiver, upon
receiving the signal $(a,i)$, finds it optimal to choose action
$a \in A$. Using the definition of $\calP_a$, this constraint can be
equivalently written as $\mu_{(a,i)} \in
\calP_a$. 
Letting $t_a(\omega) = \sum_{i =1}^B \pi(\omega, (a,i))$ for
each $a \in A$, the problem can be reformulated as
\begin{align*}
  \max_{\pi}   &\sum_{\omega \in \Omega} \sum_{ a \in A}  t_a(\omega)  v(\omega, a)\\
  \text{subject to, } &\mu_{(a,i)} \in \calP_a, \quad \text{for all $(a,i) \in S$,}\\
               &\sum_{a \in A} t_a(\omega) = \mu^*(\omega), \quad \text{for all $\omega \in \Omega$.}\\
               &  t_a(\omega) = \sum_{i=1}^B \pi\left(\omega, (a,i)\right), \quad
                 \text{for all $\omega \in \Omega$ and $a \in A$.}
\end{align*}

Let $k(a,i) = \sum_{\omega' \in \Omega} \pi(\omega', (a,i))$. Note
that $k(a,i) \geq 0$, and $\sum_{i=1}^B k(a,i) \leq 1$. Since
$\mu_{(a,i)}(\omega) = \frac{\pi(\omega, (a,i)) }{\sum_{\omega'}
  \pi(\omega', (a,i))} = \frac{\pi(\omega, (a,i))}{k(a,i)}$, we obtain
for all $a\in A$ and $\omega \in \Omega$,
\begin{align*}
  t_a(\omega) =\sum_{i=1}^B \pi\left(\omega, (a,i)\right)= \sum_{i=1}^B k(a,i)\mu_{(a,i)}(\omega) =  \sum_{i=1}^B k(a,i)\mu_{(a,i)}(\omega) + \left(1 - \sum_{i=1}^B k(a,i)\right) \zero. 
\end{align*}
Since $\mu_{(a,i)} \in \calP_a$ for all $1 \leq i \leq B$, we deduce
that $t_a \in \conv(\calP_a \cup \{\zero\})$. Hence, the problem can
be written as
\begin{align*}
  \max_{\pi}   &\sum_{\omega \in \Omega} \sum_{ a \in A}  t_a(\omega)  v(\omega, a)\\
  \text{subject to, } &t_a \in \conv(\calP_a \cup \{\zero\}), \quad \text{for all $a \in A$,}\\
               &\sum_{a \in A} t_a(\omega) = \mu^*(\omega), \quad \text{for all $\omega \in \Omega$.}\\
               &\mu_{(a,i)} \in \calP_a, \quad \text{for all $(a,i) \in S$,}\\
               &  t_a(\omega) = \sum_{i=1}^B \pi(\omega, (a,i)), \quad
                 \text{for all $\omega \in \Omega$ and $a \in A$.}
\end{align*}
The formulation \eqref{eq:sender-opt-convex} then follows from an
application of the Caratheodory's theorem, which implies that in the
preceding optimization problem, the last two constraints are redundant
(implied by the first constraint) as long as $B \geq |\Omega|$, and
the optimization can be done directly over $\{t_a\}$ instead of
$\pi$. The optimal signaling scheme $\pi$ (along with the beliefs
$\mu_{(a,i)}$) can then be obtained from the optimal $t_a$
using~\eqref{eq:scheme-from-t}, as described in the discussion
following Theorem~\ref{thm:convex}.


  \section{Proofs from Section~\ref{sec:tractable}}\label{ap:tractable}

In this section, we provide the missing proofs from
Section~\ref{sec:tractable}. Before we proceed, we recall the
definition of a Bayes-plausible measure~\citep{kamenicaG11}:
\begin{definition} A measure $\eta \in \Delta(\Delta(\Omega))$ is
  Bayes-plausible for prior $\mu^* \in \Delta(\Omega)$ if
  $\expec_\eta[ \bar{s}] = \mu^*$, where $\bar{s} \in \Delta(\Omega)$
  is distributed as $\eta$.
\end{definition}
With this definition, we begin with the proof of Lemma~\ref{lem:splitting-lemma}.
\proof{Proof of Lemma~\ref{lem:splitting-lemma}.} Consider a signaling
scheme $\pi \in \Delta(\Omega \times \Delta(\Omega))$ satisfying
$\mu_s = s$ for almost all $s \in \Delta(\Omega)$. Let
$\eta(\cdot) \defeq \pi(\Omega, \cdot) \in \Delta(\Delta(\Omega))$. By
definition, we have for any $\omega \in \Omega$,
\begin{align*}
  \mu^*(\omega) = \pi(\omega, \Delta(\Omega)) = \expec_\pi[ \ind\{\bar{\omega} = \omega\}] = \expec_\pi \left[ \expec_\pi [ \ind\{\bar{\omega} = \omega\} | \bar{s} ] \right] = \expec_{\pi} [ \mu_{\bar{s}} (\omega) ] = \expec_\pi [ \bar{s}(\omega)] = \expec_\eta[ \bar{s}(\omega)].
\end{align*}
Thus, $\eta$ is Bayes-plausible.

Next, let $\eta \in \Delta(\Delta(\Omega))$ be Bayes-plausible, and
let $\pi(\omega, ds) \defeq s(\omega) \eta(ds)$ for all
$\omega \in \Omega$ and $s \in \Delta(\Omega)$. Observe that
$\pi(\omega, \Delta(\Omega)) = \int_{\Delta(\Omega)} s(\omega)\eta(ds)
= \expec_\eta[\bar{s}(\omega)] = \mu^*(\omega)$, where the final
equality follows from the Bayes-plausibility of $\eta$. Hence, $\pi$
is a valid signaling scheme. Using Bayes' rule, for any
$s \in \Delta(\Omega)$ ($\eta$-almost surely), we have
$\mu_s(\omega) = \prob_\pi( \bar{\omega} = \omega | \bar{s} = s) =
\pi(\omega, ds)/ \pi(\Omega, ds) = s(\omega)$ for all
$\omega \in \Omega$. Thus, under $\pi$, the receiver's belief $\mu_s$
upon receiving signal $s \in \Delta(\Omega)$ satisfies $\mu_s = s$
almost surely. \Halmos\endproof

\begin{lemma}\label{lem:intermediate-reduction}
  The sender's persuasion problem~\eqref{eq:sender-opt-beliefs} is
  equivalent to the following optimization problem over
  Bayes-plausible measures $\eta$:
  \begin{align}\label{eq:sender-opt-bayes}
\begin{aligned}
  \max_{\eta \in \Delta(\Delta(\Omega))}     &  \expec_\eta \left[ \sum_{\omega \in \Omega}  \bar{s}(\omega) v(\omega, a(\bar{s})) \right]\\
  \text{subject to, } \quad     a(s) &\in \argmax_{a \in A} \rho(s,
  a), \quad \text{for all $s \in \Delta(\Omega)$,}\\
    \quad \expec_\eta [ \bar{s}(\omega)] &= \mu^*(\omega), \quad \text{for each $\omega \in \Omega$.}                        
\end{aligned}
  \end{align}  
\end{lemma}
\proof{Proof.} From Lemma~\ref{lem:splitting-lemma}, we obtain that
for each Bayes-plausible $\eta$, there exists a corresponding
signaling scheme $\pi \in \Delta(\Omega \times \Delta(\Omega))$
satisfying $\mu_s = s$ for almost all $s \in \Delta(\Omega)$, and
conversely for each such scheme $\pi$, the corresponding measure
$\eta(\cdot) = \pi(\Omega, \cdot)$ is Bayes-plausible. Furthermore,
observe that for any Bayes-plausible measure $\eta$, the sender's
expected utility under the corresponding signaling scheme
$\pi(\omega, ds) = s(\omega)\eta(ds)$ can be written as
\begin{align*}
  \expec_\pi [ v(\bar{\omega}, a(\bar{s})) ]
  = \expec_\pi \left[ \expec_\pi \left[ v(\bar{\omega}, a(\bar{s})) \big| \bar{s} \right] \right]
  &= \expec_\pi \left[ \sum_{\omega \in \Omega} \mu_{\bar{s}}(\omega) v(\omega, a(\bar{s}))\right]\\
  &= \expec_\pi \left[ \sum_{\omega \in \Omega} \bar{s}(\omega) v(\omega, a(\bar{s}))\right]
  = \expec_\eta \left[ \sum_{\omega \in \Omega} \bar{s}(\omega) v(\omega, a(\bar{s}))\right],
\end{align*}
where the third equality follows from the fact that $\mu_s = s$ for
almost all $s \in \Delta(\Omega)$, and the final equality follows from
the definition of $\eta$. Since the sender's expected utility can be
written as a function of the receiver's strategy and the probability
measure $\eta$, we obtain the reformulation in the lemma
statement.~\Halmos\endproof

\proof{Proof of Lemma~\ref{lem:main-reduction}.} Fix any
Bayes-plausible measure $\eta$ and an optimal receiver strategy
$a(\cdot)$. For each $a \in A$, define
$b_a \defeq \prob_\eta ( a(\bar{s}) =a ) \in [0,1]$ to be the
probability that the receiver chooses action $a$ under the
corresponding signaling scheme. For each $a \in A$, if $b_a = 0$, then
let $\eta_a$ be any probability measure with support contained in
$\calP_a$. Otherwise, define $\eta_a$ to the measure obtained by
conditioning $\eta$ on the event $a(\bar{s}) = a$. More precisely, we
have $\eta_a(ds) \defeq \frac{1}{b_a} \ind\{a(s) = a\} \eta(ds)$ if
$b_a >0$. Note that, by the definition of the sets $\calP_a$, the
support of $\eta_a$ is contained in $\calP_a$ for each $a \in A$. The following
equations are immediate from the definitions:
\begin{align*}
  \sum_{a \in A} b_a \eta_a  &= \eta, \quad 
                               \sum_{a \in A} b_a = 1.
\end{align*}
We let $m_a(\omega) \defeq \expec_{\eta_a} [ \bar{s}(\omega) ]$ for
each $\omega \in \Omega$. Note that if $b_a > 0$, then
$m_a(\omega) = \expec_\eta [ \bar{s}(\omega) | a(\bar{s}) = a]$. Thus,
$m_a(\omega)$ is the {\em mean-posterior belief} of the receiver that
the state is $\omega$, given that she chooses the action $a$.  From
the Bayes-plausibility of $\eta$, we obtain for each
$\omega \in \Omega$:
\begin{align*}
    \sum_{a \in A} b_a m_a(\omega) &= \sum_{a \in A} b_a \expec_{\eta_a}\left[ \bar{s}(\omega) \right] = \expec_\eta \left[ \bar{s}(\omega) \right] = \mu^* (\omega),
  \end{align*}
  where the second equality follows from the fact that
  $\sum_{a \in A} b_a \eta_a = \eta$.  Moreover, it is straightforward
  to verify that $m_a \in \conv(\calP_a)$, since $m_a$ is the mean of
  the posterior distribution $\eta_a$ with support contained in the closed set
  $\calP_a$. Finally, note that for each $\omega \in \Omega$, we have
  \begin{align*}
    \sum_{a \in A} b_a m_a(\omega)v(\omega, a)
    &=  \sum_{a \in A} b_a \expec_{\eta_a} \left[ \bar{s}(\omega) \right] v(\omega ,a)\\
    &=  \sum_{a \in A} b_a \expec_{\eta_a}\left[ \bar{s}(\omega) v(\omega ,a(\bar{s})) \right]\\
    &= \expec_\eta \left[ \bar{s}(\omega) v(\omega,a(\bar{s}))\right],    
  \end{align*}
  where the first equality uses the definition of $m_a$, the second
  equality follows from the fact that $a(\bar{s}) = a$ when
  $\bar{s} \sim \eta_a$, and the third equality follows from the fact
  that $\sum_{a \in A} b_a \eta_a = \eta$.

  Conversely, suppose we have
  $\{(b_a, m_a) : b_a \in [0,1], m_a \in \conv(\calP_a)\}_{a \in A}$
  with $\sum_{a \in A} b_am_a = \mu^*$. By the definition of the
  convex hull, $m_a \in \conv(\calP_a)$ implies the existence of
  $\{ (\mu^a_{i}, \lambda^a_i) : i = 1, \ldots, j_a\}$ such that
  $\mu^a_i \in \calP_a$ and $\lambda_i^a \geq 0$ for each $i \leq j_a$
  with $\sum_{i=1}^{j_a} \lambda^a_i=1$ and
  $m_a = \sum_{i=1}^{j_a} \lambda^a_i \mu^a_i$. Define
  $\eta \in \Delta(\Delta(\Omega))$ to be the discrete distribution
  that selects the posterior $\mu^a_i$ with probability
  $b_a\lambda^a_i$. Then, we have for all $\omega \in \Omega$,
\begin{align*}
  \expec_\eta [ \bar{s}(\omega) ]
  = \sum_{a \in A} \sum_{i=1}^{j_a} b_a \lambda^a_i \mu^a_i(\omega)
  = \sum_{a \in A} b_a  \left(\sum_{i=1}^{j_a} \lambda^a_i \mu^a_i(\omega)\right)
  = \sum_{a \in A} b_a  m_a(\omega)
  = \mu^*(\omega).
\end{align*}
This proves the Bayes-plausibility of $\eta$. Finally, define the
strategy $a(\cdot) : \Delta(\Omega) \to A$ so that $a(\mu_i^a) = a$
for each $i \leq j_a$ and $a \in A$, and for other values of $\mu$,
let $a(\mu)$ be an arbitrary element in
$\arg\max_{a \in A} \rho(\mu,a)$. Since $\mu_i^a \in \calP_a$, it is
straightforward to verify that the strategy $a(\cdot)$ is
optimal. Finally, we have for each $\omega \in \Omega$,
\begin{align*}
  \sum_{a \in A} b_a m_a(\omega)v(\omega,a)
  &= \sum_{a \in A} b_a \sum_{i=1}^{j_a} \lambda_i^a \mu_i^a(\omega) v(\omega,a)\\
  &= \sum_{a \in A} \sum_{i=1}^{j_a} \left( b_a \lambda_i^a\right) \cdot \mu_i^a(\omega) \cdot  v(\omega,a(\mu_i^a))\\
  &= \sum_{a \in A} \sum_{i=1}^{j_a} \eta(\mu_i^a) \cdot \mu_i^a(\omega)\cdot  v(\omega,a(\mu_i^a))\\
  &= \expec_\eta \left[ \bar{s}(\omega) v(\omega, \bar{s}) \right].
\end{align*}
Here, the first equation follows from the fact that
$m_a = \sum_{i=1}^{j_a} \mu_i^a \lambda_i^a$, the second equation
follows from the fact that $a(\mu_i^a) = a$, and the third equation
follows from the definition of $\eta$. This completes the proof of the
lemma.
\Halmos\endproof

\proof{Proof of Theorem~\ref{thm:convex}.} Note that for any
Bayes-plausible $\eta \in \Delta(\Delta(\Omega))$,
Lemma~\ref{lem:main-reduction} guarantees a corresponding
$\{(b_a, m_a)\}_{a \in A}$ with $b_a \in [0,1]$ and
$m_a \in \conv(\calP_a)$ satisfying~\eqref{eq:bayes-plausible} and
\eqref{eq:objective}. Conversely, for each such
$\{(b_a, m_a)\}_{a \in A}$, there exists a Bayes-plausible
$\eta$. Thus, using Lemma~\ref{lem:intermediate-reduction}, we can
reframe the sender's persuasion problem~\eqref{eq:sender-opt-bayes} as
\begin{equation*}
\begin{aligned}
  \max_{\{b_a, m_a : a  \in A\}} ~& \sum_{\omega \in \Omega} \sum_{a \in A}  b_am_a(\omega)v(\omega, a)\\
  \text{subject to, }\quad   & \sum_{a \in A} b_a m_a = \mu^*,\\
  m_a &\in \conv(\calP_a), \quad b_a \in [0,1] \quad \text{ for each $a \in A$.}
\end{aligned}
\end{equation*}
Substituting $t_a = b_am_a$ for each $a \in A$, we obtain that for
each feasible $\{(b_a, m_a)\}_{a \in A}$, we have
$t = \{t_a\}_{a \in A}$ feasible for~\eqref{eq:sender-opt-convex},
with equal objective values. Conversely, for any feasible
$t = \{t_a\}_{a \in A}$, $t_a \in \conv(\calP_a \cup \{\zero\})$
implies $t_a = b_a m_a + (1 - b_a) \zero$ for some $b_a \in [0,1]$ and
$m_a \in \conv(\calP_a)$. It follows that such
$\{(b_a, m_a)\}_{a \in A}$ is feasible for the preceding program, with
the same objective value. Thus, we obtain that the
preceding program and~\eqref{eq:sender-opt-convex} are
equivalent.~\Halmos\endproof

\proof{Proof of Proposition~\ref{prop:signal-bound}.} Since the
optimal $m_a$ lies in the set $\conv(\calP_a)$, it follows that $m_a$
can be written as a convex combination of at most $\cara(\calP_a)$
points in $\calP_a$. As detailed in the discussion preceding the
proposition statement, one can then construct an optimal signaling
scheme using such a convex combination that sends, for each $a \in A$,
at most $\cara(\calP_a)$ signals for which the receiver's optimal
action is $a$.  Hence, the total number of signals is at most
$\sum_{a \in A} \cara(\calP_a)$.

Using Caratheodory's theorem, we have
$\cara(\calP_a) \leq \dim(\calP_a) + 1$, where $\dim(H)$ is the
dimension of the smallest affine space containing $H$. Since the set
$\calP_a \subseteq \Delta(\Omega)$ lies in an affine space of
dimension $\reals^{|\Omega| -1}$, we obtain
$\cara(\calP_a) \leq |\Omega|$.  \Halmos\endproof

\section{Proofs from
  Section~\ref{sec:benefits-persuasion}}\label{ap:benefits-persuasion}

\proof{Proof of Proposition~\ref{prop:benefits-persuasion}.} We begin
by showing that the two parts of the proposition statement are
equivalent, and hence it suffices to prove the second part.

Suppose there is information the sender would share, i.e, there exists
a belief $\mu$ with
$\hat{\nu}(\mu) > \expec_{\mu}[v(\bar{\omega},a_{\mu^*}]$. Then, for
$a = a_\mu$ and $m_a = \mu$, we have
$m_a \in \calP_a \subseteq \conv(\calP_a)$, and
$\sum_{\omega \in \Omega} m_a(\omega) \left(v(\omega, a) - v(\omega,
  a_{\mu^*})\right) = \expec_{\mu}[v(\bar{\omega}, a_\mu) -
v(\bar{\omega}, a_{\mu^*})] > 0$. Conversely, suppose there is no
information the sender would share. Then, for any $\mu \in \calP_a$,
we have
$\expec_\mu[v(\bar{\omega}, a)] = \hat{\nu}(\mu) \leq \expec_\mu[
v(\bar{\omega}, a_{\mu^*})]$. Using linearity of expectation, we
obtain
$\expec_\mu[v(\bar{\omega}, a)] \leq \expec_\mu[ v(\bar{\omega},
a_{\mu^*})]$ for all $\mu \in \conv(\calP_a)$, implying that
$\sum_{\omega \in \Omega} m_a(\omega) \left(v(\omega, a) - v(\omega,
  a_{\mu^*})\right) \leq 0$ for all $m_a \in \conv(\calP_a)$ and
$a \in A$. Thus, the two parts of the proposition statement are
equivalent, and we next prove the second part.

Suppose for all $a \in A$ and $m_a \in \conv(\calP_a)$, we have
$\sum_{\omega \in \Omega} m_a(\omega)\left(v(\omega, a) - v(\omega,
  a_{\mu^*})\right) \leq 0$.  Then, for any feasible
$t = \{t_a\}_{a \in A}$, let $b_a\in [0,1]$ and
$m_a \in \conv(\calP_a)$ be such that $t_a = b_a m_a$. We have
\begin{align*}
  \sum_{a \in A} \sum_{\omega \in \Omega} t_a(\omega) v(\omega, a)
  &=   \sum_{a \in A} b_a \left( \sum_{\omega \in \Omega} m_a(\omega) v(\omega, a) \right)\\
  &\leq   \sum_{a \in A} b_a \left( \sum_{\omega \in \Omega} m_a(\omega) v(\omega, a_{\mu^*}) \right)\\
  &=  \sum_{\omega \in \Omega} \mu^*(\omega) v(\omega, a_{\mu^*}) \\
  &= \hat{\nu}(\mu^*),
\end{align*}
where in the penultimate equality we use
$\mu^* = \sum_{a \in A} b_a m_a$. Since this holds for all feasible
$t$, we obtain $V(\mu^*) =\hat{\nu}(\mu^*)$, and hence the sender does
not benefit from persuasion.

Next, suppose there exists an $a \in A$ and $m_a \in \conv(\calP_a)$
with
$\sum_{\omega \in \Omega} m_a(\omega)\left(v(\omega, a) - v(\omega,
  a_{\mu^*})\right) > 0$. If $\mu^*$ lies in the interior of
$\calP_{\mu^*}$, then there exists a $\delta$-ball $B_\delta$ around
$\mu^*$ contained in $\calP_{\mu^*}$. Let $\mu' \in B_\delta$ be such
that $\mu^* = \gamma \mu' + (1 - \gamma) m_a$ for some
$\gamma \in (0,1)$. Then, let $t_{a'} = \gamma \mu'$ for
$a' = a_{\mu^*}$, let $t_{a} = (1-\gamma)m_a$, and let
$t_{a'} = \zero$ for $a' \neq a_{\mu^*}, a$. It follows that
$t = \{ t_{a'} \}_{a' \in A}$ is feasible for
\eqref{eq:sender-opt-convex}. We have
\begin{align*}
  V(\mu^*) &\geq \sum_{\omega \in \Omega} \sum_{a' \in A} t_{a'}(\omega)
             v(\omega, a')\\
           &= \gamma \sum_{\omega \in \Omega} \mu'(\omega) v(\omega,
             a_{\mu^*}) + (1 - \gamma) \sum_{\omega \in \Omega} m_a(\omega)
             v(\omega, a)\\
           &> \gamma \sum_{\omega \in \Omega} \mu'(\omega) v(\omega,
             a_{\mu^*}) + (1 - \gamma) \sum_{\omega \in \Omega} m_a(\omega)
             v(\omega, a_{\mu*})\\
           &= \sum_{\omega \in \Omega} \mu^*(\omega) v(\omega, a)\\
           &= \hat{\nu}(\mu^*).
\end{align*}
Thus, the sender (strictly) benefits
from persuasion.~\Halmos\endproof

  \proof{Proof of Theorem~\ref{thm:full-persuasion}.} Let
  $\pi \in \Delta(\Omega \times S)$ be a signaling scheme that fully
  persuades the receiver. Then, for any $s \in S$ with $a(s) = a$, the
  induced belief $\mu_s$ lies in the set $\calP_a$ almost surely
  (w.r.t. $\pi$).  Thus, we obtain
  $\expec_\pi[ \mu_{\bar{s}} \ind\{ a(\bar{s}) =a \}] \in
  \conv(\calP_a \cup \{\zero\})$. Using the fact that
  $\mu_s(\omega) = \prob_\pi(\bar{\omega} = \omega | \bar{s} = s)$, we
  obtain, using the tower property of conditional expectation, that
  \begin{align*}
    \expec_\pi[\mu_{\bar{s}}(\omega) \ind\{ a(\bar{s}) =a \}]
    &= \expec_\pi[ \ind\{ \bar{\omega} = \omega, a(\bar{s}) =a\}]\\
    &= \expec_\pi[ \ind\{ \bar{\omega} = \omega, \bar{\omega} \in
      \Upsilon_a\}]\\
    &= \mu^*(\omega) \ind\{\omega \in \Upsilon_a\}\\
    &=  t_a^*(\omega),
  \end{align*}
  where the second equality follows from the fact that, since $\pi$
  fully persuades the receiver and $\{\Upsilon_a : a \in A\}$ is a
  partition, we have $a(\bar{s}) = a$ if and only if
  $\bar{\omega} \in \Upsilon_a$ ($\pi$-almost surely). Thus, we obtain
  that $t_a^* \in \conv(\calP_a \cup \{\zero\})$. Moreover, it is
  straightforward to verify that $\sum_{a \in A} t_a^* = \mu^*$. Thus,
  $t^*$ is feasible for \eqref{eq:sender-opt-convex}.

  Conversely, suppose $t^*$ is feasible for~\eqref{eq:sender-opt-convex}. From
  $t_a^* \in \conv(\calP_a \cup \{\zero\})$, we obtain that there
  exists $\{\mu_i^a \in \calP_a \colon i=1, \ldots, j_a\}$ and
  $\{\lambda_i^a \geq 0 \colon i = 1, \ldots, j_a\}$ such that
  $\sum_{i=1}^{j_a} \lambda_i^a \leq 1$ and
  $t_a^* = \sum_{i=1}^{j_a} \lambda_i^a \mu_i^a$. Since
  $\sum_{\omega \in \Omega} \sum_{a \in A} t_a^*(\omega) =
  \sum_{\omega \in \Omega} \mu^*(\omega) = 1$, we obtain that
  $\sum_{a \in A} \sum_{i = 1}^{j_a} \lambda_i^a = 1$. Thus, we
  $\mu^* = \sum_{ a \in A} \sum_{i=1}^{j_a} \lambda_i^a \mu_i^a$ with
  $\sum_{a \in A} \sum_{i=1}^{j_a} \lambda_i^a = 1$. This implies that
  there exists a valid signaling scheme $\pi$ that induces beliefs
  $\{\mu_i^a : i=1, \ldots, j_a, a \in A\}$. Furthermore, since
  $t_a^*(\omega) = \mu^*(\omega) \ind\{\omega \in \Upsilon_a\}$ for
  each $\omega \in \Omega$, it must be the case that
  $\mu_i^a \in \Delta(\Upsilon_a)$. Thus, under $\pi$, if the receiver
  chooses an action $a \in A$, then the realized state is in the set
  $\Upsilon_a$. This implies that $\pi$ fully persuades the
  receiver.~\Halmos\endproof

\section{Proofs from
  Section~\ref{sec:binary-persuasion}}\label{ap:binary-persuasion}
In this section, we provide the proofs of the results in
Section~\ref{sec:binary-persuasion}. We begin with the following
simple argument showing that the quasiconvexity of $\hat{\rho}(\cdot)$
implies Assumption~\ref{as:convex}.

\proof{Proof of Lemma~\ref{lem:quasiconvexity}.} The proof follows
from the fact that if $\hat{\rho}$ is quasiconvex, then
$\hat{\rho}(\gamma \mu + (1- \gamma) \mu') \leq \max\{
\hat{\rho}(\mu), \hat{\rho}(\mu')\}$ for $\gamma \in [0,1]$ and
$\mu , \mu' \in \Delta(\Omega)$. \Halmos\endproof

We next focus on the proof of Theorem~\ref{thm:linear}. The proof of
the theorem rests on two helper lemmas that characterize the geometry
of the sets $\Delta(\Omega)$ and $\conv(\calP_1)$.  The first lemma
shows that the set $\Delta(\Omega)$ can be viewed as the union of two
regions, each of which is the convex hull of a finite set of
points. Fig.~\ref{fig:conv} and Fig.~\ref{fig:conv-alt} illustrate the
geometric intuition behind this lemma.

\begin{figure}
    \centering
    \includegraphics[width = 0.33\textwidth]{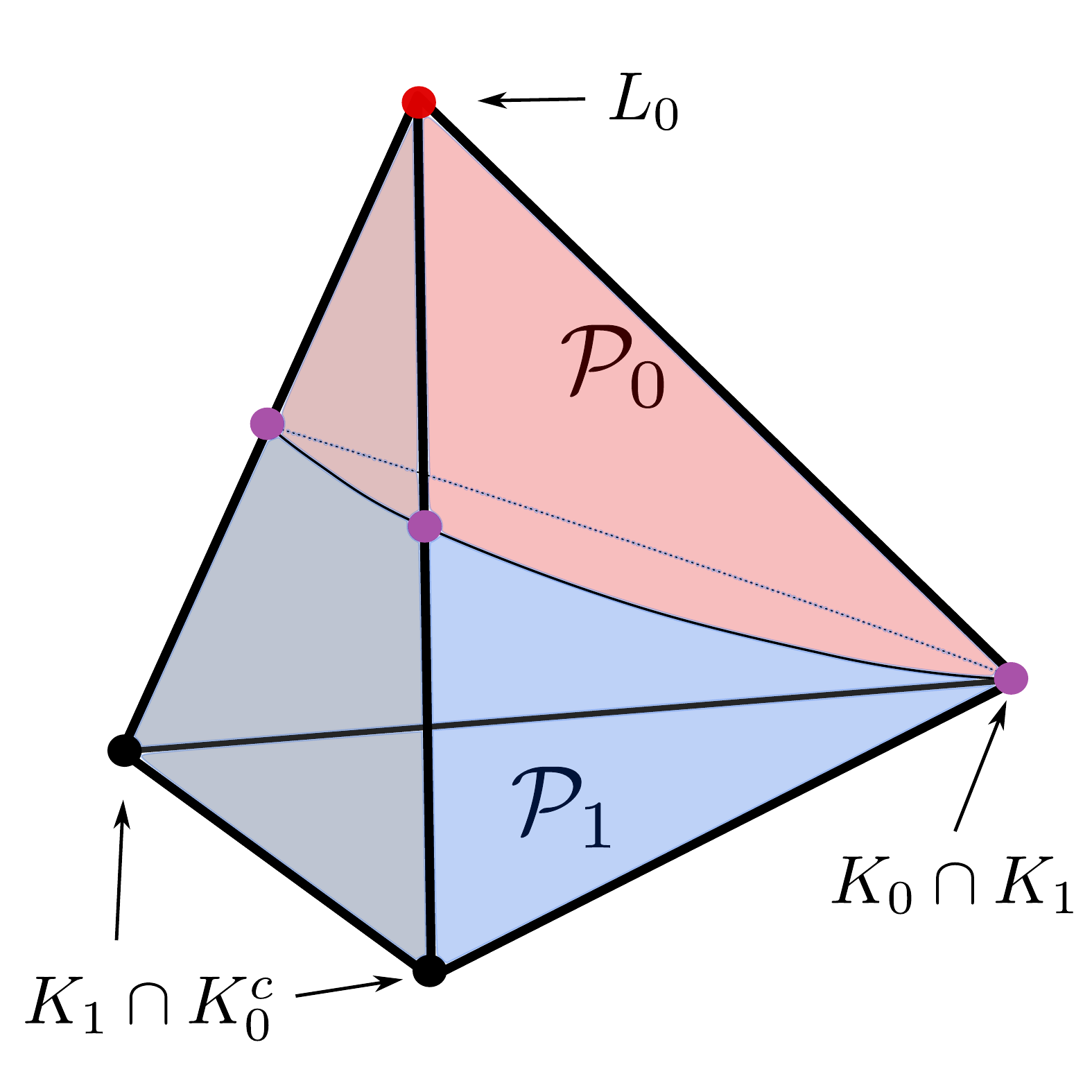}
    \caption{The sets $L_0, K_0$ and $K_1$ when $K_0 \neq L_0$. Here,
      $K_{01}$ is the set of the purple vertices.
      \label{fig:k0k1k01-1}}
\end{figure}

\begin{figure}
  \begin{subfigure}[c]{0.33\textwidth}
    \centering
    \includegraphics[width = \textwidth]{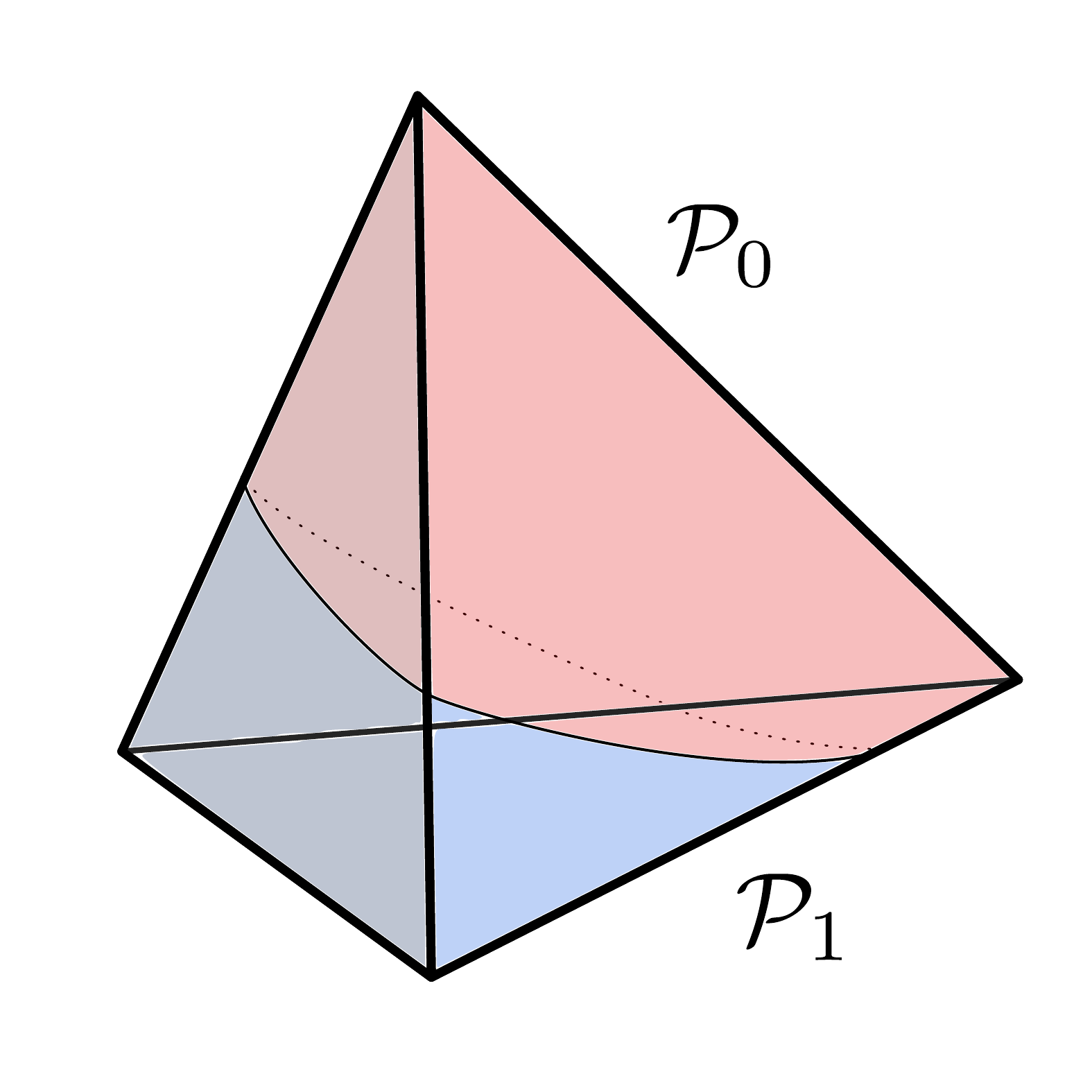}
  \end{subfigure}\begin{subfigure}[c]{0.33\textwidth}
    \centering
    \includegraphics[width = \textwidth]{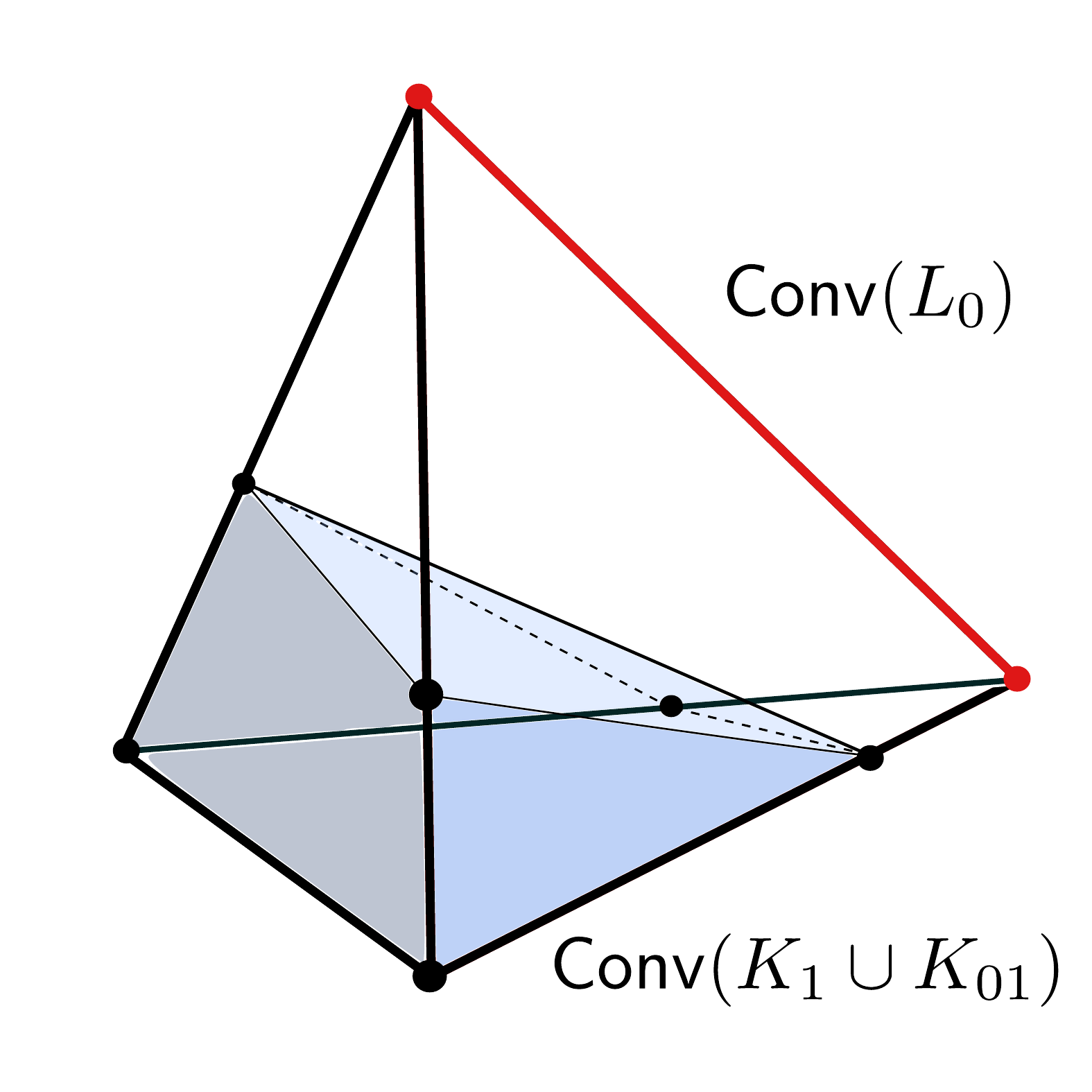}
  \end{subfigure}\begin{subfigure}[c]{0.33\textwidth}
    \centering
    \includegraphics[width = \textwidth]{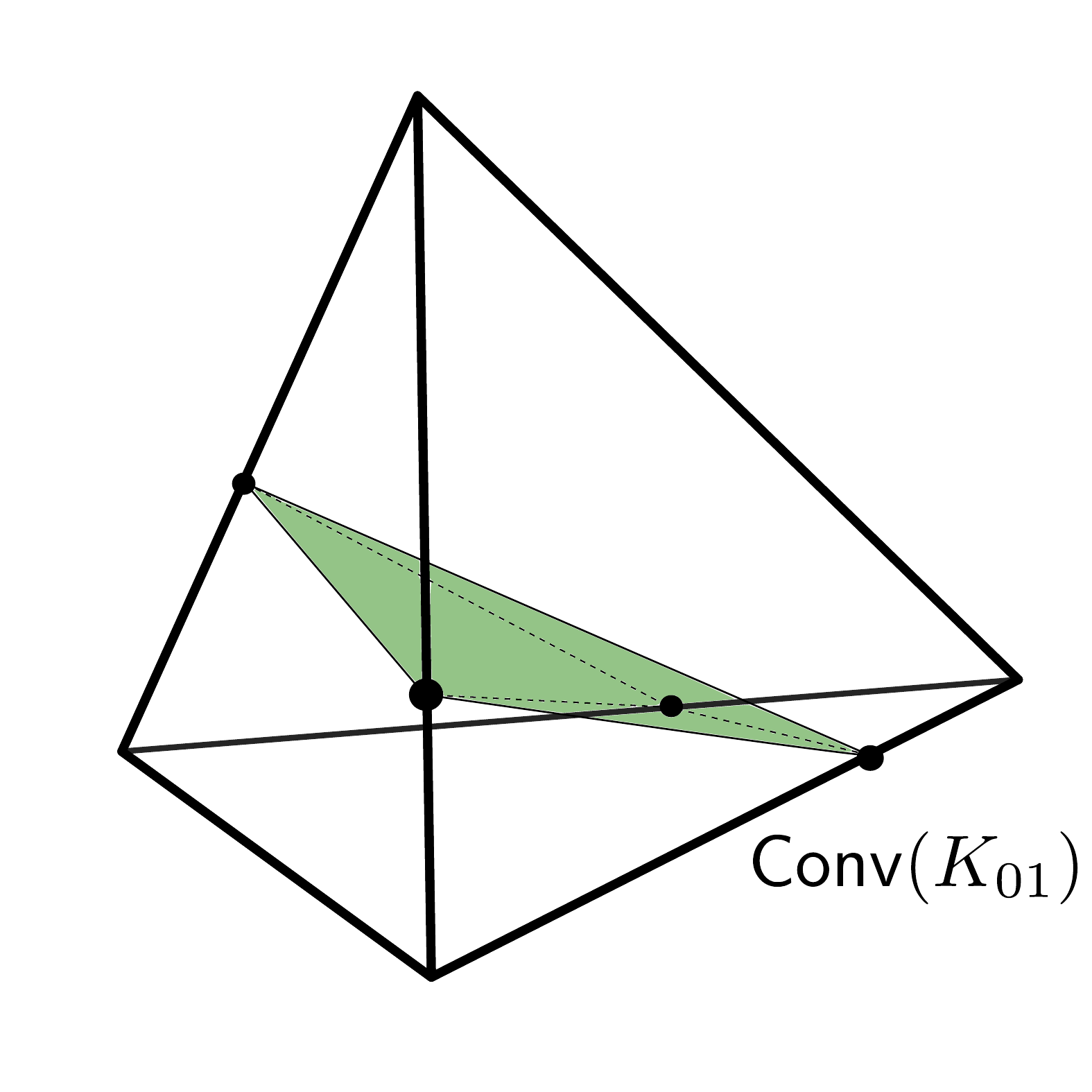}
  \end{subfigure}
  \caption{Geometry of $\Delta(\Omega)$ when $\conv(K_{01})$
      is not a hyperplane. Here, $\conv(L_0)$ is an edge of the
      simplex, shown in red.}\label{fig:conv-alt}
\end{figure}
Recall that $L_0 \defeq K_0 \cap K_1^c$ denotes the set of states for
which action $0$ is uniquely optimal for the receiver.
\begin{lemma}\label{lem:omega}
  $\Delta(\Omega) = \conv(L_0 \cup K_{01})\cup \conv(K_1 \cup
  K_{01})$.
\end{lemma}
\proof{Proof.} Let $\mu \in \Delta(\Omega)$. Since
$\Omega = L_0 \cup K_1$, we have
$\mu \in \conv(L_0 \cup K_1) = \conv(L_0 \cup K_1 \cup K_{01})$.
Consider any convex decomposition of
$\mu = \sum_{ \phi \in L_0 \cup K_1 \cup K_{01}} \alpha_\phi \phi $.
If the convex decomposition does not place positive weights either on
the elements of $L_0$ or on those of $K_1\setminus K_{01}$, then we
are done. Otherwise, let $\omega_0 \in L_0$ and
$\omega_1 \in K_1\setminus K_{01}$ be such that $\alpha_\phi > 0$ for
$\phi \in \{\omega_0, \omega_1\}$. Let
$ \mu' \defeq \frac{1}{\alpha_{\omega_0} + \alpha_{\omega_1}}(
\alpha_{\omega_0} \omega_0 + \alpha_{\omega_1} \omega_1)$, and note
that
$\mu = (\alpha_{\omega_0} + \alpha_{\omega_1}) \mu' + \sum_{\phi \in
  L_0 \setminus \{\omega_0\}} \alpha_\phi \phi + \sum_{\phi \in K_{01}
  \cup K_1 \setminus \{\omega_1\}} \alpha_\phi \phi$. Since
$\mu' \in \conv(\{\omega_0, \omega_1\})$, one can write $\mu'$ as
either a convex combination of $\omega_0$ and
$\chi(\omega_0, \omega_1)$ or a convex combination of $\omega_1$ and
$\chi(\omega_0, \omega_1)$. In either scenario, using this
decomposition for $\mu'$, we obtain a convex decomposition of $\mu$
that places positive weight on at least one fewer element of
$L_0 \cup (K_1\setminus K_{01})$. Continuing with this process, we
obtain a convex decomposition of $\mu$ that places no positive weight
either on elements on $L_0$ or on elements of $K_1\setminus K_{01}$,
yielding the lemma statement.%
\Halmos\endproof

The second lemma uses this result to establish that $\conv(\calP_1)$
is a convex polytope with extreme points in the set $K_1 \cup K_{01}$.
\begin{lemma}\label{lem:conv}
  Under Assumption~\ref{as:convex},
  $\conv(\calP_1) = \conv(K_1 \cup K_{01})$.
\end{lemma}
\proof{Proof.} 
We have $K_1 \subseteq \calP_1$, and furthermore, as
  $\cal{P}_1$ is closed, we have $K_{01} \subseteq \calP_1$. Hence, we
  obtain that $\conv(K_1 \cup K_{01}) \subseteq \conv(\calP_1)$. Thus,
  to prove the lemma statement, we must show
  $\conv(\calP_1) \subseteq \conv(K_1 \cup K_{01})$. Morever, since
  $\conv(\calP_1)$ is the smallest convex set containing $\calP_1$, it
  suffices to show $\calP_1 \subseteq \conv(K_1 \cup K_{01})$.

Let $\mu \in \calP_1$. By Lemma~\ref{lem:omega},
  $\mu \in \conv(L_0 \cup K_{01})\cup \conv(K_1 \cup K_{01})$. Suppose
  for the sake of contradition, $\mu \notin \conv(K_1 \cup
  K_{01})$. Then, it follows that $\mu \in \conv(L_0 \cup K_{01})$ and
  $\mu \notin \conv(K_{01})$. Consider any convex decomposition of
  $\mu$ as follows:
  \begin{align*}
    \mu = \sum_{\omega \in L_0} \alpha_\omega^0 \omega  + \sum_{\omega \in
    K_{01} \cap K_1}\alpha_\omega^0 \omega  + \sum_{\phi \in
    K_{01} \cap K_1^c}\alpha_\phi^0 \phi.
  \end{align*}
  Define $H_0^0 = \{ \omega \in L_0 : \alpha_\omega^0 > 0\}$,
  $H_0^1 = \{ \omega \in K_{01} \cap K_1 : \alpha_\omega^0 > 0\}$, and
  $H_0^2 = \{ \phi \in K_{01} \cap K_1^c : \alpha_\phi^0 > 0\}$.  Since
  $\mu \notin \conv(K_{01})$, it follows that
  $\sum_{\omega \in L_0 } \alpha_\omega^0 > 0$. Thus, we obtain that
  $H_0^0$ is non-empty.

In the following, we inductively define sets
  $\{H_n^i : i = 0,1,2\}$ for $n \geq 1$ as long as both $H_{n-1}^0$
  and $H_{n-1}^1$ are non-empty, such that the following four
  properties hold: (1) $H_n^0 \subseteq L_0$; (2)
  $H_n^1 \subseteq K_{01} \cap K_1$; (3) for any
  $\chi(\omega_0, \omega_1) \in H_n^2 \subseteq K_{01}$, either
  $\chi(\omega_0, \omega_1) \in K_{01} \cap K_1^c$ or
  $\omega_0 \in H_n^0$; and (4) $\mu$ is a strict convex combination
  of elements in $H_n^0 \cup H_n^1 \cup H_n^2$. Towards that end,
  first note that the sets $\{H_0^i : i =0,1,2\}$ satisfy the
  aforementioned four properties.

  Next, suppose for some $n \geq 1$ the sets
  $\{H_{n-1}^i : i =0,1,2\}$ satisfy the four properties, with both
  $H_{n-1}^0$ and $H_{n-1}^1$ being non-empty. Consider a strict
  convex decomposition of $\mu$ in terms of elements in
  $\cup_{i=0}^2 H_{n-1}^i$ with coefficients $\alpha^{n-1}_\phi > 0$
  for $\phi \in \cup_{i=0}^2 H_{n-1}^i$. Choose some
  $\omega_0 \in H_{n-1}^0$ and $\omega_1 \in H_{n-1}^1$, and let
  $\beta_{n-1} = \alpha_{\omega_0}^{n-1} / (\alpha_{\omega_0}^{n-1} +
  \alpha_{\omega_1}^{n-1})$. Observe that
  $\beta_{n-1} \omega_0 + (1- \beta_{n-1})\omega_1$ either (1) equals
  $\chi(\omega_0, \omega_1)$; or (2) is a strict convex combination of
  $\omega_0$ and $\chi(\omega_0, \omega_1)$; or (3) is a strict convex
  combination of $\chi(\omega_0, \omega_1)$ and $\omega_1$. We split
  the analysis into the three respective cases:
  \begin{enumerate}
    \item  If $\beta_{n-1} \omega_0 + (1- \beta_{n-1})\omega_1$ equals
  $\chi(\omega_0, \omega_1)$, let
  $H_n^i = H_{n-1}^i \setminus \{ \omega_i\}$ for $i \in \{0,1\}$ and
  $H_n^2 = H_{n-1}^2 \cup \{ \chi(\omega_0, \omega_1)\}$. Since
  $\alpha_{\omega_i}^{n-1} > 0$ for $i \in \{0,1\}$, we have
  $\beta_{n-1} \in (0,1)$ and hence
  $\chi(\omega_0, \omega_1) \in K_{01} \cap K_1^c$. Finally, letting
  $\alpha^n_\phi = \alpha^{n-1}_\phi$ for all
  $\phi \in \cup_{i=0}^2 H_n^i \setminus \{\chi(\omega_0, \omega_1)\}$
  and
  $\alpha^n_\phi = (\alpha_\phi^{n-1} + \alpha^{n-1}_{\omega_0} +
  \alpha^{n-1}_{\omega_1}) > 0$ for $\phi = \chi(\omega_0, \omega_1)$,
  we obtain a strict convex combination of $\mu$ in terms of elements
  of $\cup_{i=0}^2 H_n^i$.  Thus, the four properties continue to hold
  for $\{ H_n^i : i=0,1,2\}$.

\item If $\beta_{n-1} \omega_0 + (1- \beta_{n-1})\omega_1$ is a strict
  convex combination of $\omega_0$ and $\chi(\omega_0, \omega_1)$, let
  $H_n^0 = H_{n-1}^0$, $H_n^1 = H_{n-1}^1 \setminus \{ \omega_1\}$ and
  $H_n^2 = H_{n-1}^2 \cup \{ \chi(\omega_0, \omega_1)\}$. Note that
  properties (1) and (2) hold trivially, and since
  $\omega_0 \in H_n^0$, property (3) continues to hold. Using the
  strict convex combination of
  $\beta_{n-1} \omega_0 + (1- \beta_{n-1})\omega_1$ in terms of
  $\omega_0$ and $\chi(\omega_0,\omega_1)$, we obtain a strict convex
  combination of $\mu$ in terms of elements of $\cup_{i=0}^2 H_n^i$,
  and hence property (4) also holds.

\item If $\beta_{n-1} \omega_0 + (1- \beta_{n-1})\omega_1$ is a strict
  convex combination of $\chi(\omega_0, \omega_1)$ and $\omega_1$, let
  $H_n^0 = H_{n-1}^0 \setminus \{\omega_0\}$, $H_n^1 = H_{n-1}^1$ and
  $H_n^2 = H_{n-1}^2 \cup \{ \chi(\omega_0, \omega_1)\}$. Again,
  properties (1) and (2) hold trivially. Since
  $\beta_{n-1} \omega_0 + (1- \beta_{n-1}) \omega_1$ with
  $\beta_{n-1} \in (0,1)$ is a strict convex combination of
  $\chi(\omega_0, \omega_1)$ and $\omega_1$, it follows that
  $\chi(\omega_0, \omega_1) \neq \omega_1$ and hence
  $\chi(\omega_0, \omega_1) \in K_{01} \cap K_1^c$. Thus property (3)
  holds. Finally, using the strict convex combination of
  $\beta_{n-1} \omega_0 + (1- \beta_{n-1})\omega_1$ in terms of
  $\chi(\omega_0,\omega_1)$ and $\omega_1$, we obtain a strict convex
  combination of $\mu$ in terms of elements of $\cup_{i=0}^2 H_n^i$,
  and hence property (4) also holds.
\end{enumerate}



  Note that in all three cases, we have
  $|H_n^0| + |H_n^1| < |H_{n-1}^0| + |H_{n-1}^1|$. Thus, this
  inductive process stops with sets $H_n^i , i=0,1,2$ for some
  $n \geq 0$ with either $H_n^0$ or $H_n^1$ empty.  If $H_n^0$ is
  empty, then
  $\mu \in \conv(H_n^1 \cup H_n^2) \subseteq \conv(K_{01})$,
  contradicting the assumption that
  $\mu \notin \conv(K_{01} \cup K_1)$. Thus, it must be that $H_n^1$
  is empty.  Furthermore, consider any
  $\phi = \chi(\omega_0, \omega_1) \in H_n^2$ for which
  $\omega_0 \notin H_n^0$. By property (3), it must be that
  $\chi(\omega_0, \omega_1) \in K_{01} \cap K_1^c$. Choose any
  $\omega' \in H_n^0$, and define
  $\beta = \alpha_{\omega'}^n/(\alpha_{\omega'}^n +
  \alpha_\phi^n)$. Since
  $\chi(\omega_0, \omega_1) \in K_{01} \cap K_1^c$ and
  $\omega' \in H_n^0$, it is straightforward to verify that
  $\beta \omega' + (1 - \beta) \chi(\omega_0, \omega_1)$ can be
  written as a strict convex combination of $\omega_0$, $\omega'$,
  $\chi(\omega_0, \omega_1)$ and $\chi(\omega', \omega_1)$. Using such
  a strict convex combination and adding $\omega_0 \in L_0$ to the set
  $H_n^0$ (and if needed, adding $\chi(\omega', \omega_1)$ to
    $H_n^2$), we can without loss of generality assume that for any
  $\phi = \chi(\omega_0, \omega_1) \in H_n^2$ we have
  $\omega_0 \in H_n^0$. We thus obtain the following strict convex
  decomposition:
  \begin{align*}
    \mu = \sum_{\omega \in H_n^0} \gamma_\omega \omega +
    \sum_{\phi \in H_n^2} \gamma_\phi \phi,
  \end{align*}
  where $\sum_{\omega \in H_n^0} \gamma_\omega >0 $, and for which if
  $\chi(\omega_0, \omega_1) \in H_n^2$ then $\omega_0 \in H_n^0$.

This can be further rewritten as the following convex
  combination
  \begin{align*}
    \mu = \sum_{\omega \in H_n^0} \hat{\gamma}_\omega \omega
    + \sum_{\chi(\omega_0, \omega_1) \in H_n^2} (\tilde{\gamma}_{\omega_0} \omega_0
    + \tilde{\gamma}_{\chi(\omega_0, \omega_1)} \chi(\omega_0, \omega_1)),
  \end{align*}
  where if $\tilde{\gamma}_{\chi(\omega_0, \omega_1)} > 0$, then
  $\tilde{\gamma}_{\omega_0} > 0$. Note that for any such
  $(\omega_0, \omega_1)$, by definition of $\chi(\omega_0, \omega_1)$,
  the belief
  $\xi(\omega_0, \omega_1) \defeq \frac{1}{ (\tilde{\gamma}_{\omega_0}
    + \tilde{\gamma}_{\chi(\omega_0, \omega_1)}) }\left(
    \tilde{\gamma}_{\omega_0} \omega_0 +
    \tilde{\gamma}_{\chi(\omega_0, \omega_1)} \chi(\omega_0,
    \omega_1)\right)$ lies in the set $\calP_1^c$. Thus, $\mu$ is a
  convex combination of elements in $H_n^0 \subseteq L_0$ and the
  elements $\{\xi(\omega_0, \omega_1)\}$, all of which belong to
  $\calP_1^c$. From Assumption~\ref{as:convex}, we then obtain that
  $\mu$ itself is an element of $\calP_1^c$, contradicting the fact
  that $\mu \in \calP_1$. This proves that our initial assumption that
  $\mu \notin \conv(K_1 \cup K_{01})$ must be false, and hence
  $\mu \in \conv(K_1\cup K_{01})$.

Thus, $\mu \in \calP_1$ implies
  $\mu \in \conv(K_1 \cup K_{01})$, and hence
  $\calP_1 \subseteq \conv(K_1\cup K_{01})$. Thus, we conclude that
  $\conv(\calP_1) = \conv(K_1\cup K_{01})$.\Halmos\endproof



With these two helper lemmas in place, we are now ready to prove
Theorem~\ref{thm:linear}.  \proof{Proof of Theorem~\ref{thm:linear}.}
Since the objectives of the programs~\eqref{eq:sender-opt-convex}
and~\eqref{eq:linear} are identical, to prove the result it suffices
to show that optimal solution of each program is a feasible for the
other.

First, consider any optimal solution to \eqref{eq:linear}.  Since
$ \conv(L_0 \cup \{\zero\}) \subseteq \conv(\calP_0 \cup \{\zero\})$
and
$ \conv(K_1 \cup K_{01} \cup \{\zero\}) \subseteq \conv(\calP_1 \cup
\{\zero\})$, it is a feasible solution to
\eqref{eq:sender-opt-convex}.
  
Next, consider any optimal solution $t = (t_0, t_1)$ to
  \eqref{eq:sender-opt-convex} with
  $t_1 \in \conv(\calP_1 \cup \{\zero\})$ and
  $t_0 \in \conv(\calP_0 \cup \{\zero\})$. By Lemma~\ref{lem:conv}, it
  follows that $t_1 \in \conv(K_1 \cup K_{01} \cup \{\zero\})$. If
  $t_0 = \zero$, then $t_0 \in \conv(L_0 \cup \{\zero\})$ and we are
  done. Instead, suppose $t_0 \neq \zero$. Let $t_0 = b_0 m_0$ where
  $b_0 \in (0,1]$ and $m_0 \in \conv(\calP_0)$. By
  Lemma~\ref{lem:omega}, we obtain
  $m_0 \in \conv(L_0 \cup K_{01}) \cup \conv(K_1 \cup K_{01})$. If
  $m_0 \in \conv(K_1 \cup K_{01})$, then the solution
  $\hat{t} = (\hat{t}_0 , \hat{t}_1)$ with $\hat{t}_0 = 0$ and
  $\hat{t}_1 = t_1 + t_0$ is feasible for \eqref{eq:sender-opt-convex}
  and achieves larger utility for the sender than $(t_0, t_1)$,
  contradicting the latter's optimality. Hence,
  $m_0 \in \conv(L_0 \cup K_{01})$. If $m_0 \notin \conv(L_0)$, then
  $m_0 = (1 -\alpha) \nu_0 + \alpha \nu_1$ with $\nu_0 \in \conv(L_0)$
  and $\nu_1 \in \conv(K_{01})$ and $\alpha > 0$. However, we then
  have that $\tilde{t} = (\tilde{t}_0, \tilde{t}_1)$ with
  $\tilde{t}_0 = b_0 (1- \alpha) \nu_0$ and
  $\tilde{t}_1 = t_ 1 + b_0 \alpha \nu_1$ is feasible for
  \eqref{eq:sender-opt-convex} and has larger utility for the sender
  than $(t_0,t_1)$, once again contradicting the latter's
  optimality. Thus, we must have $m_0 \in \conv(L_0)$ and hence
  $t_0 \in \conv(L_0 \cup \{\zero\})$. Taken together, we obtain that
  $(t_0 , t_1)$ is feasible for \eqref{eq:linear}.  \Halmos\endproof

  Finally, we conclude this section with the proof of
  Proposition~\ref{prop:threshold-binary}.

  \proof{Proof of Proposition~\ref{prop:threshold-binary}.} First,
  note that if $v(\omega, 1) - v(\omega, 0) = v>0$ for all
  $\omega \in \Omega$, then for any feasible $t = (t_0, t_1)$
  for~\eqref{eq:linear}, we have
  $\sum_{\omega \in \Omega} v(\omega, 1)t_1(\omega) + v(\omega, 0)
  t_0(\omega) = \sum_{\omega \in \Omega} v(\omega, 0) \mu^*(\omega) +
  v \sum_{\omega \in \Omega} t_1(\omega)$. Thus, it suffices to show
  that for any $t = (t_0, t_1)$ that does not have the threshold
  structure, one can find another feasible
  $\tilde{t} = (\tilde{t}_1 , \tilde{t}_0)$ that satisfies
  $\sum_{\omega \in \Omega} t_1(\omega) < \sum_{\omega \in \Omega}
  \tilde{t}_1(\omega)$.

  Toward that end, let $t = (t_0, t_1)$ be a feasible solution
  to~\eqref{eq:linear} such that there exists
  $\omega^\dagger, \omega' \in \Omega$ satisfying
  $\omega^\dagger \prec \omega'$ with
  $t_1(\omega^\dagger) < \mu^*(\omega^\dagger)$ and $t_1(\omega') > 0$. Since
  $t_1 + t_0 = \mu^*$, we obtain $t_0(\omega^\dagger) > 0$. As
  $t_0 \in \conv(L_0 \cup \{\zero\})$, this implies that
  $\omega^\dagger \in L_0$.  By Assumption~\ref{as:monotone}, this
  implies that $\omega' \in L_0$ as well.

  Now, $t_1 \in \conv(K_1 \cup K_{01} \cup \{\zero\})$ implies that
  \begin{align*}
    t_1 = \sum_{j \in K_1} \alpha_j e_j +  \sum_{\phi \in K_{01}\setminus K_1} \alpha_\phi \phi,
  \end{align*}
  with $\alpha_j, \alpha_\phi \geq 0$ and
  $\sum_{j \in K_1} \alpha_j + \sum_{\phi \in K_{01}\setminus K_1}
  \alpha_\phi \leq 1$.  Since $t_1(\omega') > 0$, we deduce that there
  exists an $\hat{\omega} \in K_1$ such that $\alpha_\phi >0$ for
  $\phi = \chi(\omega',\hat{\omega})$. For fixed small enough
  $\epsilon > 0$, define $\tilde{t} = (\tilde{t}_0, \tilde{t}_1)$ as
  \begin{align*}
    \tilde{t}_1 &= t_1 + \epsilon \chi(\omega^\dagger,\hat{\omega}) - \epsilon' \chi(\omega',\hat{\omega})\\
    \tilde{t}_0 &= t_0 - \epsilon \chi(\omega^\dagger,\hat{\omega}) + \epsilon' \chi(\omega',\hat{\omega}),
  \end{align*}
  where $\epsilon' > 0$ satisfies
  $\epsilon' (1 - \gamma(\omega', \hat{\omega})) = \epsilon (1-
  \gamma(\omega^\dagger,\hat{\omega}))$. Clearly, we have
  $\tilde{t}_1 + \tilde{t}_0 = t_1 + t_0 = \mu^*$. Furthermore, for
  small enough $\epsilon > 0$, we have
  $\tilde{t}_1 \in \conv(K_1 \cup K_{01} \cup \{\zero\})$. Finally,
  using the definition of $\chi(\omega^\dagger, \hat{\omega})$ and
  $\chi(\omega', \hat{\omega})$ and the choice of $\epsilon'$, we have
  \begin{align*}
    \tilde{t}_0 &= t_0 - \epsilon   \gamma(\omega^\dagger,\hat{\omega}) e_{\omega^\dagger}  + \epsilon'   \gamma(\omega',\hat{\omega}) e_{\omega'}.
  \end{align*}
  Since $t_0 \in \conv(L_0 \cup \{\zero\})$ with
  $t_0(\omega^\dagger) > 0$, and $\omega' \in L_0$, we obtain that
  $\tilde{t}_0 \in \conv(L_0 \cup \{\zero\})$ for small enough
  $\epsilon >0$. Thus, putting it all together, for small enough
  $\epsilon >0$, $\tilde{t}$ is feasible for~\eqref{eq:linear}.

  Finally, we have
  \begin{align*}
    \sum_{\omega \in \Omega} \tilde{t}_1(\omega)
    &= \sum_{\omega \in \Omega} t_1(\omega) + \epsilon \gamma(\omega^\dagger, \hat{\omega}) - \epsilon' \gamma(\omega', \hat{\omega})\\
    &= \sum_{\omega \in \Omega} t_1(\omega)  + \epsilon (1- \gamma(\omega^\dagger, \hat{\omega})) \left(   \frac{\gamma(\omega^\dagger,\hat{\omega})}{1- \gamma(\omega^\dagger, \hat{\omega})}   -   \frac{\gamma(\omega',\hat{\omega})}{1 - \gamma(\omega',\hat{\omega})}  \right)\\
    &> \sum_{\omega \in \Omega} t_1(\omega),
  \end{align*}
  where we have used the fact that since $\omega^\dagger \prec \omega'$,
  we have
  $\gamma(\omega^\dagger, \hat{\omega}) > \gamma(\omega',\hat{\omega})$.
  \Halmos\endproof

\section{Proofs from Section~\ref{sec:signaling-queues}}%
\label{ap:signaling-queues}

In this section, we provide the proofs of the results from
Section~\ref{sec:signaling-queues}. To begin, let
$\Omega = \{0, 1, \ldots, C-1\}$ denote the set of queue-lengths for
which an arriving customer could possibly join the queue. (We assume
that immediately upon arrival a customer is informed whether the
queue-length equals $C$ or not.) Let $X$ denote the waiting time until
service completion faced by an arriving customer if she decides to
join the queue. Under the belief $e_n$, an arriving customer knows
that there are exactly $n$ customers already in queue, and thus $X$
has the distribution of the sum of $n+1$ independent unit exponential
random variables.

Recall the definition~\eqref{eq:mean-stdev-rho} of the differential
utility function:
$\hat{\rho}(\mu) = \tau - \left(\expec_\mu[X] + \beta
  \sqrt{\var_\mu[X]}\right)$ for some $\tau >0$ and $\beta \geq 0$. It
is straightforward to verify that
$\hat{\rho}(e_n) = \tau - \left(n+1 + \beta \sqrt{n+1}\right)$, and
hence $\hat{\rho}(e_n)$ is strictly decreasing in $n$. As
$K_1 = \{ n \in \Omega : \hat{\rho}(e_n) \geq 0\}$, and
$L_0 = \{ n \in \Omega : \hat{\rho}(e_n) < 0\}$, we obtain that
$K_1 = \{0, 1, \ldots,M\}$ and $L_0 = \{n \in \Omega: n > M\}$ for
some $M < C$.

For $n, m \in \Omega$ and $\gamma \in [0,1]$, the belief
$\mu = \gamma e_n + (1-\gamma) e_m$ assigns probability $\gamma$ to
the state with $n$ customers already in queue, and probability
$1-\gamma$ to the state with $m$ customers already in queue;
consequently, under the belief $\mu$, the distribution of $X$ is a
convex mixture (with weight $\gamma$) of its distribution under
beliefs $e_n$ and $e_m$. Thus, we obtain
$\expec_\mu[X] = 1+ m + (n-m)\gamma $ and
$\var_\mu[X] = 1+ m + (n-m)\gamma + (n-m)^2 \gamma (1-\gamma)$.
Finally, for $n \in L_0$ and $m \in K_1$,
$\gamma(n,m) = \sup\{ \gamma \in [0,1] : \hat{\rho}(\gamma e_n +
(1-\gamma) e_m) \geq 0\}$ is the largest value of $\gamma \in [0,1]$
for which $\gamma e_n + (1-\gamma)e_m \in \calP_1$, and
$\chi(n,m) = \gamma(n,m) e_n + (1- \gamma(n,m))e_m$ is the
corresponding belief. Note that under the belief $\chi(n,m)$, the
customer is indifferent between joining and leaving the queue. Using
the expression for $\hat{\rho}$ and after some straightforward
algebra, we obtain for $n \in L_0$ and $m \in K_1$,
\begin{align}\label{eq:gamma-exp}
  \gamma(n,m) = \frac{ 2  (\tau - 1 -m) + \beta^2 (n-m+1)  - \beta\sqrt{h(n,m)}}{2 (n-m) (1+\beta^2) },
\end{align}
where
$h(n,m) \defeq \beta^2 (n-m+1)^2 + 4 (\tau - 1 -m)(n-m+1) + 4
(1+\beta^2) ( 1 + m) -4 (\tau - 1 -m)^2$.

The following result implies that differential utility function
$\hat{\rho}(\cdot)$ defined in \eqref{eq:mean-stdev-rho} is convex and
monotone (Assumptions~\ref{as:convex} and~\ref{as:monotone}).
\begin{lemma}\label{lem:meanstdv}  For $\tau >0, \beta \geq 0$, let
    $\hat{\rho}(\mu) = \tau - \left(\expec_\mu[X] + \beta
      \sqrt{\var_\mu[X]}\right)$. Then,
    \begin{enumerate}
    \item The differential utility function $\hat{\rho}(\mu)$ is
      convex in $\mu$, and hence satisfies Assumption~\ref{as:convex}.
    \item Under the usual total order on $\Omega$, $\hat{\rho}(\mu)$
      satisfies the monotonicity condition in
      Assumption~\ref{as:monotone}.
    \end{enumerate} 
\end{lemma}
\proof{Proof.} From the fact that expectation is linear in the belief,
it is straightforward to show that
$\var_\mu[X] = \expec_\mu\left(X^2\right) -
\left(\expec_\mu(X)\right)^2$ is concave in the belief $\mu$. Since
$\sqrt{x}$ is concave and strictly increasing in $x$, using Jensen's
inequality, we obtain that $\sqrt{\var_\mu[X]}$ is also concave in
$\mu$. From these facts, we conclude that
$\hat{\rho}(\mu) = \tau - \left(\expec_\mu[X] + \beta
  \sqrt{\var_\mu[X]}\right) $ is convex in $\mu$. Thus, we obtain that
the set $\calP_a^c = \{ \mu : \hat{\rho}(\mu) < 0\}$ is convex, and
hence $\hat{\rho}(\cdot)$ satisfies Assumption~\ref{as:convex}.

Since $K_1 = \{0, 1, \ldots,M\}$ and
  $L_0 = \{n \in \Omega: n > M\}$, if $m \in K_1$ then $m < n$ for all
  $n \in L_0$.  Next, let $m \in K_1$ and $n, \ell \in L_0$ with
  $n < \ell$. For $\gamma \in (0,1]$, let
  $\mu = \gamma e_n + (1-\gamma) e_m$ and
  $\mu' = \gamma e_\ell + (1- \gamma)e_m$. Since
  $\expec_{\mu}[X] < \expec_{\mu'}[X]$ and
  $\var_{\mu}[X] < \var_{\mu'}[X]$, we obtain that
  $\hat{\rho}(\mu) > \hat{\rho}(\mu')$ for all $\gamma \in
  (0,1]$. Thus, we obtain that $\gamma(n, m) > \gamma(\ell,m)$.  To
  summarize, the usual order on $\Omega$ satisfies the property that
  if $n < \ell$, then either $n \in K_1$ or for all $m \in K_1$, we
  have $\gamma(n,m) > \gamma(\ell,m)$. Thus, we obtain that
  $\hat{\rho}(\cdot)$ satisfies the monotonicity condition
  (Assumption~\ref{as:monotone}).  \Halmos\endproof

\subsection{Proof of Lemma~\ref{lem:thresh}}\label{ap:proof-thresh}

Our proof follows a similar approach as in \citet{lingenbrinkI19}: we
show that any feasible solution where the customers' actions do not
have a threshold structure can be perturbed to obtain another feasible
solution corresponding to a signaling scheme with higher throughput.

\proof{Proof of Lemma~\ref{lem:thresh}.} First, note that if
$K_1 = \Omega$. i.e., $M=C-1$, then joining the queue is always
optimal. In this case, the full-information mechanism (a mechanism
inducing a threshold structure on the actions) has the highest
throughput. Hence, hereafter we focus on the case where $K_1$ is a
strict subset of $\Omega$, i.e, $M <C-1$.

Consider a feasible solution $t = (t_1, t_0)$ to the linear
program~\eqref{eq:queue-linear}. Since
$t_0 \in \conv(L_0 \cup \{\zero\})$ and $L_0 = \{M+1, \ldots, C-1\}$,
we have $t_0(k) = 0$ for all $k \leq M$. Then the
  constraint~\eqref{eq:queue-balance} implies that
  $t_1(k+1) = \lambda t_1(k)$ for all $k < M$, and hence
  $t_1(k) = \lambda^{k} t_1(0)$ for all $k \leq M$.

Let $N$ be the largest value of $n$ such that
$t_1(k) = \lambda^k t_1(0)$ for all $k \leq n$. (Note the preceding
argument implies $N \geq M$.) If either $N \geq C-2$ or
$t_1(N+2) = 0$, then we obtain that $t$ induces a threshold structure
in the customers' actions. Hence, for the rest of the proof assume
that $N < C-2$ and $t_1(N+2)>0$.  Using~\eqref{eq:queue-balance}, we
then obtain $0 < t_1(N+1) < \lambda t_1(N)$, $t_0(N+1) > 0$ and
$t_0(n) = 0$ for $n \leq N$. We now show that $t = (t_1, t_0)$ cannot
be an optimal solution to~\eqref{eq:queue-linear}. We do this by
constructing another feasible solution
$\tilde{t} = (\tilde{t}_1, \tilde{t}_0)$ that has a larger objective value
than $t$.

For some small $\epsilon >0$ and $\beta > 0$ to be chosen later,
consider $\tilde{t} = (\tilde{t}_1, \tilde{t}_0)$ with
\begin{align*}
  \tilde{t}_1 &= \frac{1}{Z} \left(t_1 - \beta \sum_{m=N+2}^{C-1} t_1(m) e_m + (\beta + \epsilon)  \sum_{m=N+2}^{C-1} t_1(m) e_{N+1}\right)
\end{align*}
and $\tilde{t}_0(0) = 0$,
$\tilde{t}_0(n) = \lambda \tilde{t}_1(n-1) - \tilde{t}_1(n)$ for all
$n > 0$, where $Z$ is chosen to satisfy~\eqref{eq:queue-normalize}; it
is straightforward to verify that 
\begin{align*}
  Z =   1 +  \lambda \epsilon\sum_{m=N+2}^{C-1}t_1(m).
\end{align*}
In particular, for $\epsilon >0$, we obtain that $Z > 1$.

We begin by showing that $\tilde{t}$ is feasible for
\eqref{eq:queue-linear} for all small enough $\epsilon, \beta
>0$. Note that $\tilde{t}$ satisfies \eqref{eq:queue-balance} and
\eqref{eq:queue-normalize} by definition. Thus, we verify the
conditions \eqref{eq:queue-leave} and \eqref{eq:queue-join}.

First, note that for $\ell \leq N$, we have
$\tilde{t}_1(\ell) = \frac{1}{Z} t_1(\ell)$.  On the other hand, we have
$\tilde{t}_1(N+1)= \frac{1}{Z} (t_1(N+1) + (\beta + \epsilon)
\sum_{m=N+2}^{C-1}t_1(m))$ and
$\tilde{t}_1(\ell) = \frac{(1-\beta )}{Z} t_1(\ell)$ for $\ell > N+1$.
Using this and the definition of $\tilde{t}_0$, it is straightforward to
verify that for small enough $\epsilon, \beta > 0$, we have
$\tilde{t}_0 \geq 0$ and $\tilde{t}_0(0) = 0$ for all $\ell \leq
N$. Further using the fact that $\sum_{\ell} \tilde{t}_0(\ell) \leq 1$,
we obtain $\tilde{t}_0 \in \conv(L_0 \cup \{\zero\})$ for all small
enough $\epsilon, \beta > 0$, and hence $\tilde{t}$ satisfies
\eqref{eq:queue-leave}.

Next, since $t_1 \in \conv(K_1 \cup K_{01} \cup \{\zero\})$, we can
write the decomposition
\begin{align*}
  t_1 &= \sum_{j=0}^M \alpha(j) e_j + \sum_{j=0}^M\sum_{i =
        M+1}^{C-1} \delta(i,j) \chi(i,j),
\end{align*}
where $\alpha(j) \geq 0$ for $j \leq M$, $\delta(i,j) \geq 0$ for
$j \leq M < i$, and
$\sum_{j=0}^{M} \alpha(j) + \sum_{j=0}^M \sum_{i=M+1}^{C-1}
\delta(i,j) \leq 1$. Thus, we have
\begin{align*}
  Z \tilde{t}_1
  &=  t_1 - \beta \sum_{m=N+2}^{C-1} t_1(m) (e_m - e_{N+1}) + \epsilon  \sum_{m=N+2}^{C-1} t_1(m) e_{N+1}\\
  &= t_1  - \beta \sum_{m=N+2}^{C-1} \sum_{j=0}^M \delta(m, j) \gamma(m, j) (e_m - e_{N+1})\\
  &\quad + \epsilon   \sum_{m=N+2}^{C-1} \sum_{j=0}^M \delta(m, j) \gamma(m, j) e_{N+1}.
\end{align*}
Now, it can be readily verified that for $0 \leq j \leq M$ and
$m > N+1$,
\begin{align*}
  \gamma(m,j) (e_m - e_{N+1}) = \chi(m,j) - \left(\frac{\gamma(m,j)}{\gamma(N+1, j)}\chi(N+1,j) + \left(1 - \frac{\gamma(m,j)}{\gamma(N+1, j)}\right) e_j\right).
\end{align*}
Thus, after some algebra, we obtain
\begin{align*}
  Z \tilde{t}_1
  &= t_1 - \beta \sum_{m=N+2}^{C-1} \sum_{j=0}^M \delta(m, j) \chi(m,j) \\
  &\quad  +\beta \sum_{m=N+2}^{C-1} \sum_{j=0}^M \delta(m, j)  \left(  \frac{\gamma(m,j)}{\gamma(N+1, j)}\chi(N+1,j) + \left(1 - \frac{\gamma(m,j)}{\gamma(N+1, j)}\right) e_j   \right) \\
  &\quad + \epsilon  \sum_{m=N+2}^{C-1} \sum_{j=0}^M \delta(m, j) \gamma(m, j) e_{N+1}.
 \end{align*}
 Now, since $t_1 \in \conv(K_1 \cup K_{01} \cup \{\zero\})$, we have
 the first term
 $t_1 - \beta \sum_{m=N+2}^{C-1} \sum_{j=0}^M \delta(m, j) \chi(m,j)
 \in \conv(K_1 \cup K_{01} \cup \{\zero\})$ for small enough
 $\beta >0$. Moreover, since $\hat{\rho}(\cdot)$ satisfies
   Assumption~\ref{as:monotone} (Lemma~\ref{lem:meanstdv}), we have
   $\gamma(N+1,j) > \gamma(m, j)$ for all $j \leq M$ and $m > N+1$.
 Hence, for all small enough $\beta > 0$, we obtain that second term
 lies in the $\conv(K_1 \cup K_{01} \cup \{\zero\})$, but not in
 $\conv(K_{01} \cup \{\zero\})$. This in turn implies that the second
 and the third term together lie in
 $\conv(K_1 \cup K_{01} \cup \{\zero\})$ for small enough
 $\beta, \epsilon >0$. Taken together, this implies that $Z\tilde{t}_1$
 (and hence $\tilde{t}_1$) lies in the cone generated by
 $\conv(K_1 \cup K_{01} \cup \{\zero\})$. Since
 $\sum_{\omega} \tilde{t}_1(\omega) \leq 1$, we conclude that
 $\tilde{t}_1 \in \conv(K_1 \cup K_{01} \cup \{\zero\})$, and hence
 $\tilde{t}$ satisfies \eqref{eq:queue-join}.

 Having shown the feasibility of $\tilde{t}$, we now verify that the
 objective value under $\tilde{t}$ is greater than that under $t$. Using
 \eqref{eq:queue-normalize} and \eqref{eq:queue-balance}, we obtain
\begin{align*}
  \sum_{\ell = 0}^{C-1} \tilde{t}_1(\ell) &= \frac{1}{\lambda} \left(1 - \tilde{t}_1(0)\right) > \frac{1}{\lambda} \left(1 - t_1(0)\right) = \sum_{\ell = 0}^{C-1} t_1(\ell).
\end{align*}
Here, the inequality follows from the fact that
$\tilde{t}_1(0) = \frac{1}{Z} t_1(0) < t_1(0)$ since $Z > 1$. Thus, the
objective of \eqref{eq:queue-linear} for $\tilde{t}$ is strictly greater
than that for $t$. Thus, it follows that $t$ cannot be optimal for
\eqref{eq:queue-linear}.

To conclude the proof, we obtain from the preceding argument that for
any optimal $t = (t_1, t_0)$, if $N$ denotes the largest value of $n$
such that $t_1(k) = \lambda^k t_1(0)$ for all $k \leq n$, then either
$N \geq C-2$ or $t_1(N+2) = 0$. In either case, we obtain that $t$
induces a threshold structure in the customers' actions.
\Halmos\endproof

\subsection{Proof of Proposition~\ref{prop:sandwich}}\label{ap:proof-sandwich}

\proof{Proof.}  The optimal signaling scheme can be found from
  the optimal solution to the linear program~\eqref{eq:queue-linear},
  using an argument similar to the discussion
  surrounding~\eqref{eq:scheme-from-t}. Moreover, from similar
  argument as in Proposition~\ref{prop:signal-bound}, it follows that
  there exists an optimal signaling scheme that uses at most
  $|\Omega| = C$ signals per action. As $\hat{\rho}$ satisfies
  Assumption~\ref{as:convex} (Lemma~\ref{lem:meanstdv}), all the
  signals that induce the customer to leave $(a = 0)$ can be coalesced
  into single signal, which we call $\leave$.  Taken together, we
  obtain that there exists a signaling scheme with the signal set
  $\{\join_1, \join_2, \dots,\join_J, \leave\}$ for some $J \leq |C|$,
  where each $\join_i$ signal induces the customer to join the queue,
  and the $\leave$ signal induces the customer to leave.

  We will now show that there exists such an optimal signaling scheme
  that satisfies the three properties listed in the proposition
  statement. For ease of exposition, we will split the proof into
  three parts, one part for each property.

  \textbf{Part 1.} From the constraints \eqref{eq:queue-leave} and
  \eqref{eq:queue-join}, we observe that the optimal signaling scheme
  can be implemented with the canonical set of join signals consisting
  of pure signals in $ K_1$ and binary mixed signals in $K_{01}$.  The
  customer's utility upon receiving a binary mixed signal in $K_{01}$
  or a signal in $K_1 \cap K_0$ is zero by definition. Thus, to prove
  that a customer's utility upon receiving each signal $\join_j$ is
  zero, we need to prove that there exists a signaling scheme that
  never sends the pure signals in $K_1 \cap K_0^c$. To show this, we
  use a perturbation argument, and show that any canonical signaling
  scheme that sends a pure signal in $K_1 \cap K_0^c$ with positive
  probability must be sub-optimal.

  Let $t = (t_0,t_1)$ be a feasible solution
  to~\eqref{eq:queue-linear} with $t_0 \neq \zero$.  If $t$ does not
  have a threshold structure, then Lemma~\ref{lem:thresh} implies that
  $t$ cannot be optimal for~\eqref{eq:queue-linear}. Thus, henceforth
  we assume that $t$ has a threshold structure, i.e., there exists an
  $m \in L_0 \cup \{C\}$ such that $t_1(\omega) > 0$ for $\omega < m$,
  $t_1(\omega)= 0$ for $\omega \geq m+1$, and $t_0(\omega)=0$ for
  $\omega \notin \{m, m+1\}$. Furthermore, since
  $\sum_{\omega \in \Omega} t_0(\omega) > 0$, we obtain that $m < C$,
  and thus one can let $m \in L_0$ with $t_0(m) > 0$.

  Now, suppose $t_1 \in \conv(K_1 \cup K_{01} \cup \{\zero\})$ but
  $t_1 \notin \conv((K_1 \cap K_0) \cup K_{01} \cup \{\zero\})$. The
  former condition implies that we can write $t_1$ as
  \begin{align*}
    t_1 = \sum_{\omega_1 \in K_1} \beta_{\omega_1} e_{\omega_1} + \sum_{\phi \in K_{01}\setminus K_1} \beta_\phi \phi.
  \end{align*}
  where
  $\sum_{\omega_1 \in K_1} \beta_{\omega_1} + \sum_{\phi \in
    K_{01}\setminus K_1} \beta_{\phi} \leq 1$. Furthermore, since
  $t_1 \notin \conv((K_1 \cap K_0) \cup K_{01} \cup \{\zero\})$, it
  follows that there exists an $\omega_1^* \in K_1 \cap K_0^c$ such
  that $\beta_{\omega_1^*} > 0$. Using these facts, we will now
  construct a feasible $\hat{t}$ with strictly higher objective, thus
  implying that $t$ cannot be optimal.

  For small enough $\epsilon>0$, let
  $\hat{t} = (\hat{t}_0, \hat{t}_1)$ be defined as
  \begin{align*}
    \hat{t}_1 &= \frac{1}{Z} \left(t_1 + \epsilon e_m\right)\\
    \hat{t}_0 &= \frac{1}{Z} \left( t_0  - \epsilon e_m + \lambda \epsilon e_{m+1} \ind\{m \neq C-1\}\right),
  \end{align*}
  where $Z$ is chosen so that $\hat{t}$
  satisfies~\eqref{eq:queue-normalize}:
  $Z = 1 + \lambda \epsilon > 1$. It is straightforward to verify that
  $\hat{t}$ satisfies~\eqref{eq:queue-balance}. Furthermore, since
  $t_0(m) > 0$, we have for small enough $\epsilon >0$,
  $t_0 - \epsilon e_m \in \conv(L_0 \cup \{\zero\})$. Moreover, if
  $m \neq C-1$, then $e_{m+1} \in \conv(L_0 \cup \{\zero\})$. This
  implies that $\hat{t}_0$ itself lies in $\conv(L_0 \cup \{\zero\})$,
  and thus satisfies~\eqref{eq:queue-leave}.

  Thus, to show the feasibility of $\hat{t}$, it remains to show
  that~\eqref{eq:queue-join} holds.  We have
    \begin{align*}
      \hat{t}_1 &= \frac{1}{Z} \left(t_1 + \epsilon e_m\right)\\
                &= \frac{1}{Z} \left( \sum_{\omega_1 \in K_1, \omega_1 \neq \omega_1^*} \beta_{\omega_1} e_{\omega_1} + \sum_{\phi \in K_{01} \setminus K_1} \beta_{\phi} \phi\right) + \frac{1}{Z} \left( \beta_{\omega_1^*} e_{\omega_1^*} + \epsilon e_m\right).
    \end{align*}
    Now, since $Z >1$, the first term lies in the set
    $\conv(K_1 \cup K_{01} \cup \{\zero\})$. Furthermore, since
    $\omega_1^* \in K_1 \in K_0^c$ and $m \in L_0$, for small enough
    $\epsilon > 0$, the second term can be written as a convex
    combination of $\omega_1^*$, $\chi(m, \omega_1^*) \in K_{01}$ and
    $\zero$. Thus, we obtain that the second term lies in the set
    $\conv(K_1 \cup K_{01} \cup \{\zero\})$ as well. Finally, since
    $\sum_{\omega \in \Omega} \hat{t}_1 (\omega) = \frac{1}{Z}
    \left(\sum_{\omega \in \Omega} t_1(\omega) + \epsilon \right) =
    \frac{1}{Z} \left( 1 - \lambda t_1(C-1) - \sum_{\omega \in \Omega}
      t_0(\omega) + \epsilon\right) < 1$ for small enough
    $\epsilon >0$, we obtain that
    $\hat{t}_1 \in \conv(K_1 \cup K_{01} \cup \{\zero\})$. Thus,
    $\hat{t}$ is feasible for \eqref{eq:queue-linear}.
    
    Finally, note that,
    \begin{align*}
      \sum_{\omega \in \Omega} \hat{t}_1(\omega) -  \sum_{\omega \in \Omega} t_1(\omega)
      &= \frac{\epsilon}{Z}   \left( 1 - \lambda \sum_{\omega \in \Omega} t_1(\omega)\right) > 0,
    \end{align*}
    where the final inequality follows from the fact that the
    throughput $\lambda \sum_{\omega \in \Omega} t_1(\omega)$ must be
    strictly less than the mean service rate, assumed to equal one.
    Thus, we obtain that $\hat{t}$ achieves a strictly higher
    objective than $t$, and hence $t$ cannot be optimal
    for~\eqref{eq:queue-linear}. Summarizing, we obtain that any
    optimal $t = (t_0, t_1)$ with $ t_0 \neq \zero$ must satisfy
    $t_1 \in \conv((K_1 \cap K_0) \cup K_{01} \cup \{\zero\})$, and
    thus under the optimal signaling mechanism, for any signal that
    induces the customer to join, the induced beliefs are in the set
    $(K_1 \cap K_0) \cup K_{01}$, implying that her utility for
    joining must be zero.

    \textbf{Part 2.}  From the proof of part (1), it follows that for
    each $j \leq J$, the signal $\join_j$ either induces a posterior
    belief $\omega^j \in K_1 \cap K_0$ or a belief
    $\chi(\omega_0^j, \omega_1^j)$ between a pair of states
    $\omega_0^j \in L_0$ and $\omega_1^j \in K_1$. In the former case,
    we let $\omega_0^j = \omega_1^j = \omega^j$. We next show that the
    set of such state-pairs forms a sandwich structure, in the sense
    stated in the proposition statement.

    Consider two signals $\join_i$ and $\join_j$ inducing beliefs on
    state-pairs $(\omega_0^i, \omega_1^i) = (n,m)$ and
    $(\omega_0^j, \omega_1^j) = (\ell, k)$ respectively, with
    $m,k \in K_1$ and $n, \ell \in K_0$. Without loss of generality,
    we assume $m \leq k$.

    Now, if $n \in K_0 \cap K_1$, then $m = n$, and using the fact
    that $\hat{\rho}(e_i)$ is strictly decreasing, we obtain that
    $m= k$, and hence $k = m = n \leq \ell$, which conforms with the
    sandwich structure. Similarly, if $\ell \in K_0 \cap K_1$, then
    $k = \ell$, and hence $m \leq k = \ell \leq n$, which again
    conforms with the sandwich structure. Thus, hereafter we assume
    that $n, \ell \in K_0 \cap K_1^c = L_0$. Hence, we obtain
    $m \leq k < n$ and $m \leq k < \ell$.  Thus, we have three
    possible cases: (1) $m \leq k < \ell \leq n$; (2)
    $m = k < n < \ell$; and (3) $m < k < n < \ell$. The first two
    cases conform with the sandwich structure stated in the
    proposition. Thus, we focus on the third case and show that such a
    signaling scheme cannot be optimal.

    Thus, suppose $m < k < n < \ell$. We next show that any strict
    convex combination of $\chi(n,m)$ and $\chi(\ell, k)$ can be
    written as either a convex combination of $e_m, \chi(\ell,m)$,
    $\chi(\ell,k)$ and $\chi(n,k)$, or a convex combination of
    $e_m, \chi(\ell,m), \chi(n,m)$ and $\chi(n,k)$ with positive
    weight on $e_m$ in either case. For notational simplicity, we
    denote $\gamma(i,j)$ by $\gamma_{ij}$. For any $\alpha \in (0,1)$,
    we obtain from straightforward (but tedious) algebra,
    \begin{align}
      &\alpha \chi(n,m) + (1- \alpha) \chi(\ell, k)\notag\\
      &= \alpha \left( \gamma_{nm} e_n + (1 - \gamma_{nm})e_m \right) + (1- \alpha) \left( \gamma_{\ell k} e_\ell + (1 - \gamma_{\ell k}) e_k \right)\notag\\
      &= \alpha (1 - \gamma_{nm}) \left(1  -   \frac{\gamma_{nm} \gamma_{\ell k}  (1- \gamma_{nk})  (1 - \gamma_{\ell m})}{\gamma_{\ell m} \gamma_{nk} (1 - \gamma_{\ell k}) (1 - \gamma_{nm})}\right) e_m     +       \alpha \frac{\gamma_{nm} \gamma_{\ell k}  (1- \gamma_{nk})}{\gamma_{\ell m} \gamma_{nk} (1 - \gamma_{\ell k})} \chi(\ell,m)\notag\\
      &\quad + \frac{\gamma_{nm} (1- \gamma_{nk})}{\gamma_{nk} (1 - \gamma_{\ell k})} \left( (1- \alpha)\frac{\gamma_{nk} (1 - \gamma_{\ell k})}{\gamma_{nm} (1- \gamma_{nk})}    -   \alpha  \right)  \chi(\ell, k) +  \frac{\alpha\gamma_{nm}}{\gamma_{nk}}  \chi(n,k)  \label{eq:either-this}   \\
      &=   (1- \alpha)  (1 - \gamma_{nm})  \frac{  \gamma_{nk}  (1 - \gamma_{\ell k}) }{ \gamma_{nm} (1- \gamma_{nk}) }  \left( 1 -       \frac{\gamma_{nm}\gamma_{\ell k}(1- \gamma_{nk}) (1 - \gamma_{\ell m})}{\gamma_{\ell m} \gamma_{nk}  (1 - \gamma_{\ell k})  (1 - \gamma_{nm}) }     \right) e_m  + (1- \alpha)\frac{(1 - \gamma_{\ell k})}{(1- \gamma_{nk})} \chi(n,k)  \notag\\
      &\quad +   \left(\alpha     -   (1- \alpha)  \frac{ \gamma_{nk}(1 - \gamma_{\ell k})}{\gamma_{nm}(1- \gamma_{nk})} \right)  \chi(n,m)  + (1- \alpha)  \frac{\gamma_{\ell k}}{\gamma_{\ell m}}  \chi(\ell,m). \label{eq:or-that}
    \end{align}
    In Lemma~\ref{lem:log-supermodular}, we show that for
    $m < k < n < \ell$,
    $0 < \frac{\gamma_{nm}\gamma_{\ell k}(1- \gamma_{nk}) (1 -
      \gamma_{\ell m})}{\gamma_{\ell m} \gamma_{nk} (1 - \gamma_{\ell
        k}) (1 - \gamma_{nm}) } < 1$.  Thus, the coefficient of $e_m$
    in both~\eqref{eq:either-this} and~\eqref{eq:or-that} is
    positive. If
    $\frac{\alpha}{1- \alpha} \leq \frac{ \gamma_{nk}(1 - \gamma_{\ell
        k})}{\gamma_{nm}(1- \gamma_{nk})}$, then the coefficient of
    $\chi(\ell,k)$ in~\eqref{eq:either-this} is non-negative. On the
    other hand, if
    $\frac{\alpha}{1- \alpha} \geq \frac{ \gamma_{nk}(1 - \gamma_{\ell
        k})}{\gamma_{nm}(1- \gamma_{nk})}$, then the coefficient of
    $\chi(n,m)$ in~\eqref{eq:or-that} is non-negative. Thus, taken
    together, we obtain that for any $\alpha \in (0,1)$, the convex
    combination $\alpha \chi(n,m) + (1- \alpha) \chi(\ell, k)$ can be
    written as either a convex combination of
    $e_m, \chi(\ell,m), \chi(\ell, k)$ and $\chi(n,k)$ or a convex
    combination of $e_m, \chi(\ell,m), \chi(n,m)$ and $\chi(n,k)$,
    with positive weight on $e_m$ in either case.

    Now, since $m < k$ with $k \in K_1$, it follows that
    $m \in K_1 \cup K_0^c$.  This in turn implies that, using
    arguments as in the proof of Part (1), one can appropriately
    combine the weight on $e_m$ with an $\omega \in L_0$ to construct
    an additional signal which induces a posterior belief
    $\chi(m, \omega)$, leading to an increased probability of customer
    joining the queue. Thus, if $m < k < n < \ell$, the signaling
    scheme cannot be optimal.

    To conclude, we obtain that for any pair of signals $\join_i$ and
    $\join_j$ with $\omega_1^i \leq \omega_1^j$, it must be that
    $\omega_1^i \leq \omega_1^j \leq \omega_0^j \leq \omega_1^i$,
    implying the sandwich structure stated in the proposition
    statement.
    
    \textbf{Part 3.} Finally, we show that in the optimal signaling
    scheme, for $i < j$, we have
    $\expec[X|\join_i] \leq \bE[X|\join_j]$ and
    $\var[X|\join_i] \geq \var[X|\join_j]$. Since
    $\expec[X|\join_i] + \beta \sqrt{\var[X|\join_i]} =
    \expec[X|\join_j] + \beta \sqrt{\var[X|\join_j]} = \tau$ from Part
    (1), it suffices to prove
    $\expec[X|\join_i] \leq \expec[X|\join_j]$.

    To prove this, let $m, k \in K_1$ and $n, \ell \in K_0$ be such
    that $\join_i$ induces posterior beliefs over the state-pair
    $\{m,n\}$ and $\join_j$ induces beliefs over the pair
    $\{k, \ell\}$. The sandwich structure from Part (2) implies that
    $m \leq k \leq \ell \leq n$. Suppose $k < \ell$, implying that
    $\ell, n \in L_0$. We have
    $\expec[X|\join_i] = \gamma_{nm} (n+1) + (1- \gamma_{nm}) (m+1) =
    1 + \gamma_{nm}n + (1- \gamma_{nm}) m$ and
    $\expec[X|\join_j] = 1 + \gamma_{\ell k}\ell + (1- \gamma_{\ell
      k}) k$.  In Lemma~\ref{lem:gamma-monotone}, we show that for
    $m \in K_1$ and $n \in L_0$, the function
    $\gamma_{nm} n + (1 - \gamma_{nm}) m$ is non-decreasing in $m$
    (for fixed $n$) and non-increasing in $n$ (for fixed $m$). Thus,
    we have for $m \leq k < \ell \leq n$,
    \begin{align*}
      \expec[X|\join_i] = \gamma_{nm}n + (1- \gamma_{nm}) m \leq \gamma_{nk}n + (1- \gamma_{nk}) k \leq \gamma_{\ell k}\ell + (1- \gamma_{\ell k}) k = \expec[X|\join_j].
    \end{align*}
    On the other hand, if $k = \ell$, then $k= \ell \in K_1 \cap K_0$,
    implying that $n \in L_0$, and hence $m < k$ and
    $\gamma_{nk} = 0$. Once again, we have
    $\expec[X| \join_i] = \gamma_{nm} n + (1- \gamma_{nm})m \leq
    \gamma_{nk} n + (1 - \gamma_{nk}) k = k =
    \expec[X|\join_j]$.\Halmos\endproof

    We use the following two lemmas in the proof of
    Proposition~\ref{prop:sandwich}.
  \begin{lemma}\label{lem:log-supermodular} Let
    $\hat{\rho}(\mu) = \expec_\mu[X] + \beta \sqrt{\var_\mu(x)}$, and
    $m,k \in K_1$ and $n, \ell \in L_0$ be such that
    $m < k < n < \ell$. Then, we have
    \begin{align*}
      0 < \frac{\gamma_{nm}\gamma_{\ell k}(1- \gamma_{nk}) (1 -
      \gamma_{\ell m})}{\gamma_{\ell m} \gamma_{nk} (1 - \gamma_{\ell
      k}) (1 - \gamma_{nm}) } < 1.
    \end{align*}
  \end{lemma}
 \proof{Proof.}  Since $n, \ell \in L_0$, it follows from
  definition that the ratio in the lemma statement is positive. Thus,
  it remains to show that the ratio is strictly less than one. To
  simplify the notation, let
  $f(n,m) \defeq \log\left( \frac{\gamma_{nm}}{1 -
      \gamma_{nm}}\right)$. Then, the upper-bound condition in the
  lemma statement is equivalent to requiring
  \begin{align*}
    f(n,k) -  f(n,m) > f(\ell,k) -  f(\ell,m),
  \end{align*}
  i.e., that $f$ has decreasing differences. Now, using the expression
  for $\gamma_{nm}$ from~\eqref{eq:gamma-exp}, we obtain
    \begin{align*}
      f(n,m) &= \log\left( \frac{\gamma_{nm}}{1 -\gamma_{nm}}\right)\\
             &= \log\left( \frac{ 2  (\tau - 1 -m) + \beta^2 (n-m+1)  - \beta\sqrt{h(n,m)}}{2 (n-m) (1+\beta^2) - 2  (\tau - 1 -m) - \beta^2 (n-m+1)  + \beta\sqrt{h(n,m)}    }  \right).
    \end{align*}
    We use the preceding expression to extend the definition of $f$ to
    non-integer values of $m$ and $n$. Then, by a straightforward
    calculation, it follows that
    \begin{align*}
      \frac{\partial^2  f(n,m)}{\partial n \partial m}  = - \frac{2 \beta (\beta^2 +1) (n-m)}{ (h(n,m))^{3/2}} < 0.
    \end{align*}
    Thus, by the mean value theorem, there exists
    $\xi_1 \in [n, \ell]$ and $\xi_2 \in [m,k]$ with
    \begin{align*}
      \left(f(\ell,k) - f(\ell,m)\right) - \left(f(n,k) - f(n,m)\right) &= \left. \frac{\partial^2  }{\partial x \partial y} f(x,y) \right|_{x= \xi_1, y = \xi_2} (\ell - n) (k-m) <0. \Halmos
    \end{align*}
    \endproof

\begin{lemma}\label{lem:gamma-monotone} For $m \in K_1$ and
  $n \in L_0$, let
  $g(n,m) \defeq \gamma_{nm} n + (1 - \gamma_{nm}) m$. Then, $g(n,m)$
  is non-decreasing in $m$ (for fixed $n$), and non-increasing in $n$
  (for fixed $m$).
\end{lemma}
\proof{Proof.} Using~\eqref{eq:gamma-exp}, we obtain
\begin{align*}
  g(n,m) &= m + \gamma_{nm} (n-m) = m + \frac{ 2  (\tau - 1 -m) + \beta^2 (n-m+1)  - \beta\sqrt{h(n,m)}}{2 (1+\beta^2) },
\end{align*}
which implies
\begin{align*}
  \frac{\partial g(n,m)}{\partial n} &= \frac{\beta \left( \beta\sqrt{h(n,m)} + \beta^2 (m-n-1) +  2m -2\tau +2\right)}{2 (1+\beta^2)\sqrt{h(n,m)}},\\
  \frac{\partial g(n,m)}{\partial m} &= \frac{\beta \left(\beta\sqrt{h(n,m)} + \beta^2(n-m-1) +  2n -2\tau +2\right)  }{2 (1+\beta^2)\sqrt{h(n,m)}}.
\end{align*}
Using the definition of $h(n,m)$, we obtain
\begin{align*}
  \beta^2 h(n,m) - \left(\beta^2 (m-n-1) + 2m -2\tau +2\right)^2
  &= - 4 (1+\beta)^2 \left(  (\tau - m -1)^2  - \beta^2 (m +1)   \right)  \leq 0,\\
  \beta^2 h(n,m) - \left(\beta^2 (n-m-1) + 2n -2\tau +2\right)^2
  &= - 4 (1+\beta)^2 \left(  (\tau - n -1)^2  - \beta^2 (n +1)   \right)  \geq 0,
\end{align*}
where we make use of the fact that
$ m+1 + \beta \sqrt{m+1} \leq \tau < n+1 + \beta \sqrt{n+1}$ since
$m \in K_1$ and $n \in L_0$. Thus, we get
\begin{align*}
  \beta \sqrt{h(n,m)} &\leq | \beta^2 (m-n-1) + 2m -2\tau +2| = - \left(\beta^2 (m-n-1) + 2m -2\tau +2 \right)\\
  \beta \sqrt{h(n,m)} &\geq | \beta^2 (n-m-1) + 2n -2\tau +2 | \geq - \left(\beta^2 (n-m-1) + 2n -2\tau +2 \right),
\end{align*}
where the first line follows because
$\beta^2 (m-n-1) + 2m -2\tau +2 \leq 0$ as $\tau \geq m+1$ and
$m \leq n$. Taken together, we obtain
$\frac{\partial g(n,m)}{\partial n} \leq 0$ and
$\frac{\partial g(n,m)}{\partial m} \geq 0$.~\Halmos\endproof

  \section{Examples of Risk-conscious Utility}%
\label{ap:deviations-eum}

In this section, we discuss the evidence for systematic deviations
from expected utility maximization (EUM), using the examples of
risk-conscious utility provided in
Table~\ref{table:risk-conscious-examples}.

%
%
%

\subsection{Cumulative Prospect Theory}

As discussed in the introduction, behavioral economists have
documented a number of ``paradoxes'' that cannot be explained by
expected utility maximization. In particular, the Allais paradox
\citep{allais79} suggests that individuals choose a certain small gain
over a small chance of a large gain (risk-averse), but choose a small
chance of a large loss over a certain small loss (risk-seeking). They
also perceive events relative to a reference point, and overweight
extreme events but underweight typical events. Cumulative prospect
theory, introduced by \citet{tverskyK92}, incorporates this idea of
\textit{rank-dependent expected utility} by weighting the subjective
probability over outcomes differently from the true distribution, that
is, they replace the true weight $\mu(\omega)$ in the expected utility
$\sum_{\omega} \mu(\omega) u(\omega,a)$ with the subjective weight
$f_{\omega}(\mu)$, yielding a risk-conscious utility of the form
$\rho(\mu, a) = \sum_{\omega \in \Omega} f_\omega(\mu) u(\omega, a)$.

\subsection{Mean-standard-deviation utility}

The expression for mean-standard-deviation (mean-stdev) utility is
$\expec_{\mu}[u(\omega,a)] - \beta
\sqrt{\var_{\mu}(g(\omega,a))}$. There are two components to the
utility: the receiver wants the first metric $u$ to have high mean,
and the second metric $g$ to have low variability.

This functional specification of utility is often used when the agent
is uncertainty-averse but is more likely to base her decision on a few
summary statistics of the distribution (in this case, the first two
moments) rather than the entire distribution. The canonical Markovitz
portfolio theory in finance \citep{markowitz52, markowitz87} considers
the utility as a function of mean return and volatility captured by
standard deviation of return. There is a rich literature on traffic
assignment and routing with stochastic travel times
\citep{nikolovaS14, cominettiT16, lianeasNS18} using the mean-stdev
utility that also deals with strategic behavior of agents like our
work. The distribution of waiting time in queues in practical settings
is also often summarized with the mean and the standard deviation of
meeting time: see, e.g., \citet{angKBPA15} for emergency department
wait time prediction.

\subsection{Maximin Utility}


The expression for maximin utility is
$\rho(\mu, a) = \min_{\theta} \expec_{\mu}[u(\omega,a;\theta)]$. This
utility formulation arises when the agent faces uncertainty about the
utility parameter $\theta$, and seeks to hedge against the worst-case
realization.
%
%
%
The notion of using the maximin criterion for decision making has a
long history; see, e.g., \citet{french88} for a discussion. More
specifically, \citet{armbrusterD15} and \citet{huM15} propose and
analyze decision making problems with maximin utility in the form
stated here, and many other subsequent works in decision theory and
finance adopt it~\citep{postK17, huBM18,guoX21}.

\subsection{Risk Measures}

A risk measure is a numerical metric assigned to an uncertain event to
capture a potential loss in decision making problems under
uncertainty. Risk measures are often designed to capture the ``tail''
of the distribution of losses, so they are necessarily nonlinear in
the distribution (belief). Commonly used risk-measures are the
Value-at-Risk
$\mathsf{VaR} = \min\{t \in \mathbb{R}: \prob_{\mu}(\ell(\omega,a) >
t) \leq 1-\alpha\}$, and the Conditional Value-at-Risk
$\mathsf{CVaR} = \expec_{\mu}[ \ell(\omega,a) | \ell(\omega,a) >
\tau]$, where $\ell(\omega,a)$ is the loss under state $\omega$ and
action $a$.

%
%
%
%
%
%
%


\end{APPENDICES}

\ACKNOWLEDGMENT{The second and the third authors gratefully
  acknowledge support from the National Science Foundation under
  grants CMMI-1633920 and CMMI-2002156. Portions of this work were
  done when the second author was with Cornell University. A
  preliminary version of this work appeared as a one-page abstract at
  WINE 2019, and as a poster at the Workshop on Behavioral EC (2019);
  we thank the anonymous reviewers at these conferences for their
  feedback.}

\bibliographystyle{informs2014}
\bibliography{sources}

\end{document}